\documentclass{aa}  

\usepackage[dvipsnames]{xcolor}
\usepackage{color}

\usepackage{natbib}
\usepackage{graphicx}
\usepackage{txfonts}
\usepackage{amsmath}

\usepackage{chemmacros}

\usepackage{lscape}
\usepackage{ulem}

\usepackage{hyperref}

\hypersetup{
    colorlinks=true,
    linkcolor=MidnightBlue,
    filecolor=MidnightBlue,      
    urlcolor=MidnightBlue,
    citecolor=MidnightBlue
}

\colorlet{myred}{red!60!gray}
\colorlet{mygreen}{green!60!gray}
\colorlet{myblue}{blue!60!gray}

\newcommand{\arj}[1]%{\textcolor{blue}{{#1}}}
{\textbf{#1}}

\newcommand{\msun}{M$_\odot$}
\def\kms {km~s$^{-1}$}

\newcommand{\micron}{\textmu m}
\begin{document}

   \title{The molecular chemistry of Type Ibc supernovae and diagnostic potential with the James Webb Space Telescope}
   
    \titlerunning{The molecular chemistry of Type Ibc supernovae}

   \author{S. Liljegren\inst{1}
   \and
   A. Jerkstrand\inst{1} 
   \and 
   P. S. Barklem\inst{2} 
   \and
   G. Nyman\inst{3}
   \and
   R. Brady\inst{4}
   \and 
   S. N. Yurchenko\inst{4}
}
   \institute{The Oskar Klein Centre, Department of Astronomy, Stockholm University, Albanova 10691, Stockholm, Sweden \label{inst1}
   \\ \email{sofie.liljegren@astro.su.se}
        \and 
        Theoretical Astrophysics, Department of Physics and Astronomy, Uppsala University, Box 516, SE-751 20 Uppsala, Sweden \label{inst2}
        \and 
        Department of Chemistry and Molecular Biology, University of Gothenburg, SE-41296 Gothenburg, Sweden\label{inst3}
       % Max Planck Institute for Astrophysics, Karl-Schwarzschild-Straße 1, 85748 Garching, Germany \label{inst2}
        %\and
        \and
        Department of Physics and Astronomy, University College London, Gower Street,WC1E 6BT London, United Kingdom \label{inst4}
        }

   \date{Received; accepted }
% \abstract{}{}{}{}{} 
% 5 {} token are mandatory
 
\abstract
  {A currently unsolved question in supernova (SN) research is the origin of stripped-envelope supernovae (SESNe). Such SNe lack spectral signatures of hydrogen (Type Ib), or hydrogen and helium (Type Ic), indicating that the outer stellar layers have been stripped during their evolution. The mechanism for this is not well understood, and to disentangle the different scenarios' determination of nucleosynthesis yields from observed spectra can be attempted. However, the interpretation of observations depends on the adopted spectral models. A previously missing ingredient in these is the inclusion of molecular effects, which can be significant.
}
   {We aim to investigate how the molecular chemistry in SESNe affect physical conditions and optical spectra, and produce ro-vibrational emission in the mid-infrared (MIR). We also aim to assess the diagnostic potential of observations of such MIR emission with JWST.}
  % the inclusion of molecular processes influences the spectral synthesis of SNe, and using novel models to explore if molecular emission is possible to observe and the possible constraints such models provide.}
{We coupled a chemical kinetic network including carbon, oxygen, silicon, and sulfur-bearing molecules into the nonlocal
thermal equilibrium (NLTE) spectral synthesis code SUMO. We let four species - CO, SiO, SiS, and SO - participate in NLTE cooling of the gas to achieve self-consistency between the molecule formation and the temperature. We applied the new framework to model the spectrum of a Type Ic SN in the 100-600d time range.}
%Here we present state-of-the-art SNe spectral synthesis models of nebular-phase type Ic, for the first time self-consistently including the coupling between the molecular formation, molecular cooling effects, and radiative transfer. 
{Molecules are predicted to form in SESN ejecta in significant quantities (typical mass $10^{-3}$ $M_\odot$) throughout the 100-600d interval. The impact on the temperature and optical emission depends on the density of the oxygen zones and varies with epoch. For example, the [O I] 6300, 6364 feature can be quenched by molecules from 200 to 450d depending on density. The MIR predictions show strong emission in the fundamental bands of CO, SiO, and SiS, and in the CO and SiO overtones.} 
%In our standard model of a typical SESNe, the most abundant molecules are CO, SiO, SiS, and SO, of which $\sim 10^{-2} -10^{-4}$M$_{\odot}$ has formed at 600 days.
%Spectral synthesis shows that the molecular emission from the ro-vibrational lines will be prominent in the infra-red region, and will be observable with the James Webb Space Telescope.  
%A small parameter space exploration indicates that molecular line observations could also potentially be used to pin down the density of the oxygen-rich zones and to constrain the physical conditions in the ejecta.
   {Type Ibc SN ejecta have a rich chemistry and considering the effect of molecules is important for modeling the temperature and atomic emission in the nebular phase. Observations of SESNe with JWST hold promise to provide the first detections of SiS and SO, and to give information on zone masses and densities of the ejecta. Combined optical, near-infrared, and MIR observations can break degeneracies and achieve a more complete picture of the nucleosynthesis, chemistry, and origin of Type Ibc SNe.}
  % Observations from the next generation of telescopes in combination with state-of-the-art spectral synthesis models provide and important window to understanding the fates of the most massive stars.  }

   \keywords{supernovae: general - astrochemistry - molecular processes}

   \maketitle
%
%-------------------------------------------------------------------

\section{Introduction}
% When massive stars ($M_{ZAMS} \gtrsim 8$~\msun) reach the end of their lives they explode as core-collapse supernovae (CCSNe). 
% These events enrich the cosmos in elements produced in both hydrostatic and explosive nucleosynthesis. 
% Understanding galactic chemical evolution of elements from carbon to nickel relies on understanding this element ejection. 
% CCSNe are also important for understanding the formation of neutron stars and black holes, which are leftovers after the violent explosions. 
% Recently, some indications have also emerged that part of the elements they eject condense into dust after a few years or decades, and may define a major source of dust in the universe\arjc{(REF)}. 

When massive stars ($M_{ZAMS} \gtrsim 8$~\msun) reach the end of their lives, they explode as core-collapse supernovae (CCSNe). 
These events enrich the cosmos in elements produced in both hydrostatic and explosive nucleosynthesis, and leave behind exotic remnants such as black holes and neutron stars. Likely part of the SNe ejecta condenses into different dust species after a few years or decades and may define a major source of dust in the Universe \citep[see][for a review]{Sarangi2018}. 
Understanding galactic chemical evolution of elements as well as dust then relies on understanding this element ejection. 

An open question in SN research is the stellar origin of Type Ibc SNe,  collectively named stripped-envelope SNe (SESNe). 
This is a subtype of CCSNe which lacks the spectral signatures of either hydrogen (Type Ib) or both hydrogen and helium (Type Ic). The indication is then that the outer hydrogen envelope, or both the hydrogen and helium envelopes, have been stripped away during the star's evolution. The mechanism for this is, however, not well understood. Two main scenarios have been proposed. The progenitors of Type Ibc SNe may be massive single stars, $M_{ZAMS}\gtrsim 25$~\msun, that lose their outer layers due to strong stellar winds \citep{conti_owr_1975}. The progenitors may also be lower-mass stars,  $M_{ZAMS} \lesssim 20-25$~\msun, that have their envelopes stripped through interaction with a binary companion \citep{nomoto_typeibc_1995}. 
Over the years, results from light curve modeling \citep[e.g.,][]{Ensman1988,Woosley1994,Bersten2014,Dessart2020Ibc, Ertl2020}, progenitor studies \citep{Smartt2015,Eldridge2016,Yoon2017}, and rate constraints \citep{smith_ccsne_2011} have indicated that the second channel is likely significant. The emergence of the transitional IIb type as a common SN class \citep[rate of $\sim$10\% of all CCSNe, e.g.,][]{Shivvers2017} has further strengthened this.  The posed question has therefore morphed toward trying to answer what contribution the single-star channel makes, and what different outcomes are possible for massive Wolf-Rayet stars at the end of their evolution. 
% Kunc2015 nebular

To determine the detailed properties and origin of an SN, spectral studies are necessary. In particular, once the photosphere has receded after a few months, observations reveal the inner region of the exploded star, including its nucleosynthesis. Studies in this so-called nebular phase are, however, challenging. Although sustained by radioactivity, the SN continuously fades, limiting the epochs over which it can be observed and the signal-to-noise ratio (S/N) of the collected spectra. The physics of the spectral formation is complex, involving fast differential expansion, a radioactive environment with strong nonlocal
thermal equilibrium (NLTE) effects, and energy transfer from MeV particles down to eV or sub-eV gas \citep[see][for a recent review]{jerkstrandSpectraSupernovaeNebular2017}.
%To disentangle the two scenarios the nucleosynthesis yields of type Ibc SNe, and therefore their progenitor masses, need to be determined from observations. However, there are two main complications; the objects of interest are rare and faint, which means that detailed observations are hard to make, and, additionally, the interpretation of such observations is difficult. The large velocities, low densities, and $\gamma$-rays from radioactive decay mean that the gas is out of local thermodynamical equilibrium (LTE), and the level populations, the chemistry,  and the radiative transfer need to be modeled with great care and detail when synthesizing the spectra of these objects \citep[see][for a recent review]{jerkstrandSpectraSupernovaeNebular2017}.

To complicate things further, at some point the SN will start to form molecules, and later, dust. Direct evidence for this comes from observations of molecular and dust emission at IR wavelengths, and also indirectly by dust obscuration of the optical atomic emission. Ground-based spectral data generally extend up to $\sim2.5$\micron, which includes the first overtone emission of CO, but no other molecular emission. 
For Type Ibc SNe, a handful of such observations have been published \citep{gerardyCarbonMonoxideType2002,hunterExtensiveOpticalNearinfrared2009,rhoNearInfraredOpticalObservations2021}. 
Observations of the emission from other molecules requires MIR spectroscopy of the 2-16 \micron~region where the bulk of molecular emission lies. %, after the molecular formation has occurred at $\sim$100 days, which historically has been lacking for all types of supernovae. 
MIR observations of SNe are typically not possible from the ground, with the single exception of the very close-by Type II SN 1987A \citep[see e.g.,][]{aitken10MuSpectral1988,rocheSiliconMonoxideSupernova1991}. The {\it Spitzer Space Telescope} was used to study a few nearby ($<$10 Mpc) Type II SNe \citep[see e.g.,][]{kotakSpitzerMeasurementsAtomic2006,kotakCoreCollapseSupernovae2009}. 
In both the case of SN1987A and in these Type II SNe observed with Spitzer, molecular emissions from CO and SiO was identified.
For SESNe, there are so far no MIR spectral observations, although the James Webb Space Telescope holds promise to change this.

Regarding modeling, there are studies focused on the formation processes of molecules \citep[e.g,][]{petuchowskiCOFormationMetalrich1989,leppMoleculesEjectaSN1990,liuCarbonMonoxideSN1992,liuOxygenTemperatureSN1995,liuFormationDestructionSilicon1996,gearhartCarbonMonoxideFormation1999}, or both molecules and dust \citep{claytonCondensationCarbonRadioactive1999,claytonCondensationCarbonRadioactive2001,cherchneffChemistryPopulationIII2009,cherchneffChemistryPopulationIII2010,biscaroMoleculesDustCassiopeia2014,biscaroMoleculesDustCassiopeia2016,sluderMolecularNucleationTheory2018}.
These studies have provided valuable insights into the chemistry of SNe and produced quantitative results for molecule and dust masses for different SN types. These calculations have, however, been limited to using parameterized temperatures and have not performed radiative transfer to make predictions for the molecular emission. In radiative transfer modeling predicting SN spectra, on the other hand, emission and cooling by molecules have been ignored or treated in a parameterized fashion \citep[e.g.,][]{dessartTypeIIPlateauSupernova2013,jerkstrandProgenitorMassType2012,jerkstrandNebularSpectraSN2014}.

The coupling of molecular chemistry with spectral synthesis is therefore so far a missing ingredient on the modeling side. 
%A previously missing key ingredient in spectral models has been the inclusion of molecular effects, which can be significant. 
%Once molecules form, molecular ro-vibrational line emission produces strong emission in the infra-red (IR) spectrum of SNe \citep{meikleSpectroscopySupernova1987A1989,kotakCoreCollapseSupernovae2009}. 
Both from the observed strength of the molecular emission \citep[e.g.,][]{meikleSpectroscopySupernova1987A1989,kotakCoreCollapseSupernovae2009}, and the only two studies calculating molecular cooling so far \citep{liuOxygenTemperatureSN1995,liljegrenCarbonMonoxideFormation2020} molecules appear capable of cooling the ejecta significantly, perhaps by several thousand degrees. This would strongly affect the optical spectrum which for many SNe is the only spectral region observed. 
%Molecules will also affect the physical conditions, which leads to indirect effects on other parts of the spectra; material will be locked up in molecules and even a small amount ($\sim 10^{-3}-10^{-2}$\msun) can effectively cool the surrounding by several thousand degrees \citep[see e.g.,][]{liuOxygenTemperatureSN1995,liljegrenCarbonMonoxideFormation2020}, resulting in weaker emission of some atomic species. 
Because there is an intricate feedback loop between molecule formation and temperature \citep{liljegrenCarbonMonoxideFormation2020},
%Molecule formation depends on the gas temperature, and once formed, molecules can be very efficient coolants. Ideally, 
the cooling and formation processes need to be modeled self-consistently and simultaneously. 

To address this need, we here present the first SN spectral synthesis models where detailed prescriptions of the molecular formation and cooling processes are coupled to the radiative transfer.
This is a continuation of the work presented in \citet{liljegrenCarbonMonoxideFormation2020}, where the methodology foundation was outlined. 
Here we extend on this work by including molecular formation in multizone models, adding silicon and sulfur-bearing molecules in addition to carbon-bearing ones, and calculating the molecular destruction by Compton electrons using self-consistent methods. In contrast to L20, where a highly simplified test model was presented to investigate different relevant processes, the models here are meant to represent realistic SN spectra, including the molecular emission in the IR region. 
Such simulations can consequently be used both to interpret available optical and near-infrared (NIR) observations and future mid-infrared (MIR) observations with, for example, JWST and ELT.
%predict the next generation of either ground-based telescopes, such as \fixme{which telescope? vlt??}, and space telescopes, such as the James Webb Space Telescope, will possibly be able to observe. 
% CO, SiO, SiS
% Standard model
% Prediction for future JWST observations
% Compare with 

To explore the impact and diagnostic potential of molecule formation in Type Ibc SNe, we here start with a case study of a Type Ic model in the 100-600d interval. With this model, we investigate the type of molecules that form, at what epochs, and in which quantities. We study the influence of molecule formation on the temperature and the resulting spectra throughout the UVOIR range.
% We additionally perform a small parameter space study with this model, to gauge how different densities influence the results.
These results can be used to predict what is possible to observe with upcoming instrumentation, and what specific spectral regions and phases are of most interest to collect data for.

This paper is structured as follows: Sect.~\ref{sect:mod_methods} gives an overview of the modeling methods used. 
Sect.~\ref{sect:evol_molmass} presents the molecular mass results. Sect.~\ref{sect:cooling} discusses the cooling effects of the molecules. Sect.~\ref{sect:comp_comtpon-distruction} presents the Compton electron destruction rates, and Sect.~\ref{sect:midir_spec} is dedicated to a detailed description of the resulting IR spectra for the standard model. In Sect. \ref{sec:lineshapes} an investigation on how line shapes depend on physical conditions is presented.  In Sect.~\ref{sect:sens_study} we make a parameter space investigation, varying the density of the standard model.
Sect.~\ref{sect:optical_spectra} discusses how the optical spectra are influenced by these changes in model density, and in Sect.~\ref{sect:comp_obs} we compare our results with observations. In Sect.~\ref{sect:obs_jwst} we investigate the observational potential with JWST. Summary and conclusions are presented in Sect. \ref{sec:summary}.

%\arjc{Should review previous molecule modeling (Cherchneff etc) - a paragraph with references basically.}

\section{Modeling methods}
\label{sect:mod_methods}

We have implemented a treatment of the formation and NLTE cooling of molecules in the SN spectral synthesis code \textsc{Sumo} \citep{jerkstrand44TipoweredSpectrumSN2011,jerkstrandProgenitorMassType2012}. The physical modeling approach is described in this section, while a detailed description of the model molecules is available in Appendix~\ref{app:model_mol}. Appendix~\ref{app:reactions} contains the molecular reactions rates used.

\subsection{SUMO}

\textsc{Sumo} is an NLTE spectral synthesis code, which employs detailed physical treatments of all key steps in the spectral formation \citep{jerkstrand44TipoweredSpectrumSN2011,jerkstrandProgenitorMassType2012,jerkstrandSpectraSupernovaeNebular2017}. 
It takes a hydrodynamic supernova ejecta model as input, with density and composition as functions of velocity, and calculates the physical conditions and emergent spectrum by solving for the temperature, the statistical equilibrium of ion abundance and level populations, and the radiation field.  
\textsc{Sumo} is specialized to epochs post the diffusion phase and can provide both snapshot steady-state solutions at specified phases as well as fully time-dependent solutions \citep{pognan2021}. Currently, the atomic treatment considers 22 elements between hydrogen and nickel, and a few r-process elements, each with multiple ionic species.% as well as 4 molecular species (CO, SiO, SiS and SO), are included in the full radiative calculations. % AJ Jumping the gun here, describe what code is doing previous to this publication.

\subsection{Molecules}

\begin{table}
\caption{Table of the neutral molecules in \textsc{Sumo}. We note that the singly ionized counterparts of these molecules are also included. For details see text.}
\centering
\def\arraystretch{1.2}
\begin{tabular}{ll}
\hline \hline
\multicolumn{2}{c}{Full inclusion} \\
\hline
CO, SiO, SiS, SO&\\
\hline
\multicolumn{2}{c}{Partial inclusion} \\
\hline
\ch{C2}, \ch{O2}, \ch{S2}, \ch{H2}, \ch{H3}, \ch{CO2}, \ch{C2O} &  \\
\hline
\label{tab:mol_included}
\end{tabular}
\end{table}

Molecular species are divided into two groups. In the first group, full inclusion, the species is included both in the formation processes and in the temperature and radiative transfer calculations (through NLTE solutions for its levels). In the second group, partial inclusion, it is included only in the formation process calculations. This division is done to avoid unnecessary, expensive NLTE calculations for rare molecules (or dipole-free ones, e.g., \ch{S2}) which have no impact on thermal conditions. The included molecules and their categories are listed in Table~\ref{tab:mol_included}.

The molecular abundances are calculated using a chemical network approach. For full inclusion group species, the impact on the temperature and spectrum are then solved for by treating the molecules on the same footing as the atoms, with complete coupling to temperature and radiation field. We treat transfer in the rovibrational lines in the Sobolev limit. %, however these likely have quite minor effects on the radiation field. 
The molecular electronic UV excitations are, however, not included - at nebular times the temperatures are too low for these to play any thermal role. 
Four molecular species; carbon monoxide (CO), silicon monoxide (SiO), silicon monosulfide (SiS), and sulfur monoxide (SO), receive this full treatment. 
We choose these because they are the most common and abundant species formed in the SN ejecta. 

For the molecules in the partial inclusion group, we calculate the molecular abundances in the same chemical network calculation as previously mentioned molecules. 
In other parts of \textsc{Sumo}, however, the partially included molecules act as passive components, that is to say they do not influence either the temperature or spectrum, only acting as storage sinks for the component atoms. 
%The reason for excluding a full treatment for these molecules is that they typically form in only trace amounts (>$10^{-6}$\msun), and therefore do not affect the thermal conditions, or lack suitable transitions to do so (e.g. \ch{S2} which is a dipole-less molecule).

%This division is made mainly due to two reason; either only trace amounts of a specific molecule is formed and, therefore, have very little influence on their surroundings, or data needed to for full inclusion, such as complete line lists in the infrared or cross-sections to calculate the Compton electron destruction rates, are not available. Notably, data for di-sulfur (\ch{S2}) is not available and thus, while abundant, it is only part of the partial treatment.  However, as \ch{S2} is an dipole-less molecule it will likely have very weak lines in the infrared, and not the type of impact on the spectra or temperature as the other abundant molecular species. 

\subsubsection{Chemical network}

\begin{table*}
\caption{Collisional reactions, pertaining to molecules, included in this work. Here double letters (e.g., AB) represent a molecule, while single letters (e.g., A or B) represent an atom, although some of these reactions can occur between two molecules as well. 
% Reactions with one entry indicate that it is a reaction specific to molecular chemistry (e.g. radiative association, which always leads to the creation of molecules). Reactions with several different examples indicate that such a reaction has an included atomic analogy to the reaction involving molecules (e.g. charge exchange, where the transfer of charge can occur between molecules, molecules of atoms, or atoms). 
}
\centering
\def\arraystretch{1.2}
\begin{tabular}{rll}
\hline \hline
\multicolumn{3}{c}{Thermal collision reactions} \\
\hline
\ch{AB + C} &\ch{ -> A + BC}& Neutral-neutral exchange\\
\rule{0pt}{3ex} 
\ch{AB^+ + C} &\ch{ -> A + BC^+} & Ion-neutral exchange\\
\rule{0pt}{3ex} 
\ch{AB^+ + C} &\ch{ -> AB + C^+}& Charge transfer\\
% \ch{A^+ + B} &\ch{ -> A + B^+}& \\
\rule{0pt}{3ex} 
\ch{A + B} &\ch{-> AB + h$\nu$} & Radiative association             \\
\hline
\multicolumn{3}{c}{Recombination reactions} \\
\hline
\ch{AB^+ + e^- } &\ch{-> AB + h$\nu$} & Radiative recombination            \\
% \ch{A^+ + e^- } &\ch{-> A + h$\nu$} & \\
\rule{0pt}{3ex} 
\ch{AB^+ + e^- } &\ch{-> A + B} & Dissociative recombination         \\
\hline
\multicolumn{3}{c}{Nonthermal reactions} \\
\hline
\ch{ AB + e^-_{C} } &\ch{-> AB^+ + e^- + e^-_{C}} & Ionization by Compton electrons   \\
% \ch{ A + e^-_{C} } &\ch{-> A^+ + e^- + e^-_{C}} &  \\
\rule{0pt}{3ex} 
\ch{ AB + e^-_{C} } &\ch{-> A + B + e^-_{C}} & Dissociation by Compton electrons \\
\hline
\label{tab:rec}
\end{tabular}
\end{table*}

To describe molecule formation and destruction, and calculate the molecular abundances, we use a chemical kinetic network approach.
We here make an overview of the methodology.
For an in-depth description, details and discussion see \citet{liljegrenCarbonMonoxideFormation2020}.

The conditions under which molecules form in supernovae are harsh; as the ejecta moves at a few percent of the speed of light the material quickly moves into a low-density regime, and both the population of Compton electrons and the strong radiation field can ionize or dissociate the molecules that do form. 
%The evolution of all included species needs to be tracked explicitly, by model the chemical reactions that take place.  As such a reaction occurs the number density for an included species can either increase (if the reaction creates that species) or decrease (if the reaction destroys that species. AJ: Det här är lite för obvious, kan skippa.

Whereas in \citet{liljegrenCarbonMonoxideFormation2020} the chemistry was modeled in the steady state approximation, we here employ a time-dependent formalism (both for temperature and the molecular and atomic abundances), solving \citep[e.g.,][]{Cherchneff1992}
\begin{equation}
    \frac{dn_i}{dt} + \frac{3n_i}{t} = C_i - n_i D_i,
\end{equation}
where $C_i$ is the total volumetric creation rate of particles of type $i$ (cm$^{-3}$s$^{-1}$) and $D_i$ is the total destruction rate per particle (s$^{-1}$). We also treat the temperature equation in its time-dependent form \citep{kozma_late_1998, jerkstrandSpectralModelingNebularphase2011,pognan2021}.

By comparing solutions in steady state versus time-dependent mode, we could assess the differences due to these treatments. We found only quite small differences in general. The only significant difference was the CO abundance curves at the latest epochs. As the O$^+$ charge transfer reaction switches off at a critical temperature, the destruction rate plummets by a factor 10-100. In steady state, this leads to an immediate increase of the CO abundance by the same factor. However, the formation time scale is several days at the low prevailing densities at 500-600d, so in time-dependent model it takes 100-200d to reach the new equilibrium with a factor 10-100 higher abundance.

The importance of time-dependent terms in a particular rate equation depends on the \emph{total} formation and destruction rates for that species. Parts of a network may be affected but not others. For the main molecular species here, we find that, apart from the late CO phase described above, these total rates are always high compared to the rates of change of density and radioactive power; thus the steady state mostly holds over the modeled time period.

%{\textcolor{red}{We make some analysis of steady-state vs time-dependent solutions in Sec. X.}}
%}

%In \textsc{Sumo} the molecular abundances are solved from steady-state equations balancing the creation and destruction rates of each species. 
%is described in steady state as
%\begin{equation}
%\label{eq:roc}
%    \frac{d(\text{[X]})}{dt} = \sum \mathcal{C}_i - %\sum \mathcal{D}_i = 0
%,\end{equation}
%where $\mathcal{C}_i$ and $\mathcal{D}_i$ are the rates of processes that create and destroy the species, respectively. AJ: Den här ekvationen tillför inte speciellt mycket då C och D inte direkt länkas någonstans. 
%Consequently, the abundance of a species can be controlled by several different formation and destruction reactions.
%The set of Eqs. \eqref{eq:roc} for all included species makes a large network of interconnected chemical reactions, which is solved using a Newton-Raphson solver.
We include several different thermal and nonthermal processes, an overview of such reactions relevant to molecular formation is given in Table~\ref{tab:rec}. 
The rates for these processes depend on the number density of the reactants, and the efficiency of the reactions, described by a temperature-dependent rate coefficient $k(T)$ (except for destruction by Compton electrons, which is discussed in Sect.~\ref{sect:comp_mod}). 

The rate coefficients for the thermal and recombination reactions are here expressed in a modified Arrhenius form as:
\begin{equation}
\label{eq:arr}
    k(T)= \alpha \times \left( \frac{T}{300~\mbox{K}} \right)^{\beta} \times \exp({-\gamma / T})
,\end{equation}
where $\alpha$, $\beta,$ and $\gamma$ are parameters specific to each reaction.
The unit of $k(T)$ depends on the number of reactants, being s$^{-1}$ for unimolecular reactions and cm$^{3}$s$^{-1}$ for bimolecular reactions. The reaction database in this work is an extension of the one used in \citet{liljegrenCarbonMonoxideFormation2020}, now including also sulfur and silicon-based species and reactions. 
The primary sources are the UMIST Database for Astrochemistry\footnote{\href{http://udfa.ajmarkwick.net/}{www.astrochemistry.net}}(\citealt{mcelroyUMISTDatabaseAstrochemistry2013} ), the online database KIDA\footnote{\href{https://kida.astrochem-tools.org/}{https://kida.astrochem-tools.org/}}(\citealt{wakelamKINETICDATABASEASTROCHEMISTRY2012}) and the chemical kinetic database NIST\footnote{\href{https://kinetics.nist.gov/}{https://kinetics.nist.gov/}} (\citealt{NIST}). 
Some reactions, for example the radiative association \ch{C + O -> CO} (\citealt{gustafssonRadiativeAssociationRate2015}), were updated with newer rates. 
Our reaction database was also supplemented by reaction rates from \citet{cherchneffChemistryPopulationIII2009} and \citet{sluderMolecularNucleationTheory2018}.
%where the molecular formation was investigated in similar SNe environments to this work. 
Readers can refer to Appendix~\ref{app:reactions} for a complete reference list. 

\subsubsection{Ionization energies}
For the full inclusion in the \textsc{Sumo} framework, the ionization energies of the included molecular species are needed. 
These were available for CO, \ch{CO+}, SiO, SiS and SO, but had to be calculated for \ch{SiO+}, \ch{SiS+} and \ch{SO+}. For details see Appendix \ref{app:ion_en}.

\subsubsection{Dissociation by Compton electrons}
\label{sect:comp_mod}
Gamma rays, produced by the radioactive decay in SNe, will repeatedly Compton scatter on free and bound electrons to create a population of high-energy Compton electrons (referred to as $\rm{e}_{\rm{C}}^{-}$ in this work).  
Collisions between these Compton electrons and molecules is a significant and sometimes dominant destruction process for molecules. 
For a diatomic molecule AB the dominant channel is typically ionization:
\begin{align}
\label{eq:compton1}
\ch{ AB + e^-_{C} &-> AB^+ + e^- + e^-_{C}},
\end{align}
followed by the weaker dissociation channel:
\begin{align}
\label{eq:compton2}
\ch{ AB + e^-_{C} &-> A + B + e^-_{C}}.
\end{align}
Other reactions (e.g., \ch{ AB + e^-_{C} -> A + B+ + e- + e^-_{C}}) are less important and disregarded in this work. 
% \label{eq:compton3}
% \ch{ AB + e^-_{C} &-> A^+ + B + e^- + e^-_{C}}. \\
% \label{eq:compton4}
% \ch{ AB + e^-_{C} &-> A + B^+ + e^- + e^-_{C}},
%however the dominant reactions are typically the ionization channel, \eqref{eq:compton1}, followed by the dissociation channel, \eqref{eq:compton4}.
In previous works, for example \citet{liuOxygenTemperatureSN1995,cherchneffChemistryPopulationIII2009,sluderMolecularNucleationTheory2018,liljegrenCarbonMonoxideFormation2020}, such rates were estimated based on work by \citet{liuOxygenTemperatureSN1995}, where the destruction of CO was investigated and presented as a relationship between the (zone) power $L$, the number of particles $N_{tot}$ and the mean energy per ion pair $W$ (or molecule pair in the case of dissociation):
\begin{equation}
\label{eq:liu_des}
    k_{C} = \frac{L}{N_{tot} W}.
\end{equation}
For a fixed composition (49.5\% O I, 49.5\% C I, 1\% CO I), \citet{liuOxygenTemperatureSN1995} calculated $W=34$ eV for CO ionization.
Here we perform direct calculations of the Compton destruction rates for the reactions~\eqref{eq:compton1}, for the most abundant molecular species. 
This is possible as SUMO contains a Compton electron distribution solver \citep[see][for details]{kozma_gamma-ray_1992}, which combined with electron impact cross-sections for the relevant molecules, can yield the destruction rates for these channels. 
For reaction~\eqref{eq:compton1}, the cross sections used are listed in Appendix~\ref{app:el_col}.
For reaction~\eqref{eq:compton2}, cross-section are available for CO, but not for the other molecules (to the authors' knowledge). 
We estimate the rate of reaction~\eqref{eq:compton2} for SiO, SiS and SO by scaling the calculated rates for reaction~\eqref{eq:compton1} with the same factor that differentiates the CO rates for reaction \eqref{eq:compton2} and ~\eqref{eq:compton1} (which is $\sim$0.2). 
%Cross-sections for atomic species are discussed in \fixme{which ref?}.
In Sect. \ref{sect:comp_comtpon-distruction} we compare our calculated rates with those estimated by \citet{liuOxygenTemperatureSN1995}.

\subsubsection{Photoionization}

In this work, photoionization is included for the atoms and atomic ions as usual in \textsc{Sumo}, however, the process is ignored for the molecular species. 
Previous work indicates that this is a subdominant process for molecules \citep{cherchneffChemistryPopulationIII2009}. 
We also found this to be the case, when experimenting with a hydrogenic approximation for the relevant cross-sections of the molecules. 
We hope to include molecular photoionization in subsequent work, however, some necessary cross-section data is still not available.

%but do not attempt it here in any of the models. 

\subsection{Molecular spectra and cooling}
Molecular emission has been observed to be significant in the near-to-mid IR region in Type II SNe, and may be the case for Type Ibc SNe as well, if molecules form in a considerable amount. 
Additionally, it is known from previous work \citep{liuOxygenTemperatureSN1995,liljegrenCarbonMonoxideFormation2020} that molecules can add significant cooling channels, if abundant enough, lowering the temperature by several thousand degrees. 
We model the NLTE (i.e., not assuming local thermodynamical equilibrium, LTE) populations and cooling for four molecular species: CO, SiO, SiS and SO.

The emission in the IR region comes from ro-vibrational lines, that is from transitions between rotational levels in one vibrational state to another.
For the investigated physical conditions, the rotational level populations within a vibrational level remain close to thermal equilibrium \citep[see discussion in][]{liuCarbonMonoxideSN1992}; the radiative lifetimes of such levels are long compared to the time-scales of pure rotational collisions with either neutral atoms or electrons.
Vibrational populations are, however, not in LTE, especially not at the later times \citep[see discussion in][]{liuSiliconMonoxideSN1994}. 
We therefore calculate the ro-vibrational population by solving the equations of statistical equilibrium, and ensure relative LTE within a vibrational level by putting the collision strengths for the pure rotational transitions to large values.
%assuming that the main excitation mechanism is electron collisions \fixme{why?} and the de-excitation occurs through electron collisions and radiative processes.

For each level with vibrational number $\nu$ and rotational number $j$, we have (somewhat simplified, ignoring exchange with other ions through ionizations and/or recombinations):
\begin{equation}
    n_{\nu,j} \sum_{\nu', j'} P_{\nu, j \rightarrow \nu', j'} =   \sum_{\nu', j'} n_{\nu',j'} P_{\nu', j' \rightarrow \nu, j},
\end{equation}
where the left-hand side represents the outgoing rate from the level $n_{\nu,j}$, and the right-hand side is the incoming rate from all levels $n_{\nu',j'}$. 
$P_{\nu, j \rightarrow \nu', j'}$ is the transition probability, which depends on the Einstein coefficients $A$ and the Sobolev escape probability $\beta$ for spontanous emission, the radiative photoexcitation/deexcitation rates $R$, and the collision-induced transitions $C$ as: 
% \arjc{Det blir meckigt att försöka skriva in B*J rates här, SUMO använder inte dessa utan räknas Monte Carlo events. Dessutom ska beta in i sådan uttryck också..skriva om med en $R$ för radiative rates blir nog bättre..}
\begin{align}
    \label{eq:deex}
    P_{\nu, j \rightarrow \nu', j'} &= A \beta_{\nu, j \rightarrow \nu', j'} + R_{\nu, j \rightarrow \nu', j'} + C_{\nu, j \rightarrow \nu', j'} &, \nu > \nu' \\
    \label{eq:ex}
    &= R_{\nu, j \rightarrow \nu', j'} + C_{\nu, j \rightarrow \nu', j'} &, \nu' > \nu.
\end{align}
Eq.~\eqref{eq:deex} represents the de-excitation probability, and Eq.~\eqref{eq:ex} represents the excitation probability, for depopulating level $\nu,j$. 
% In the investigated environment $B_{\nu, j \rightarrow \nu', j'}J_{\nu, j \rightarrow \nu', j'}$ are typically negligible \arjc{Hur vet vi det?}.

% \arjc{Skriv något om de sensitivity/convergence test du gjorde.}

For these calculations, we adopt the energy levels and A-values from \citet{liRovibrationalLineLists2015} for CO, from \citet{bartonExoMolLineLists2013} for SiO and from \citet{upadhyayExoMolLineLists2018} for SiS. 
This data is available online through the Exomol collaboration\footnote{\href{https://www.exomol.com}{www.exomol.com}}.
The line list for SO was received through private communication with Ryan Brady and Sergei N. Yurchenko, based on in-prep work to be published \citep{brady22}. 

The collision-induced transitions $C$ are here due to collisions between molecules and thermal electrons, and therefore depend on the electron density $n_e$ as 
\begin{equation}
\label{eq:e_coll}
    C = q(T) \times n_e,
\end{equation}
where $q(T)$ is the transition rate coefficient, which depends on the electron temperature. 
We  use data from \citet{risticRateCoefficientsResonant2007} for CO. % and electron collision de-excitation rates between vibrational levels, and calculate the excitation rates with detailed balance.
% Additionally, since the rotational levels are in LTE, we can assume that collisional excitation and de-excitation only occurs between the same rotational levels in different vibrational bands, and that $q(\nu, j \rightarrow \nu', j) = q(\nu \rightarrow \nu')$, i.e. that the rate between the vibrational band is conserved \arjc{Ganska oklart, kan det förklaras ytterligare?}. 
To the authors' knowledge, equivalent data is not available for any of the other molecular species investigated here.For our NLTE molecules, SiO, SiS and SO, we therefore used the CO data as well, for lack of a better alternative. For details on how we extrapolate rate coefficients for transitions without available data, and details about the model molecules, see Appendix~\ref{app:model_mol}. 

% For the collision with electrons, data was available for CO and electron collision de-excitation, for vibrational transitions to the ground level, i.e. $q(\nu=1 \rightarrow \nu =0)$, $q(\nu=2 \rightarrow \nu =0)$, etc \citep{risticRateCoefficientsResonant2007}. 
% To account for other such transitions, where no data was available, the scaling law for collision-induced transition probabilities from \citet{chandraCollisionalRatesVibrotational2001} was applied:
% \begin{equation}
%     q(\nu \rightarrow \nu') = \frac{\nu\left(2\nu' + 1\right)}{2\nu-1} \times q(1 \rightarrow 0).
% \end{equation}
% There are other such scaling laws \citep[see the discussion in][]{thiRadiationThermochemicalModels2013}, however, we found through experimentation that the different methods produced similar results.
We included levels up to the vibrational quantum number $\nu = 7$ and rotational quantum number $j = 80$. Through experimentation we found that higher levels had no significant impact on the temperature or spectra. We include ro-vibrational radiative transitions up to $\Delta \nu = 5$.

\subsection{The supernova model}

\begin{table*}
\caption{Properties of the standard Ic ejecta model used (total mass 5 \msun). %Mass fractions below $10^{-9}$ are set to 0. 
}             
\label{table:standard_model}      
\centering 
\def\arraystretch{1.2}
\begin{tabular}{lccccc}
\hline \hline
Zone &1&2&3&4&5\\
&Fe/He&Si/S&O/Si/S&O/Ne/Mg&O/C \\ \hline
Mass (\msun)&0.064&0.18&0.62&2.7&1.4 \\ 
\textit{f} &  0.4833  &  0.1570  &  0.0609    & 0.1915 & 0.1072 \\ 
$n(200d)$ (cm$^{-3}$) & $7.9\times 10^6$ & $5.3\times 10^7$ & $8.7\times 10^8$ & $9.8\times 10^8$& $1.1\times 10^9$ \\
\hline
\multicolumn{6}{c}{Mass fractions} \\
\hline
\ch{^{56}Ni + ^{56}Co}&0.77&0.075&4.8$\times 10^{-7}$&1.3$\times 10^{-7}$&3.9$\times 10^{-8}$ \\ 
%\ch{^{57}Co}&0.031&1.7$\times 10^{-3}$&4.4$\times 10^{-6}$&1.7$\times 10^{-7}$&2.1$\times 10^{-8}$ \\ 
%\ch{^{44}Ti}&3.4$\times 10^{-4}$&1.8$\times 10^{-5}$&1.7$\times 10^{-6}$&0&0 \\ \hline
%H&7.5$\times 10^{-8}$&1.3$\times 10^{-9}$&0&0&0 \\ 
He&0.12&1.1$\times 10^{-5}$&1.9$\times 10^{-6}$&1.5$\times 10^{-6}$&2.6$\times 10^{-5}$ \\ 
C&2.7$\times 10^{-6}$&6.4$\times 10^{-7}$&8.6$\times 10^{-5}$&8.1$\times 10^{-3}$&0.2 \\ 
N&2.8$\times 10^{-7}$&0&2.2$\times 10^{-5}$&4.2$\times 10^{-5}$&1.5$\times 10^{-5}$ \\ 
O&7.9$\times 10^{-6}$&6.7$\times 10^{-6}$&0.55&0.74&0.75 \\ 
Ne&8.6$\times 10^{-6}$&1.7$\times 10^{-6}$&1.5$\times 10^{-4}$&0.16&0.034 \\
Na&4.8$\times 10^{-7}$&7.4$\times 10^{-7}$&5.5$\times 10^{-6}$&2.2$\times 10^{-3}$&2.0$\times 10^{-4}$ \\ 
Mg&1.9$\times 10^{-5}$&1.7$\times 10^{-4}$&0.014&0.063&7.1$\times 10^{-3}$ \\ 
Al&6.7$\times 10^{-6}$&2.2$\times 10^{-4}$&4.5$\times 10^{-4}$&6.3$\times 10^{-3}$&1.0$\times 10^{-4}$ \\ 
Si&2.3$\times 10^{-4}$&0.37&0.26&0.013&8.9$\times 10^{-4}$ \\ 
S&1.9$\times 10^{-4}$&0.39&0.15&3.9$\times 10^{-4}$&2.3$\times 10^{-4}$ \\ 
Ar&1.5$\times 10^{-4}$&0.06&0.022&8.1$\times 10^{-5}$&8.6$\times 10^{-5}$ \\ 
Ca&1.7$\times 10^{-3}$&0.042&5.0$\times 10^{-3}$&2.7$\times 10^{-5}$&2.8$\times 10^{-5}$ \\ 
Sc&2.1$\times 10^{-7}$&2.7$\times 10^{-7}$&2.7$\times 10^{-7}$&1.5$\times 10^{-6}$&7.8$\times 10^{-7}$ \\ 
Ti&1.1$\times 10^{-3}$&5.8$\times 10^{-4}$&4.8$\times 10^{-5}$&7.9$\times 10^{-6}$&8.0$\times 10^{-6}$ \\
V&1.3$\times 10^{-5}$&1.5$\times 10^{-4}$&3.8$\times 10^{-6}$&5.2$\times 10^{-7}$&3.2$\times 10^{-7}$ \\
Cr&1.6$\times 10^{-3}$&7.5$\times 10^{-3}$&3.3$\times 10^{-5}$&1.2$\times 10^{-5}$&1.3$\times 10^{-5}$ \\
Mn&2.1$\times 10^{-6}$&3.3$\times 10^{-4}$&2.3$\times 10^{-6}$&2.8$\times 10^{-6}$&2.1$\times 10^{-6}$ \\
Fe&9.7$\times 10^{-4}$&0.047&4.7$\times 10^{-4}$&5.7$\times 10^{-4}$&5.8$\times 10^{-4}$ \\
Co&3.7$\times 10^{-8}$&3.1$\times 10^{-9}$&1.7$\times 10^{-6}$&1.6$\times 10^{-4}$&1.7$\times 10^{-4}$ \\
Ni&0.037&2.7$\times 10^{-3}$&1.1$\times 10^{-3}$&6.6$\times 10^{-4}$&6.8$\times 10^{-4}$ \\ \hline
\end{tabular}
\end{table*}

To represent a typical Type Ic SN  we define a standard model based on a $M_{ZAMS}$=25~\msun~progenitor core-collapse supernova from \citet{woosleyNucleosynthesisRemnantsMassive2007}, with the hydrogen and helium layers removed. The progenitor has a CO core of 6.8 $M_\odot$, of which 1.8 $M_\odot$ forms a compact object and 5.0 $M_\odot$ is ejected in the supernova. As in \citet{jerkstrandNebularSpectraSN2014,jerkstrandLatetimeSpectralLine2015}, we divide the ejecta into five main compositional zones - Fe/He, Si/S, O/Si/S, O/Ne/Mg and O/C, further specified in Table~\ref{table:standard_model} This will be referred to as the standard model throughout this work. By choice of the inner zone mass the $^{56}$Ni mass in the model is 0.06 $M_\odot$. The zones are macroscopically (but not microscopically) mixed throughout a region between $0 -3500$ \kms, following the treatment in \citet{jerkstrandLatetimeSpectralLine2015}. We assume a distance of 10 Mpc for the model.

There are no robustly determined major differences between Type Ic, Ib and IIb ejecta. The thin hydrogen layer in Type IIb SNe has no influence on any of the inner ejecta properties or spectra \citep{jerkstrandLatetimeSpectralLine2015}. Type Ib/IIb SNe seem to have similar ejecta masses and velocities as Ic SNe \citep[e.g.,][]{Prentice2019}. As long as the helium in these SNe does not mix into the metal zones and quench the chemistry by He+ reactions, we therefore expect to see little differences in models with our without a He layer, and the model here should be representative for Ib/IIb SNe as well, apart from He and H lines.

Further following the method in \citet{jerkstrandLatetimeSpectralLine2015}, we assume a uniform density for the three oxygen zone (O/Si/S,  O/Ne/Mg, and O/C zones), with a lower density for the two innermost zones (Fe/He and Si/S). This differentiation is based on expansion of the $^{56}$Ni-rich zones during the first days due to trapped radioactive heating.
The density structure of the model is set to obey
\begin{equation}
\label{eq:den_struc}
    \rho_{rel} = (1-10-\chi-\chi-\chi),
\end{equation}
% (i.e. the density of Si/S zone is 10 times that of the Fe/He zone etc.), 
where the numbers/symbols on the right hand side represent the five zones so that the density of for example zone 4 is $\rho_4 = \chi \times \rho_1$. 
From previous work on nebular spectra of stripped-envelope SNe values of $\chi\sim 30-200$ have been indicated \citep{jerkstrandLatetimeSpectralLine2015}, although the number is not very well constrained, see for example discussion in \citet{dessart2021}. 
For the standard model we use $\chi$=100, making the density of the O/Si/S,  O/Ne/Mg and O/C zones 100 times larger than that of the Fe/He zone. 
This clumping structure can equivalently be described by the density of the zone relative to the constant density in a model where all the ejecta is distributed uniformly throughout the expansion volume.
Relative to the density in such uniform model, $\chi=100$ corresponds to an overdensity of 2.8 for the O-zones, an underdensity of 0.026 for the Fe/He zone, and an underdensity of 0.23 for the Si/S zone. 
In Sect.~\ref{sect:sens_study} we make a small parameter space exploration for different $\chi$ values.
The lowest $\chi$ model there corresponds to an O-zone overdensity of 1.8, and the largest to an overdensity of 14.

%\fixme{Filling factor of a given zone is the fraction of the expansion volume that the zone occupies}

From the density structure the filling factors, which is the measure of clumping used in \textsc{Sumo}, can be calculated.
The filling factor of a given zone is defined as the fraction of the expansion volume that the zone occupies, and the sum of all filling factors $f_i$ must equal one 
\begin{equation}
    \sum_1^{N_{zone}}f_i = 1. 
\end{equation}
Additionally, the volume and filling factors of two zones can be related using the density structure as
\begin{equation}
    \frac{M_j}{f_j \times \rho_j} = \frac{M_{j+1}}{\rho_{j+1} \times f_{j+1}}, j \in (1, N_{zone}-1), 
\end{equation}
\begin{equation}
\begin{aligned} 
    \rightarrow &f_j - \left( \frac{\rho_{j+1}}{\rho_j} \frac{M_j}{M_{j+1}} \right) f_{j+1} = \\
    &f_j - \left( \frac{ \chi_{j+1}}{ \chi_{j}} \frac{M_j}{M_{j+1}} \right) f_{j+1} = 0.
\end{aligned}
\end{equation}
This forms a system of linear equations that is solved for $f_i$.
The $f_i$ values for the standard model are available in Table~\ref{table:standard_model}.

\section{Molecular formation results}
\label{sect:evol_molmass}

\begin{figure*}
\centering
\includegraphics[width=0.99\textwidth]{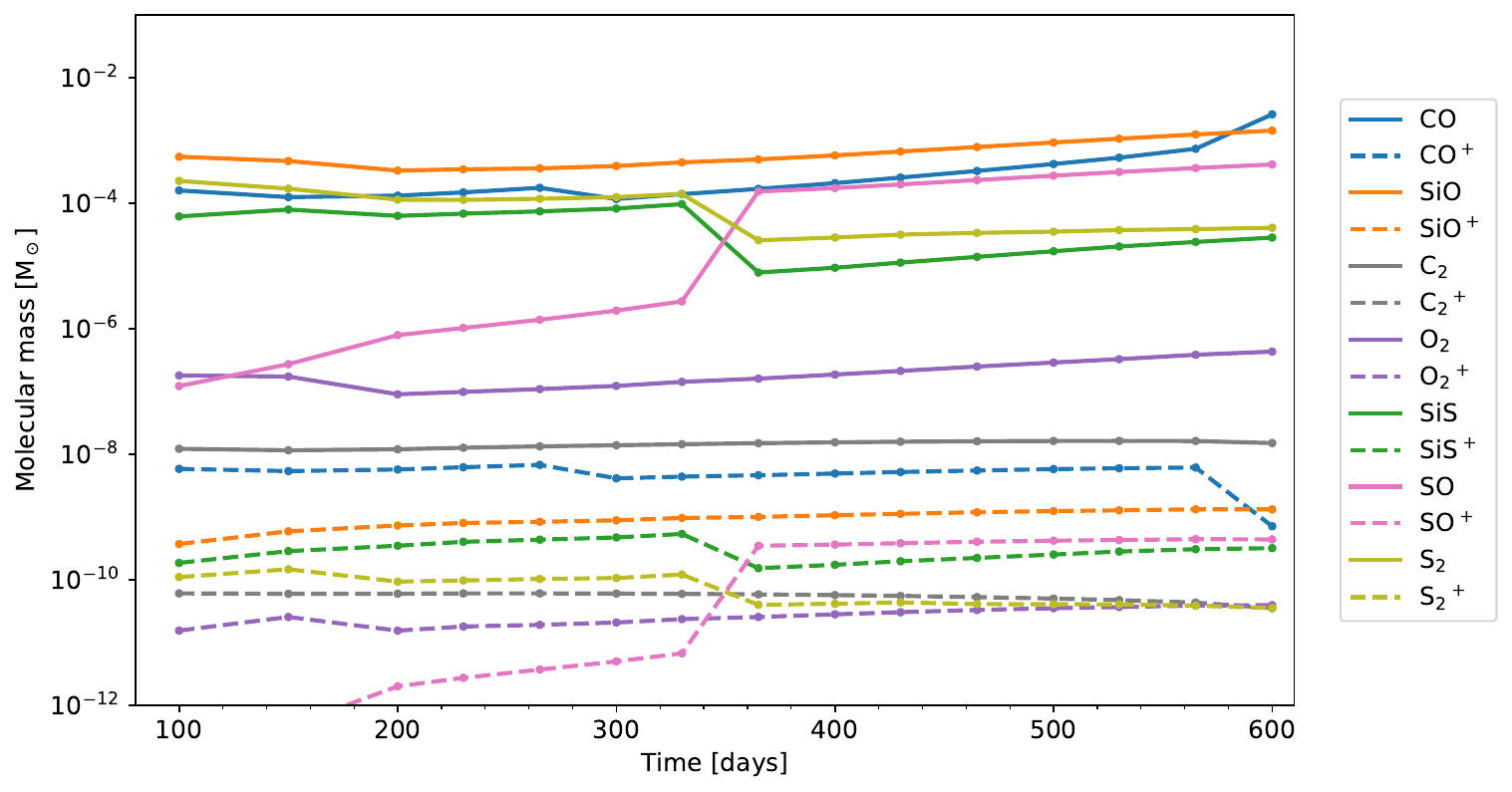}
\caption{Total mass formed for different molecular species as function of time. Solid lines are neutrals and dashed lines ionized molecules.} % AJC: All molecules in the network are seen here except 2: CO2 and C20.
\label{fig:molmass_tot}
\end{figure*}

  \begin{figure*}
  \centering
  \includegraphics[width=0.49\textwidth]{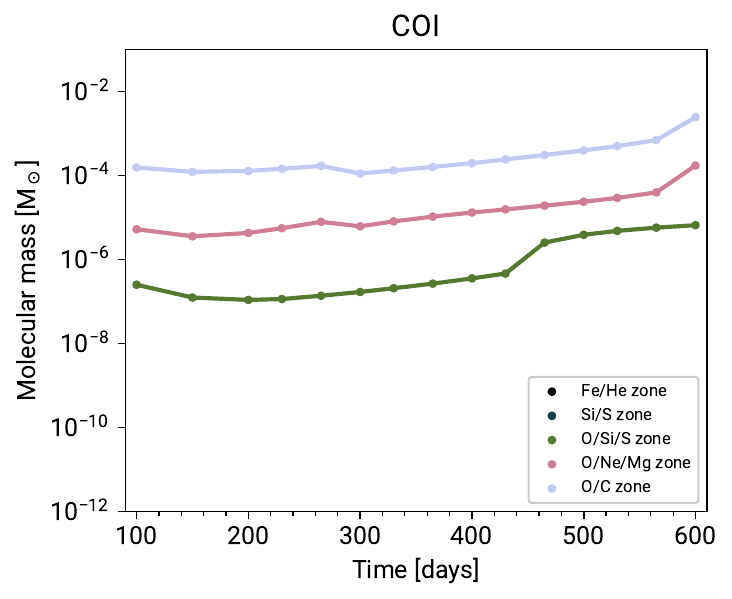}
  \includegraphics[width=0.49\textwidth]{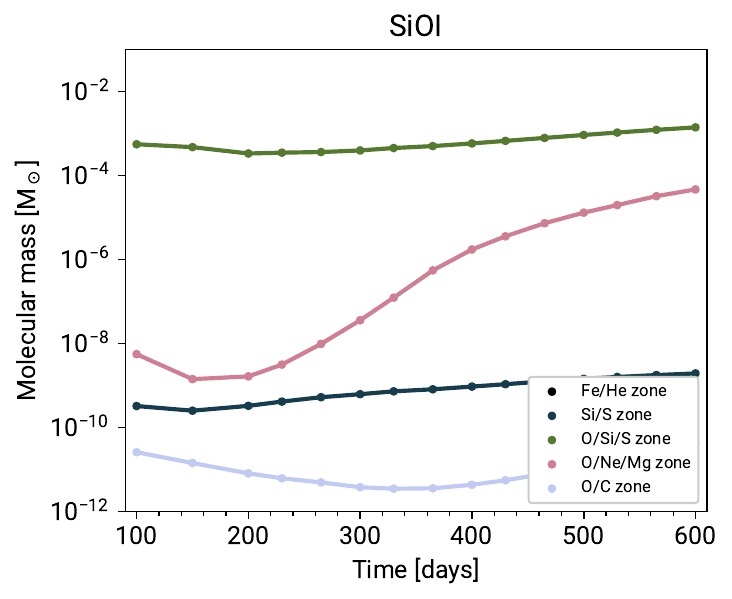}
  \includegraphics[width=0.49\textwidth]{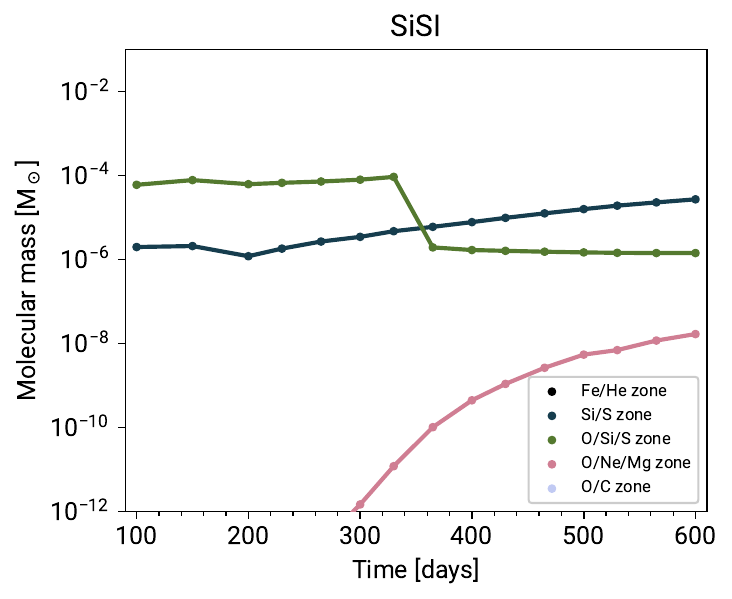}
  \includegraphics[width=0.49\textwidth]{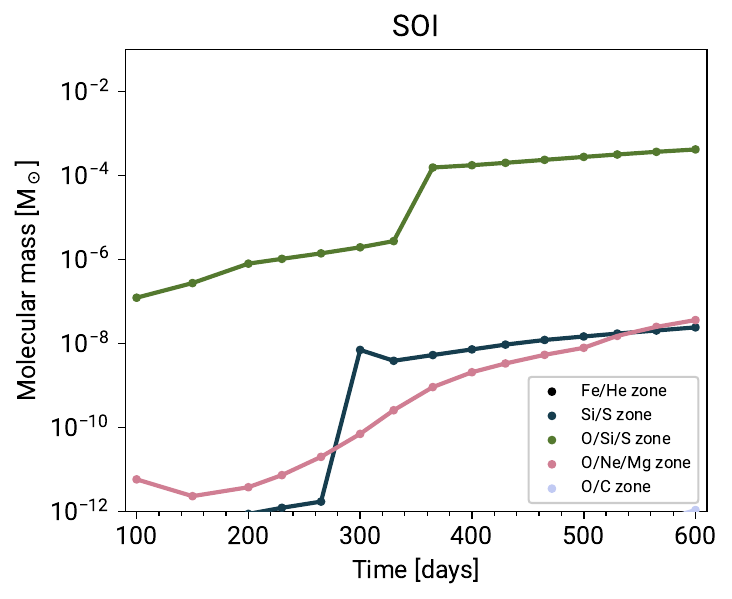}
      \caption{The masses of the four molecular species included in the cooling and radiative transfer calculations as function of time. The different lines show the different masses in each zone.}
         \label{fig:molmass_zone}
  \end{figure*}

We model the supernova between 100-600 days after explosion, at time intervals of 50 days. Earlier epochs than 100 days are difficult and time-consuming to converge with full NLTE atomic and molecular physics. Such epochs also still may have diffusion effects taking place, and relatively high opacities makes emergent diagnostics complex to interpret. Later than 600 days virtually no observations of stripped-envelope SNe exist, and issues with dust formation (which we do not model) become pressing.

Fig.~\ref{fig:molmass_tot} shows the time evolution of the amount of molecules formed in the standard model. The five most abundant molecules to form are carbon monoxide (CO), silicon monoxide (SiO), silicon monosulfide (SiS), sulfur monoxide (SO), and disulfur (\ch{S2}).
Of these, CO, SiO,  and SO increase with time and their highest masses are reached at the last epoch of 600 days. SiS and \ch{S2}, on the other hand, have their highest masses early on and decline at later times.

For all molecules, the abundances are relatively constant between 100-300 days; the model gives $\sim$$10^{-3}$~\msun~for SiO, $\sim$$10^{-4}$~\msun~for \ch{S2}, CO and SiS, and $\sim$$10^{-7}-10^{-6}$ \msun~for SO.  
After some rapid changes around 330 days, the abundances then again settle, now with significantly more SO but less SiS and \ch{S2}. 
At the last epoch an acceleration in the growth of CO is seen.
The final masses at 600 days are $2\times 10^{-3}$~\msun~of \ch{CO}, $3\times 10^{-3}$~\msun~of SiO, $7\times 10^{-4}$~\msun~of SO, and $3\times 10^{-5}$~\msun~of SiS and \ch{S2}. The fraction of atoms locked in molecules for the relevant species at this final time from a few per-mile for S (0.2\%), Si (0.4\%) and O (0.5\%) to a percent for C (1\%).
This is a smaller fraction found than by for example \citet{sarangiChemicallyControlledSynthesis2013}, however, the amount of molecules formed partly depends on the assumed density of a model and will therefore vary for different types of SNe (previously investigated in \citet{biscaroMoleculesDustCassiopeia2014} and discussed here in Sect.\ref{sect:sens_study}).

Other molecules are much less abundant; about $10^{-7}$~\msun~of \ch{O2} and $10^{-8}$~\msun~of \ch{C2} forms, both constant in time, and other included species form less than $10^{-8}$~\msun. 
The reactions governing the formation and destruction balance of each molecule, and their evolution over the simulated times, are discussed below.

\subsection{Carbon monoxide}

Initially the total mass of CO is $\sim$$10^{-4}$~\msun, increasing to $\sim$2$\times 10^{-2}$~\msun~after 450 days.
CO predominantly forms in the O/C zone, with a small amount (always less than 10\% of the total) forming in the O/Ne/Mg zone (Fig. \ref{fig:molmass_zone}).

As seen in Fig. \ref{fig:co_reaction_rates} in Appendix \ref{app:reaction_rate_plots}, CO is primarily formed by radiative association, 
\begin{equation}
    \ch{C + O -> CO + h$\nu$}.
\end{equation}
Other prominent creation channels are by recombination,
\begin{equation}
    \ch{CO+ + e- -> CO + h$\nu$}, 
\end{equation}
through neutral exchange reaction with \ch{C2}
\begin{align}
\ch{C2 + O &-> C + CO},  \label{eq:c2co}
\end{align}
and through charge exchange between O and \ch{CO+} 
\begin{equation}
    \ch{CO+ + O -> CO + O+}.
\end{equation}
%and through recombination
Up until 450 days a major destruction pathway is through the charge exchange with \ch{O+}, 
\begin{equation}
    \label{eq:co+o+}
    \ch{CO + O+ -> CO+ + O} .
\end{equation}
This reaction is endothermic and turns off when the gas gets too cool, consequently resulting in a growth of the CO abundance after 550 days once this destruction channel is quenched. 
This is discussed further in \citet{liljegrenCarbonMonoxideFormation2020}. 

Other important destruction channels are collisions with Compton electrons, 
\begin{equation}
    \ch{CO + e^-_C -> CO^+ + e^- + e^-_C} \text{ or } \ch{C + O + e^-_C},
\end{equation}
and through the reverse reaction of \eqref{eq:c2co}. % and \eqref{eq:o2co}.
These results broadly agree with the findings in \citet{liljegrenCarbonMonoxideFormation2020}, where CO formation was investigated in a Type IIP supernova using a smaller network model.
% After 400 days, the temperature drops enough to quench a major destruction channel of CO (discussed in Sect. \fixme{add} in \fixme{ref}). 

\subsection{Silicon monoxide}

The amount of SiO that forms in the standard model is $\sim$$10^{-3}$~\msun~throughout the simulated time interval. As seen in Fig.~\ref{fig:molmass_zone}, SiO is almost exclusively formed in the O/Si/S zone. The SiO mass in the O/Ne/Mg zone does grow with time, ending at almost $10^{-4}$~\msun~at 600 days, but this still a small fraction of the total mass. 
%Trace amounts of SiO also form in the Si/S and O/C zones. 

The main production channel is at all epochs (Fig. \ref{fig:s_reaction_rates}) through radiative association, 
\begin{equation}
    \ch{Si + O -> SiO + h$\nu$}.
\end{equation}
Other important channels are through neutral exchange with \ch{O2},
\begin{equation}
    \label{eq:o2sio}
    \ch{O2 + Si -> SiO + O},
\end{equation}
and through recombination,
\begin{equation}
    \ch{SiO+ + e- -> SiO + + h$\nu$}.
\end{equation}
The major destruction reactions are Compton electron destruction and  %as 
%\begin{equation}
%    \ch{SiO + e^-_C -> SiO^+} \text{ or } \ch{Si + O}
%\end{equation}
the reverse reaction of Eq.~\eqref{eq:o2sio}. 
%\begin{equation}
%    \label{eq:o2sior}
%    \ch{SiO + O -> O2 + Si}.
%\end{equation}
While the SiO mass is mostly stable over the simulated times, it is still connected to the behavior of the sulfur-bearing species in the O/Si/S zone (discussed in Sect.~\ref{sect:sulf}); SO and SiS.
%acting as sinks for O and Si respectively, which can influence how much SiO can potentially form, and both SO and SiO has formation and destruction channels that depend on \ch{O2} (Eqs.~\eqref{eq:o2sio} and \eqref{eq:o2sso}).

\subsection{Silicon monosulfide, sulfur monoxide and disulfur}
\label{sect:sulf}

  \begin{figure}
  \centering
  \includegraphics[width=\hsize]{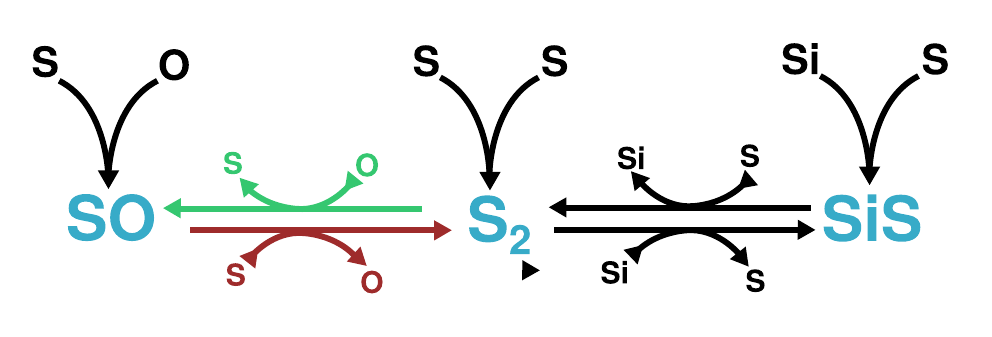}
      \caption{Graphical representation of the part of the chemical network that connects the sulfur-bearing molecular species. }
         \label{fig:sosis}
  \end{figure}

There is an intricate relationship between the sulfur-bearing molecular species \ch{SiS}, \ch{SO}, and \ch{S2}, which makes their behavior complex and temperature-dependent. 
This is illustrated in Fig.~\ref{fig:sosis}, which contains the part of the chemical network that is relevant to this connection. 
Each of the species can form through radiative association %as
%\begin{align}
%    \ch{S + O &-> SO + h $\nu$} \\
%    \ch{S + S &-> S2 + h $\nu$} \\
%    \ch{Si + S &-> SiS + h $\nu$}
%\end{align}
and through recombination.% as 
%\begin{align}
%    \ch{SO+ + e- &-> SO + h $\nu$} \\
%    \ch{S2+ + e- &-> S2 + h $\nu$} \\
%    \ch{SiS+ + e- &-> SiS + h $\nu$}.
%\end{align}
There are strong neutral exchange reactions which connect these three molecules to each other. SO is connected to \ch{S2} by
\begin{align}
\label{eq:so2s2}
    \ch{SO + S} \rightleftharpoons \ch{S2 + O},
\end{align}
and, in turn, \ch{S2} to SiS through
\begin{align}
    \ch{S2 + Si} \rightleftharpoons \ch{SiS + S}.
    \label{react:s2si}
\end{align}
%The abundances of these three species are consequently dependent on each other. 
Initially significant amounts of \ch{S2} and \ch{SiS} form, $\left(1-3\right)\times 10^{-4}~$\msun, while SO is much less abundant (around 5$\times 10^{-7}$~\msun). 
After 300 days there is a swift decline of the molecular masses of \ch{S2} and \ch{SiS}, simultaneously as the \ch{SO} mass increases to $\sim$2$\times 10^{-4}$~\msun, and then slowly continues to increase for the rest of the evolution. 
This is due to reaction~\eqref{eq:so2s2}, marked red in Fig.~\ref{fig:sosis}, being endothermic (in the S$_2$ forming direction) and becoming inefficient at low temperatures. A temperature drop from 3000 K to 2000 K (the relevant temperatures at around 300 days) will result in a decrease of an order of magnitude for this reaction rate. 
For the reverse reaction of \eqref{eq:so2s2}, marked in green in Fig.~\ref{fig:sosis}, data for the rate is only available for lower temperatures.
This pathway between SO and \ch{S2} is then effectively quenched, while the pathway between \ch{S2} and SO turns on, when the gas cools at later times. 
The overall effect is that the net flow from \ch{S2} to SO becomes larger. 
The consequence is then a higher SO abundance at later phases, as the other involved reaction rates are not very sensitive to temperature, and a lower \ch{S2} and SiS abundance, as more of it is turned into SO. 

The molecule SO is connected to \ch{O2} by 
\begin{equation}
    \label{eq:o2sso}
    \ch{O2 + S} \rightleftharpoons \ch{SO + O}.
\end{equation}
This then further connects the behavior of the sulfur species to the SiO abundances through reaction \eqref{eq:o2sio} and its reverse reaction.
The rates, and therefore the importance, of these reactions are however lower than the reactions \eqref{eq:so2s2} and \eqref{react:s2si} previously discussed.
An important destruction process for all three molecules is collisions with Compton electrons.% as
%\begin{align}
%    \ch{SO + e^-_C &-> SO^+} \text{ or } \ch{S + O} \\
%    \ch{S2 + e^-_C &-> S2^+} \text{ or } \ch{S + S} \\
%    \ch{SiS + e^-_C &-> SiS^+} \text{ or } \ch{Si + S}. 
%\end{align} AJ: behöver inte skrivas ut.

Initially, SiS and \ch{S2} are formed primarily in the O/Si/S zone.
After 300 days, when the production of SO becomes much more efficient, these species are instead mostly formed in the Si/S zone: the lack of oxygen in this zone prevents the previously discussed effects involving O and O$_2$ that quenches SiS and \ch{S2} in the O/Si/S zone.
%Very small amounts also forms in the O/Ne/Mg zone. 
SO forms almost exclusively in the O/Si/S zone at all simulated times. 

\subsection{Comparison to previous results}

These are the first model calculations for molecule formation in a Type Ic supernova. It is of interest to compare the results with earlier models for Type II SNe. Because both ejecta structures and methodologies differ between works, it is not straightforward to identify the specific driving factors for differences, but a discussion can be had on this. We focus the comparison on CO and SiO. 

% CO
In the Type II model of \citet{liljegrenCarbonMonoxideFormation2020}, the CO abundance had an initial rapid growth phase, then settled at a quasi-constant abundance after 200-300d. It was demonstrated that both the turn-off time and the settling value were sensitive to the zone density and the amount of radioactive power absorbed. For the standard model, the settling value was a few times $10^{-3}$ $M_\odot$.

The density in the standard model here is about a factor of 10 lower, for a fixed time, due to the higher expansion velocities in a Type Ic SN. This means that our first epoch of 100d corresponds, density-wise, to a bit more than 200d ($\rho \propto t^{-3}$) in a Type II SN and we are already on the ``settled'' plateau here by 100d. The CO masses here are lower, $10^{-4}-10^{-3}$ $M_\odot$ (Fig. \ref{fig:molmass_tot}) - the lower densities lead to less efficient CO cooling which, as analyzed in L20, in turn, leads to less CO. The formation curves here are qualitatively similar to the ``no cooling'' variants in L20.

In the Type II model of a $M_{ZAMS}=20$ $M_\odot$ star in \citet{cherchneffChemistryPopulationIII2009}, CO forms rapidly and then settles on a quasi-constant value after about 200d. The settling value, a few times $10^{-1}$ $M_\odot$, is quite a bit higher than in the L20 model, due to some combination of different ejecta models, chemical network, and temperature evolution. Only at significantly later times ($\gtrsim$ 500d in the standard Type Ic model here, which would correspond to $\gtrsim$ 1000d in a Type II model) are large CO masses formed in our models. In the $M_{ZAMS}=15$ $M_\odot$ Type II model of \citet{sarangiChemicallyControlledSynthesis2013}, the CO forms over a longer time-scale, settling at a similar final value as in \citet{cherchneffChemistryPopulationIII2009} but only after about 600-700d. 

For SiO, both \citet{cherchneffChemistryPopulationIII2009} and \citet{sarangiChemicallyControlledSynthesis2013} obtain an abundance peak quite early on (150-200d) and then a monotonic decrease, having almost completely vanished by 500d. The Si and O instead get locked up in O$_2$, SO, and silicate dust precursors. In our Type Ic model the SiO abundance is declining already from the first epoch of 100d. It then starts to build up again by about 350d, together with SO, as SiS rapidly declines. The chemistry is complex, with SiS, SO, and O$_2$ all playing roles for the SiO abundance - and in addition, dust precursors which are included in the \citet{sarangiChemicallyControlledSynthesis2013} model but not in ours. Obtaining observations of the time evolution of SiO, SiS, SO, and silicate dust emission with JWST would provide important tests and diagnostics of the silicate chemistry of the supernova ejecta.

% a activation energy of around 11600K leads to this reaction being inefficient at lower temperatures, and the reaction rate decreases with an order of magnitude between 3000K and 2000K. 

% between 3000K and 2000K the rate coefficiant 

% Before 400 days, the most abundant species is SiO at around 3$\times 10^{-3}$ \msun, and continue to increase its mass to 7.3$\times 10^{-3}$\msun at 600 days. 

% This leads to a significant increase in the molecular mass of CO, to around 2$\times 10^{-2}$ \msun at 600 days. 
% We find molecular masses of 9$\times 10^{-4}$\msun and 5$\times 10^{-4}$\msun for SiS and SO, respectively, at 600 days. 

% The panels in Fig. \ref{fig:molmass_zone} shows the molecular masses in the different zones, for CO, SiO, SiS and SO. 
% As seen most of the molecular formation takes place in either zone 3, for SiO, SiS and SO, or in zone 5, for CO. 
% These two zones should then be most impacted by molecular cooling, discussed in Sect. \fixme{ref}. 

% Small amounts of molecules also form in zone 2 (mainly SiS with traces of SiO and SO) and zone 4 (mainly  CO and SiO, with traces of SiS and SO).

%\fixme{\section{The impact of time dependence}}

\section{Molecular cooling effects}
\label{sect:cooling}
  \begin{figure}
  \centering
  \includegraphics[width=\hsize]{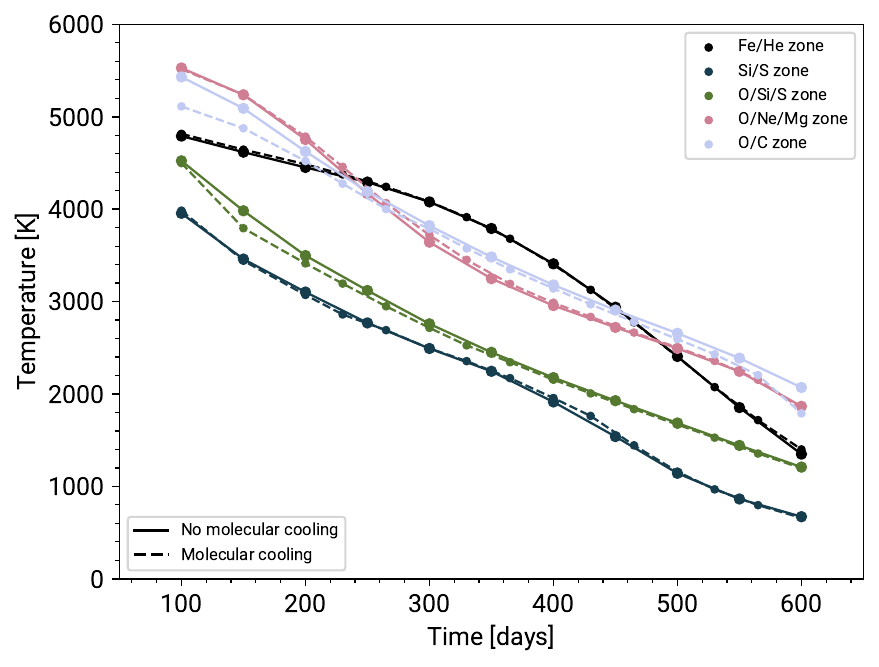}
      \caption{Temperature evolution for the standard model, with molecules included (dashed lines), and excluded (solid lines), for the different zones.}
         \label{fig:temp}
  \end{figure}
  
   \begin{figure*}
  \centering
   \includegraphics[width=0.99\textwidth]{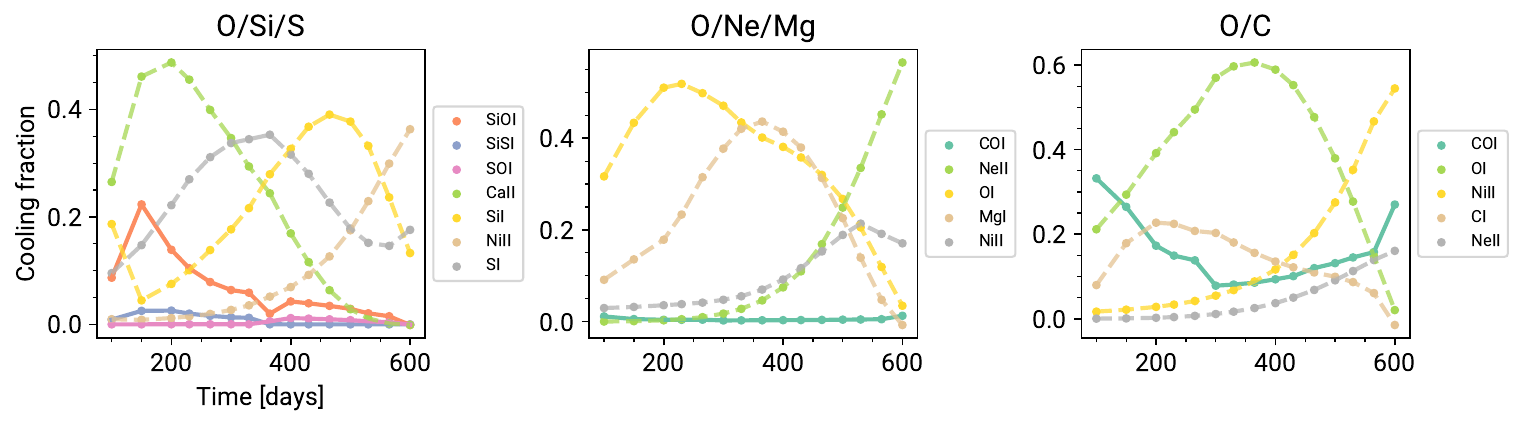}
      \caption{Fraction of the cooling due to different species, for the three oxygen zones. Species are plotted if they contribute more than 5\% of the cooling at any of the modeled epochs. We note that the same species can cool at one phase, and heat at another phase (i.e., have a negative cooling fraction). }
         \label{fig:coolfrac}
  \end{figure*}

   \begin{figure*}
  \centering
   \includegraphics[width=0.99\textwidth]{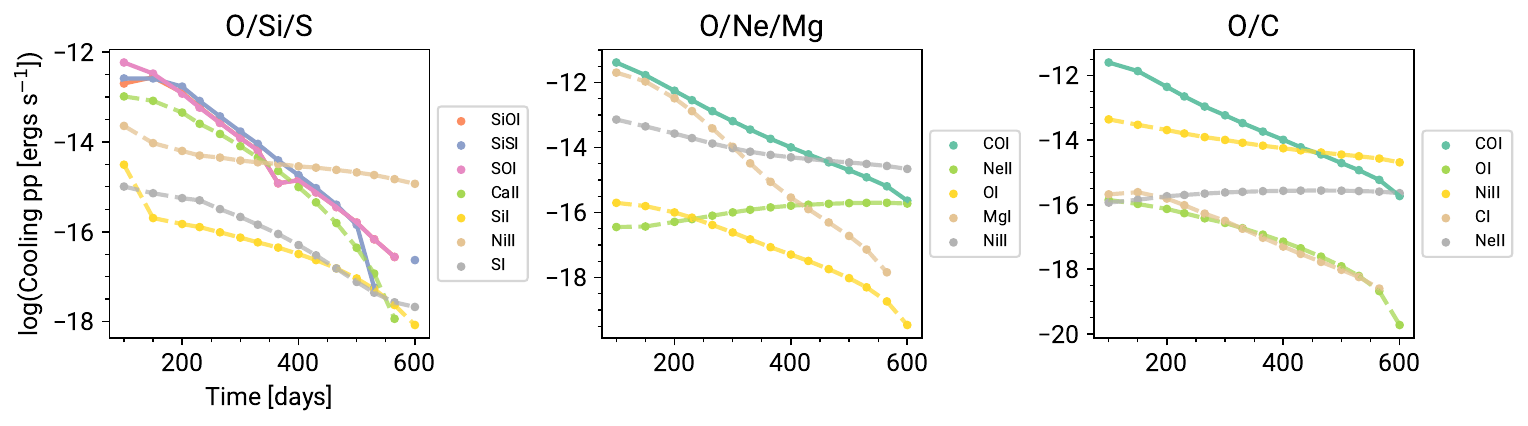}
      \caption{Cooling per particle, for the species plotted in Fig.~\ref{fig:coolfrac}. Only epochs when the species cool (rather than heat) are included.}
         \label{fig:coolpp}
  \end{figure*}

% The presence of molecules can have a large cooling effect, as previously explored in \fixme{liljegren+20}. 
Fig.~\ref{fig:temp} shows the time evolution of the temperature in the different zones, for the standard model with molecular cooling included (dashed lines) and excluded (solid lines). It is immediately clear that our standard model suggests that temperatures in Type Ic SNe are, with a few exceptions, not significantly affected by molecular cooling. The strongest effects occur in the O/Si/S zone and the O/C zone, which are the ones with the highest molecular masses.

In the O/Si/S zone, there are molecular effects on the temperature early on, between 100-300 days. The contributions from the most important cooling species (both molecular and atomic) are shown in the left panel of Fig.~\ref{fig:coolfrac}.
The initial molecular cooling is primarily done by SiO, which accounts for up to 20\% of the cooling, and to a lesser degree SiS (up to 5\%).
With molecular cooling, this zone has a temperatures up to 200 K lower compared to the molecule-free model, at early phases. The effect then lessens with time, and after 300 days there are negligible differences between the models with and without molecular cooling. %postsubmit:\arjc{Somewhere, discuss J15 treatment.}

A similar pattern can be seen in the O/C zone (right-most panel in Fig.~\ref{fig:coolfrac}); at the early stages, there is significant CO cooling (up to $35\%$ of the total cooling), which leads to a $\sim$400 K lower temperature in this zone. However, by 300 days the molecular cooling contribution are less than $10\%$. 
From around 450 days, CO starts to become important again and a temperature gap arises, reaching a few 100 K at 600d. The trends in both of these zones are similar; initially, there are prominent molecular cooling effects, which then lessen with time. As the abundances of CO and SiO, the most important molecular coolants, are essentially constant before for the first few hundred days  (Fig. \ref{fig:molmass_tot}), a decrease of  these species is not the cause of them becoming less relevant coolants with time. 
Rather, this behavior can be explained by looking at the evolution of cooling per particle, which is shown in Fig.~\ref{fig:coolpp}. The figure shows that the decline in cooling per particle for the molecules is steeper than those for the atomic species. 
% the cooling per particle done by molecules have to decrease quicker than that of the important atomic coolants. 
% Fig.~\ref{fig:coolpp} shows the cooling done per particle with time; the decline of the molecular coolants is steeper than those of the atomic coolants. 
Consequently, if the number of particles of each species is approximately constant, molecules will become less important when compared to atomic coolants. 

For the other zones, the molecular influence on the temperature is minor for most of the modeled epochs.
In the Si/S zone, where the only significant molecule formed is SiS, at $10^{-6}-10^{-5}$ $M_\odot$, the temperature effect is always small, $\lesssim$50~K.
In the O/Ne/Mg zone, as seen in Fig.~\ref{fig:molmass_zone}, there is an increasing amount of molecules forming with time, mainly CO and SiO. 
The masses are still relatively small, however, and only at the last modeled phase do we see any temperature effect. At this time (600 days) CO causes about 10\% of the cooling, which leads to a temperature decrease of about 90 K. 

The Fe/He zone has negligible molecular formation and is consequently not impacted by molecular cooling. 

Our results broadly support the parameterized ansatz in \citet{jerkstrandLatetimeSpectralLine2015}, where it was assumed that molecules have neglegible thermal effect in the Fe/He, Si/S and O/Ne/Mg zones in a stripped-envelope supernova. However, in some of those models, strong molecular cooling of the O/Si/S and O/C zones were assumed. The results here indicate that this can generally not be assumed for all epochs. 
%postsubmit
%\arjc{Move to a discussion section?}

While a temperature difference of $\lesssim$ 1000 K may not seem very large, the atomic emission lines have exponential dependencies on temperature and can thus be significantly affected by also quite small relative temperature changes. In Sect.~\ref{sect:optical_spectra} we explore the impact on optical atomic emission by the molecular cooling and the sensitivity to the gas density.

\section{Destruction by Compton electrons}
\label{sect:comp_comtpon-distruction}
  \begin{figure}
  \centering
  \includegraphics[width=\hsize]{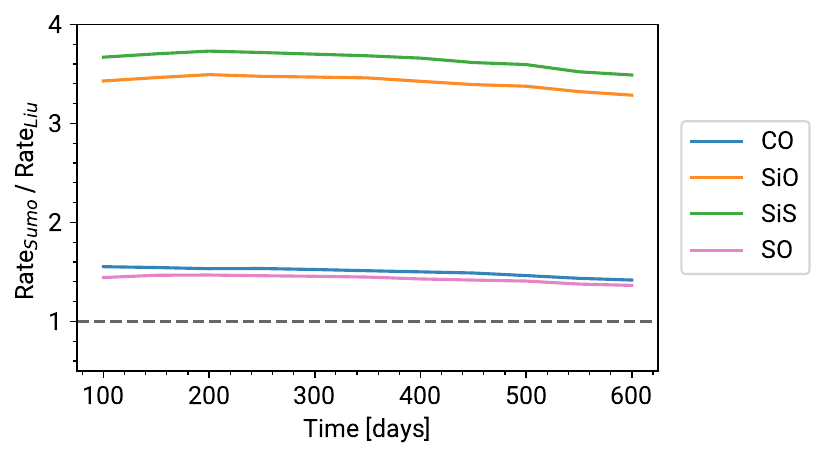}
      \caption{The ratio between the Compton destruction rate calculated with \textsc{Sumo}, for our standard model, and the one estimated using the \citet{liuOxygenTemperatureSN1995}
      prescription, which was calculated for CO. }
         \label{fig:compt_comp}
  \end{figure}

As discussed in previous sections, collisions with nonthermal electrons is an important destruction mechanism for all the investigated molecules. 
In this work, we directly calculate the Compton ionization rate of reaction~\eqref{eq:compton1}, and from this infer the Compton dissociation rate of reaction~\eqref{eq:compton2}, as described in Sect.~\ref{sect:comp_mod}.
In previous works these rates have often been estimated using the Eq.~\eqref{eq:liu_des} prescription, for which \citet{liuOxygenTemperatureSN1995} reported $W$ values for different Compton destruction processes of CO.
Typically these $W$ value are assumed for molecules other than CO as well \citep[see e.g.,][]{liuFormationDestructionSilicon1996, cherchneffChemistryPopulationIII2009,sluderMolecularNucleationTheory2018,liljegrenCarbonMonoxideFormation2020}.

Fig.~\eqref{fig:compt_comp} shows the ratios between our calculated rates for ionization by Compton electrons and the rate estimated by using Eq.~\eqref{eq:liu_des} with the $W$ value \citet{liuOxygenTemperatureSN1995} estimated for CO for this process ($W = 34$ eV). For SiO, SiS and SO we compare the two rates using results from the O/Si/S zone, and for CO we use results from the O/C zone, as these are the zones where the different molecules are most abundant. 

For CO the two estimates agree to within a factor 2 for all the investigated phases in the standard model, with the $W=34$ eV estimate consistently yielding a lower rate by about 35\%. 
We can conclude that using the \citet{liuOxygenTemperatureSN1995} formula gives a quite satisfactory estimate of the Compton electron destruction rate for CO. %if direct calculations are not possible. 
For SiS and SiO our calculated rates are around a factor 3.5 larger compared to what Eq.~\eqref{eq:liu_des} gives with $W =34$ eV.
The destruction rate of SO, similar to CO, is also higher in our calculation, but not by as much. 

The differences between our calculations of the Compton destruction rate and that of \citet{liuOxygenTemperatureSN1995} can be explained by a few key differences in methods and data used. 
First, they calculate their rates for a fixed composition of 49.5\% O~I, 49.5\% C~I and 1\% CO~I, while our rate is calculated during a \textsc{Sumo} run, which uses the \textsc{Sumo} composition and ionization solutions.
Second, for CO we use a different, more recent cross-section \citep{itikawa_cross_2015} than that of Liu and Dalgarno \citep{liuElectronEnergyDeposition1994}, which likely accounts for some of the difference. Third, using the Liu and Dalgarno CO estimate for other molecules will result in uncertainties. As seen in Appendix~\ref{app:el_col}, the cross-sections do differ between the molecules, which will yield differences in the rates. 

% The differences in rates are due to the differences in the cross-sections (see Appendix~\ref{app:el_col}).
% \fixme{modern cross sections}

Calculating backward, we can infer what the corresponding $W$ values for CO, SiO, SiS and SO, would be for our standard model, these are listed in Table~\ref{tab:compt_wi}). Eq.~\eqref{eq:liu_des} does not, however, exactly reproduce the behavior of our calculated Compton rates even for these values as there is a slight downward slope in the ratio of the two for any used $W$.
Consequently, using Eq.~\eqref{eq:liu_des} with the updated $W$ values still yields an error of up to 10\% when compared to the calculated Compton destruction rates. While our reported $W$ values are derived for the standard model, results from the parameter investigation described in Sect.~\ref{sect:sens_study} indicate that using these yields gives satisfactory estimates of the Compton ionization rates for the included molecules. 
We find at most deviations of around 20\% between Compton destruction rates using our estimated $W$ values and rates calculated with \textsc{Sumo}.

\begin{table}
\caption{$W$ values for different species. }             % title of Table
\label{tab:compt_wi}      % is used to refer this table in the text
\centering      % used for centering table
\def\arraystretch{1.2}
\begin{tabular}{c c c }        % centered columns (4 columns)
\hline\hline                 % inserts double horizontal
Species & Source & $W$ (eV) \\ \hline
CO & \citet{liuOxygenTemperatureSN1995} & 34 \\
CO & This work & 23 \\
SiO & This work & 10 \\
SiS & This work & 9 \\
SO & This work & 24
\end{tabular}
\end{table}

% However, the Liu prescription will not capture all the relevant physics.
% This is particularly obvious by the kink seen in Fig.~\ref{fig:compt_comp} at 450 days. 
% At this time, large changes occur in the model; the temperature-dependent destruction reaction \eqref{eq:o2} for CO turns off, which leads to a quick increase in the CO mass. 
% This in turn leads to a more efficient cooling, lowering the temperature by almost 1000K.
% These changes to the physical environment consequently leads to changes in our calculated $\gamma_{nt}$ rate, which is not captured by the simplified Liu presciption. 

\section{Mid-infrared spectra}
\label{sect:midir_spec}
  \begin{figure*}
  \centering
  \includegraphics[width=\hsize]{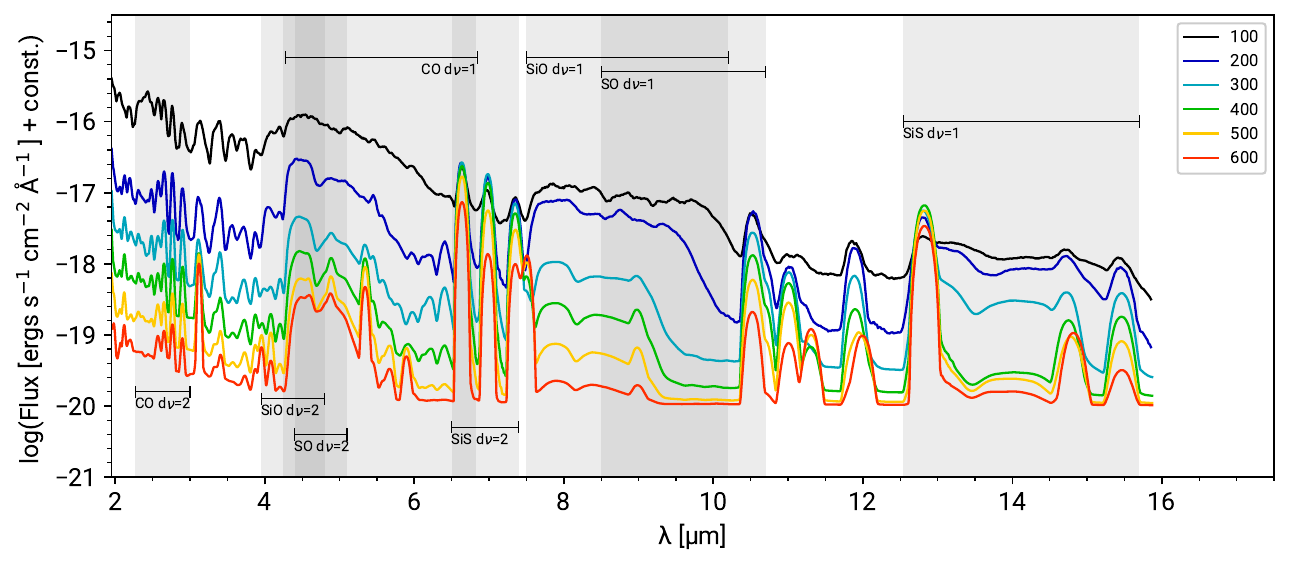}
      \caption{Model spectra in the MIR, where each line is the spectra at different times in days indicated by the right label, for the standard model. The fundamental (d$\nu=1$) and first overtone (d$\nu=2$) molecular bands are indicated. Unblended molecular emission has a light gray background, blends from two bands mid-gray, and blends from three bands dark gray. The flux here is on a log-scale.}
         \label{fig:irspec}
  \end{figure*}

Fig.~\ref{fig:irspec} shows the mid-infrared spectra of the standard model, at different times. 
%The signatures of the molecules included in the radiative transfer calculations are indicated. 
For all simulated phases there are strong molecular emissions in several spectral regions. The details of the spectral signatures from the different molecular species are discussed below.  For a discussion of the atomic MIR emission in SNe, see \citet{jerkstrandProgenitorMassType2012}. 
Possible future observations of the model IR spectra are further discussed in Sect.~\ref{sect:obs_jwst}.

\subsection{CO}

  \begin{figure*}
  \centering
  \includegraphics[width=\hsize]{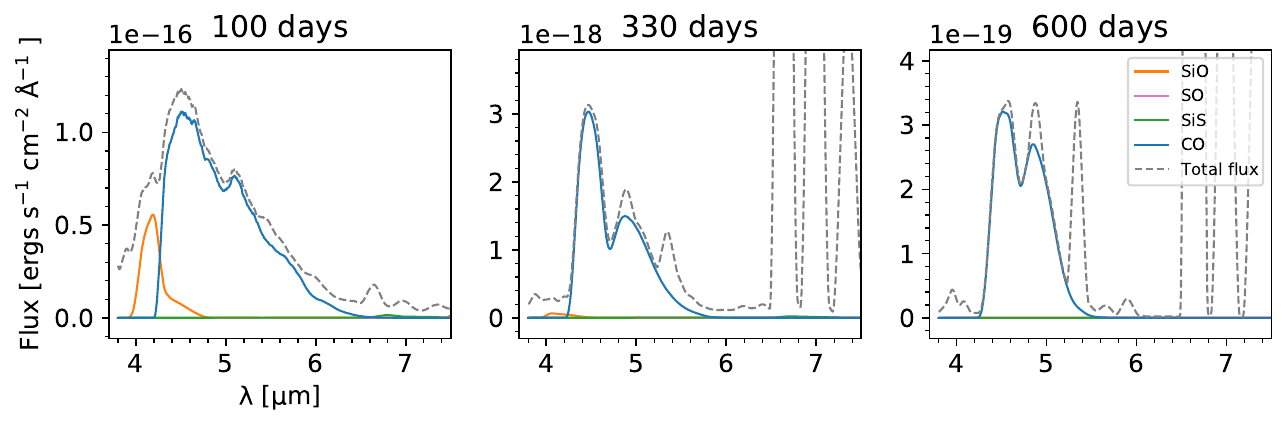}
      \caption{The blend between the fundamental bands of CO (blue) and the first overtone of SiO (green), SO (pink) and SiS (green). Only SiO will have significant contribution from the first overtone bands; both SO and SiS first overtone emission is weak at all simulated times. The flux here is on a linear scale.}
         \label{fig:cosio}
  \end{figure*}

Carbon monoxide has a fundamental band emission that extends between $\sim$4.3 and $\sim$6.8 \micron, with a peak at around 4.6 \micron. 
This is the most luminous molecular emission at all the investigated times, and, except for the very late phases, additionally the most luminous emission in the MIR spectral range.
The CO fundamental band should consequently be the easiest MIR emission to observe. 
This emission has been observed for a few Type II supernovae, most notably SN 1987A \citep[see e.g.,][]{meikleSpectroscopySupernova1987A1989,rankNickelArgonCobalt1988,wooden_airborne_1993-1}, and with the Spitzer Space Telescope for SN 2004di \citep{kotakEarlyTimeSpitzerObservations2005}, SN 2004et \citep{kotakDustTypeIIPlateau2009}, and SN 2005af \citep{kotakSpitzerMeasurementsAtomic2006}. 
The Spitzer observations consist of the red tail of the band, as the spectral window starts at 5~\micron.
Any photometric measurements from this region will likely also be dominated by CO emission, which is discussed in for instance \citet{catchpoleSpectroscopicPhotometricObservations1988,andrewsPhotometricSpectroscopicEvolution2011,ergonOpticalNearinfraredObservations2014}. 

% \arjc{Förklara band shape, säg något om den är temperaturkänslig. Optiska djup?}
% \fixme{vet inte riktigt vad jag ska skriva här; lägre temperatur leder till mindre excitation -> tunnare linjer. Shape ändrar sig också, men är osäker på varför. borde vi göra experiment på detta, kanske med nlte och visa?}

The rapid growth in CO mass seen in the model after 550 days (Fig.~\ref{fig:molmass_tot}) leads to a plateauing of the CO luminosity decline between 500-600d, with a possible regrowth at yet later times. 
Observational monitoring with 50-100 days cadence would be of importance to test such predictions and constrain the ejecta properties, as the transition time depends on the ejecta density.
%If such a quick increase in molecular mass does occur, this would then be possible to observe with monitoring. 

It should be noted that the CO fundamental band overlaps with  the overtones of SiO and SO (and marginally with SiS), as seen in Fig.~\ref{fig:cosio}. In the model here, only the SiO first overtone is predicted to be visible of these, with both SO and SiS overtones being very weak compared to the CO emission. For a discussion about the SiO first overtone see Sect.~\ref{sect:spec_sio}.

The CO first overtone ranges from $\sim$2.2~\micron~to $\sim$3~\micron~with a peak at around 2.3~\micron. 
While weak compared to the fundamental band, it contributes around 50\% of the flux in this wavelength region, and is visible especially in the early phases (100-200 days). 
%and also in the \fixme{very late phases (after 450 days) when the CO mass has increased significantly. }
The CO overtone is the most observed molecular feature for all types of supernovae, as its wavelength is positioned so that it possible to observe with ground-based telescopes. For Type II SNe the first overtone has been observed in the following: SN 1987A \citep{catchpoleSpectroscopicPhotometricObservations1988}, SN 1995ad \citep{spyromilioCarbonMonoxideSupernova1996}, SN 1998S \citep{gerardyDetectionCODust2000}, SN 1998de \citep{spyromilioCarbonMonoxideType2001}, SN 1999em \citep{spyromilioCarbonMonoxideType2001}, SN 2002hh \citep{pozzoOpticalInfraredObservations2006}, SN 2004di \citep{kotakEarlyTimeSpitzerObservations2005}, and SN 2011dh \citep{ergonTypeIIbSN2015}. 
For Type Ibc there are another handful of observations available: SN 2000ew \citep{gerardyCarbonMonoxideType2002}, SN 2007gr \citep{hunterExtensiveOpticalNearinfrared2009}, SN 2013eg \citep{droutDoublepeakedSN2013ge2016}, and SN 2020oi \citep{rhoNearInfraredOpticalObservations2021}.
It should be noted that CO is not universally observed in type Ibc SNe; for instance \citet{pandey_2021} reports the absence of CO first overtone features up to 319 days in SN 2012au.
% \citep[see e.g.][]{catchpoleSpectroscopicPhotometricObservations1988,gerardyDetectionCODust2000,pozzoOpticalInfraredObservations2006,kotakEarlyTimeSpitzerObservations2005,droutDoublepeakedSN2013ge2016}. \arjc{Split references over Type II and Type Ibc SNe. Write out SN names. Make sure list is complete.} 
In Sect.~\ref{sect:comp_obs} we compare the luminosity in the CO first overtone emission in our model with the observed luminosity for Type Ibc SNe.

\subsection{SiO}
\label{sect:spec_sio}

  \begin{figure*}
  \centering
  \includegraphics[width=\hsize]{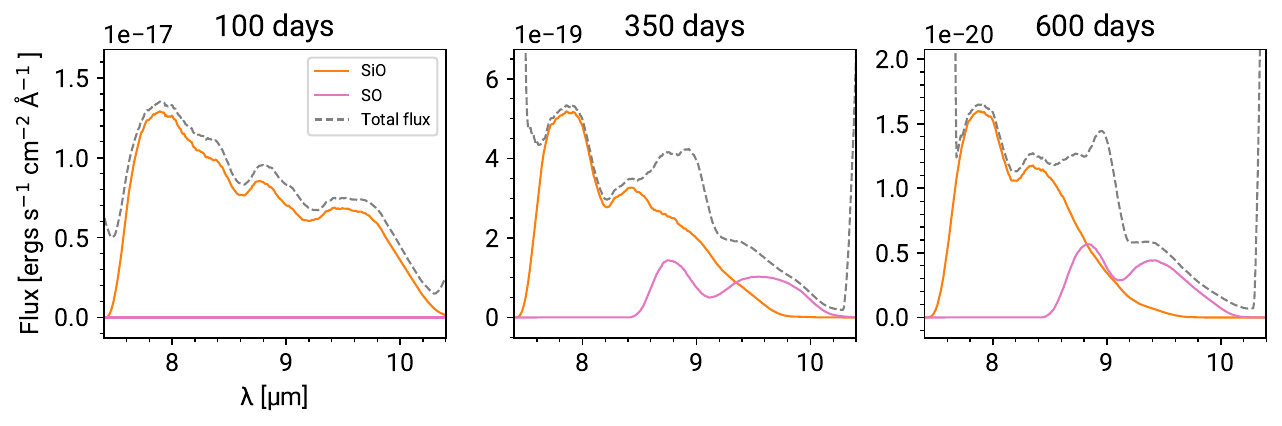}
      \caption{The blend between the fundamental bands of SiO (orange) and SO (pink), at different times. }
         \label{fig:sioso}
  \end{figure*}

The second strongest molecular emission in the model is from the fundamental band of SiO, which ranges from about 7~\micron~to 10~\micron, with a peak at about 7.9~\micron. 
The SiO fundamental band is initially almost as luminous as its CO counterpart, but gets noticeably weaker compared to the CO fundamental band at later stages, when the CO mass increases after 450 days. 
The fundamental band of SiO has been observed for SN 1987A \citep{aitken10MuSpectral1988,rocheSiliconMonoxideSupernova1991} and additionally in two Type II SNe; SN 2004et \citep{kotakCoreCollapseSupernovae2009} and SN 2005af \citep{,kotakSpitzerMeasurementsAtomic2006}.
It has not been observed in any Type Ibc SN.

As seen in Fig.~\ref{fig:sioso}, which shows three different time instances, there is a blend between the SiO fundamental band and the fundamental band of SO. Initially, when the SO abundance is low, all the emission in this range comes from SiO. However, when the SO molecular mass increases the emission eventually becomes comparable to that of SiO. While blended, there is still about 1~\micron~separating the peaks of the SiO and SO fundamental bands, so the two contributions can potentially be distinguished. 
%In particular, SO makes itself known by broad peaks at 8.8 and 9.6 \micron, whereas SiO gives peaks at 7.9 and 8.4 \micron.

The SiO overtone lies between $\sim$3.95~\micron~and $\sim$4.8~\micron~with a peak at $\sim$4.1~\micron. 
This has partial overlap with the blue wing of the strong CO fundamental band, and consequently only part of the SiO emission, between 3.95 and $\sim4.2$~\micron, is readily visible (Fig. \ref{fig:cosio}). 
With time this emission rapidly dims as the gas cools; identification prospects are therefore best at the early epochs.
The SiO first overtone was tentatively identified in SN 1987A by \citet{meikleSpectroscopySupernova1987A1989}, however there was no clear sign of it in the Kuiper observations \citep{wooden_airborne_1993-1}. Thus there is yet no unambiguous detection of this emission in any supernova.

\subsection{SO}

The fundamental band emission of SO extends from $\sim$8.5~\micron ~to $\sim$10.7~\micron, with a peak at around $\sim$8.7~\micron.
This emission becomes visible once the mass of SO starts to become significant at about 300 days. As mentioned in Sect.~\ref{sect:spec_sio}, there is an overlap between the SiO fundamental band and the SO fundamental band, shown for three different time instances in Fig.~\ref{fig:sioso}. 
As seen, at later times the emission from SO is visible and distinct from that of SiO due to the separation of around 1~\micron.  These results indicate that if SO forms in the amount predicted by the model, it should be possible to observe and separate from SiO emission. The best time window for this is 300-600d.

The first overtone of SO ranges from $\sim$4.4~\micron~to $\sim$5.1~\micron, which overlaps completely with the strong emission of the fundamental band of CO. Therefore, the SO first overtone will likely not be possible to observe, and in our model it is always much weaker than the CO band.
Rovibrational SO emission has not yet been identified in any supernova.  There are, however, observations of pure rotation transitions in the radio range for SN 1987A at an age of $\sim$10 000 days, which has been attributed to SO \citep{matsuuraALMASpectralSurvey2017}. 
% This detection gives direct evidence that SO indeed forms in SNe. \arjc{Check for any further followup papers.}
% From these lines an SO mass of $\sim 4 \times 10^{-5}$\msun~was estimated, which is about an order of magnitude smaller than the mass we find at 600 days in our standard model. 

% \arjc{Fick Matsuura några constraints på massa, densitet?} %These results indicate that SO does form in these environments.

\subsection{SiS}

\label{sect:spec_sis}

  \begin{figure*}
  \centering
  \includegraphics[width=\hsize]{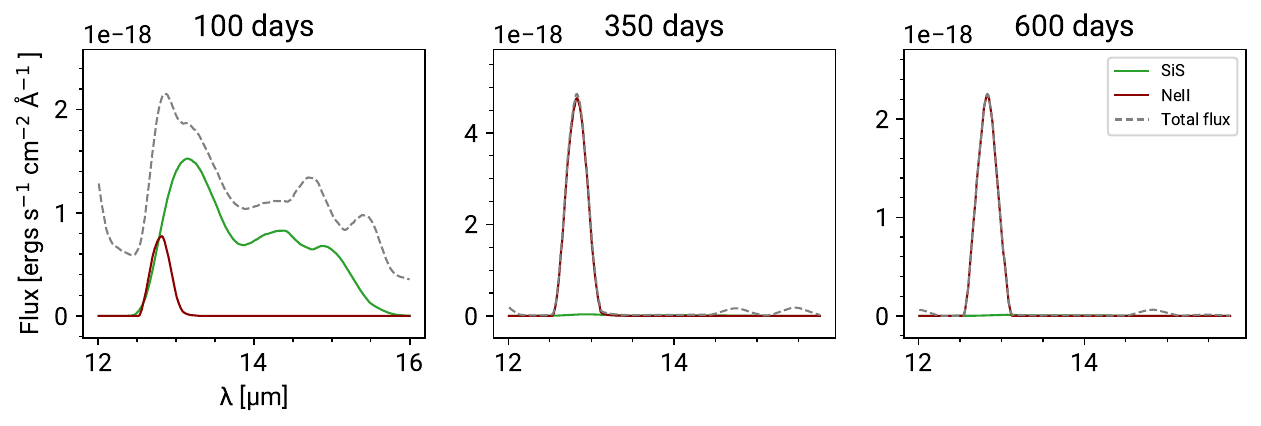}
      \caption{The blend of the SiS fundamental band (green) and [Ne II] 12.81 $\mu$m (brown), at three different times. At times >300 days the SiS band is not visible anymore, swamped by the neon line. }
         \label{fig:sis}
  \end{figure*}

SiS has a fundamental band between $\sim$12.5~\micron~and $\sim$15.6~\micron~with a peak at $\sim$13.2~\micron, shown at three different times in Fig.~\ref{fig:sis}. Emission here is distinct up until 300 days, at which time the SiS mass decreases significantly. This SiS band is weaker than the fundamental bands of CO and SiO at all phases. 
The model suggests, nevertheless, that this emission could be possible to observe as there is no blending with other molecular emission, and blending with atomic emission (mainly Ne II emission at 12.8~\micron~and two Co II lines at 14.7 and 15.4~\micron) should be possible to distinguish.

The first overtone band of SiS is positioned between $\sim$6.2~\micron~and $\sim$7.8~\micron, with a peak at $\sim$6.8~\micron. 
Due to blending with strong atomic lines of Ni II at 6.6~\micron, Ar II at 6.9~\micron~and Ni III at 7.3~\micron, this emission is not distinct in model spectra at any phases and would be hard to detect.
% If these results accurate, it would then  be hard to observe the SiS first overtone. 
%Any identification of SiS then has to come from observations of the fundamental band of SiS. 

SiS has so far not been identified in any supernova. 
Identification has been made difficult by the challenges to observe far out into the MIR, where the fundamental band is positioned, and by this band being relatively weak compared to the SiO and CO bands.
% The possibility to observe out intoEvolution of the line shapesEvolution of the line shapes the mid-IR for such object has so far been lacking, which would be a potential explanation of this. 

\section{Model properties and line shapes}
\label{sec:lineshapes}

 \begin{figure*}
  \centering
  \includegraphics[width=0.99\textwidth]{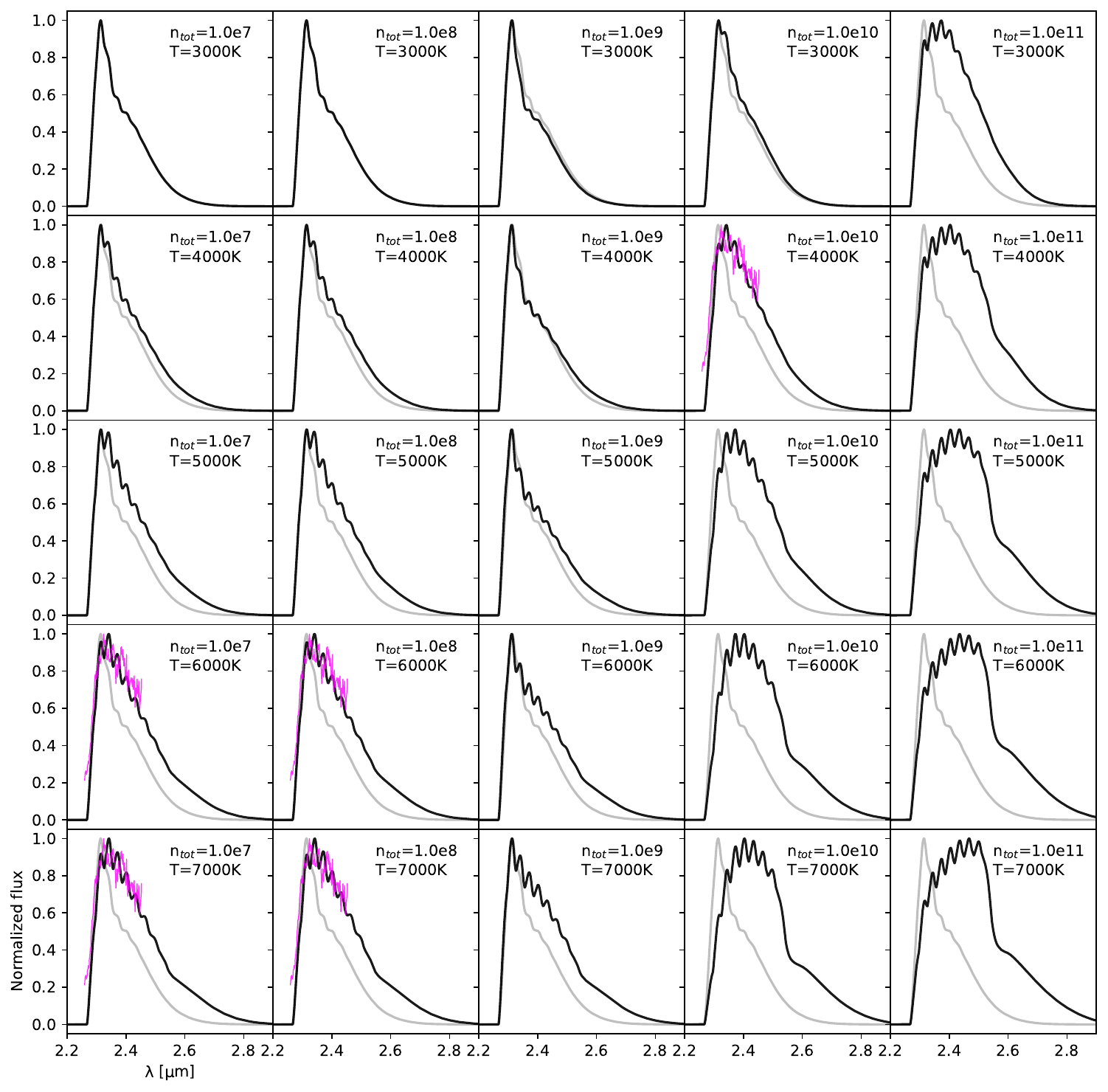}
      \caption{Line shapes of the CO $d\nu=2$ line, for a grid of temperatures and densities, at 100 days. The black line is for the specified density and temperature in each panel. The gray line is for the lowest density and temperature and is used as a reference. The pink lines are observations from \citet{hunterExtensiveOpticalNearinfrared2009}, plotted in panels where the model results are a good fit to the observations. In these simplified simulations we use $n_{CO}=10^{-3}n_{tot}$ and $n_e=0.1 n_{tot}$.  }
         \label{fig:shape_overview}
  \end{figure*}

It may be possible to derive SN characteristics from the line shapes of molecular bands. 
The line shape will depend on the excitation structure of the molecule and this, in turn, depends on different properties, most importantly the temperature and the density. %All these change during the evolution of a SN.
A higher total density typically also means a higher electron density, which leads to higher collisional excitation rates for the molecules, as the collision-induced transition rate depends on $n_e$ (Eq.~\ref{eq:e_coll}).
A higher temperature also leads to a more broadly excited molecule, and emission over a broader wavelength range. 
%The transition probability $q(T)$ for electron collisions increases with temperature, as well as the Boltzmann factor used to calculate the reverse reaction. 

In the standard model, however, temperature and density have a complex relationship; higher density leads to more molecules being formed, which in turn can lead to a more efficient molecular cooling and a lower temperature. 
Additionally, both the temperature and the density will impact how close to LTE a specific species is. 
To disentangle the effects of temperature and density on the shape of the bands we here calculate NLTE molecular spectra for a grid of different temperatures and densities.
One parameter is the total density $n_{tot}$. We then set ratios of electron density $n_e = 0.1 \times n_{tot}$ and molecule density $n_{mol}=10^{-3} \times n_{tot}$, which corresponds roughly to the standard model results (which is 11\% ionized, and CO makes up $6\times 10^{-4}$ of the total density at 100 days).
It should be noted that these results are not from full \textsc{Sumo} simulations, as the temperature and molecular formation are not calculated self-consistently, and we assume that the lines do not interact (self-absorption is included by the Sobolev formalism but we ignore line-to-line absorption). 
These types of calculations will, however, give an indication of how the shape of the bands depends on density and temperature when varied independently. 

Fig.~\ref{fig:shape_overview} shows the CO first overtone, at 100 days, for a range of different densities and temperatures. 
For reference, at this time instance the total number density of the O/C zone, where the majority of CO is formed, is $8.8 \times 10^{9}$ cm$^{-3}$ and the temperature is 5110 K. 
The black lines of each panel are the resulting bands for the specified temperature and density, and the gray lines are the band shape of the lowest temperature and lowest density plotted as a reference. All lines have been normalized to have peak flux unity. 

By comparing the resulting black lines with the gray reference lines we can deduce that increasing temperature and increasing density have similar effects. 
The line shapes become broader and higher-order band heads (e.g., 3-1, 4-2) become more prominent as higher vibrational bands are more populated.
There is then a degeneracy, and changes to the line shape can be driven by either of these factors.
While this broadening is larger for the investigated range of densities, the density range is 4 orders of magnitude in comparison to a factor 2 range in temperature. 
Increasing the density also shifts the maximum of the band to longer wavelengths.
This is the tendency when increasing the temperature as well, however the effect is smaller for the investigated parameter space. Thus, band head position may be linked to density constraints.

The pink curve in the figures represents the observed CO first overtone in the Type Ic supernova SN2007gr, from \citet{hunterExtensiveOpticalNearinfrared2009}, at 137 days after explosion (which are the observations with best signal-to-noise available).
For comparison to the model results, the observations have similarly been normalized, and are over-plotted on the density and temperature combinations that give model spectra agreeing reasonably well with the observations. 
Several parameter combinations reproduce the line shape of these observations to similar degree. This demonstrates the degeneracy between temperature and density, as the observations can be fit with either a model with low density at high temperature, or with higher density but lower temperature. 
Information from these line shapes thus probably need to be complemented with other diagnostics. For other molecules the trend is similar.
While the details might differ, due to the inherent differences between the various molecules, increasing the temperature or the density leads to a broader line in both cases. 
Typically, very high densities will also shifts the line maximum to longer wavelengths, as for the CO $dv=2$ line.

It should be noted that these calculations assume that the lines in the band are optically thin; this likely does not hold for any of the fundamental bands of the included molecules at early times ($\lesssim$200 days). 
Additionally, some of the lines in the CO $d\nu=1$ band become optically thick again after 500 days, with the large increase in CO mass. 
This makes the line shape of these bands more complex and requires full \textsc{Sumo} calculations to predict more accurately.

% \arjc{En diskussion om vi har har ideer för vad observationer ger oss, eller vilka som behövs. Är tex spektralformen annorlunda för olika densitet, grundton/överton ratios? Kanske kan man plotta linjeprofilerna normaliserade till molekylmassan för att få fram skillnader i form..}

% \fixme{du har ett samband mellan: 

% densitiet <-> linjeform (high dens-> high e-dens -> more excited molecule -> diff shape i.e. wider lines)

% temperatur <-> linjeform 1) (higher temp -> more excited molecule -> different shape) 2) (higher temp -> more ionized -> higher e-dens -> more excited mol)

% densitet <-> temperatur (high dens -> more molecules -> more efficient cooling)

% temperatur + densitet <-> nlte (higher dens or temp -> closer to nlte)

% sjukt svårt att disentagla vad som händer när man ändrar en grej när man kör hela sumo-körningar, eftersom det är så icke-linjärt, dvs ändra sak a -> ändring i sak b, men oklart om det beror på sak a eller på sekondära effekter, e.g. ändring i densitet leder till ändrad form men är det pga ändring i densitet eller ändring i temperatur?
% borde kanske göra tester med nlte.f90 där allting hålls konstant förutom en grej då, om man vill vara super-konkret. annars är det svårt att säga om man jämnför äpplen med päron. generella slutsatser som att en mer exiterad molekul leder till en bredare linje kan man nog skriva utan att skämmas.
% banden är också en blandning av flera olika linjer, där vissa kan vara betydligt mer optiskt tjocka än andra. 
% }

\section{Parameter space investigation}
\label{sect:sens_study}

 \begin{figure*}
  \centering
  \includegraphics[width=0.49\textwidth]{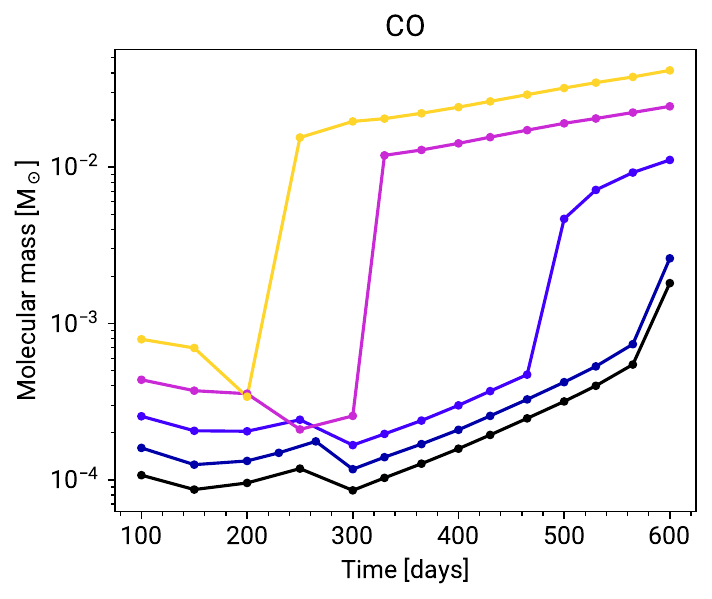}
  \includegraphics[width=0.49\textwidth]{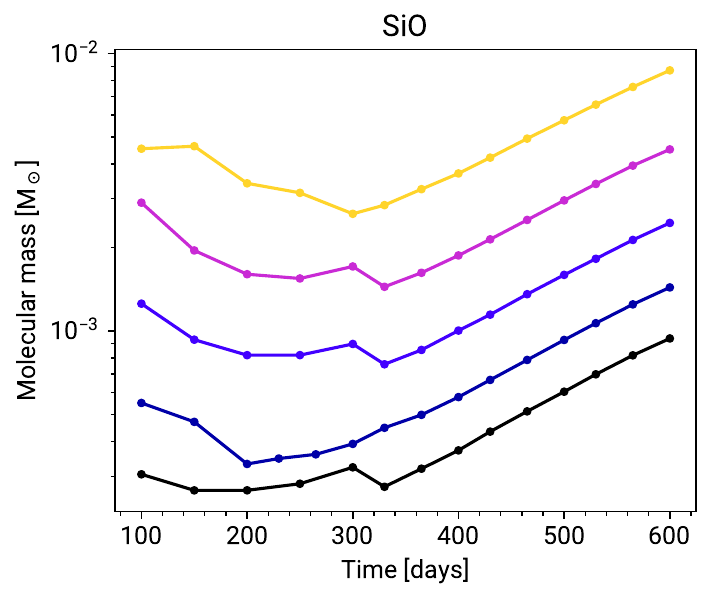}
  \includegraphics[width=0.49\textwidth]{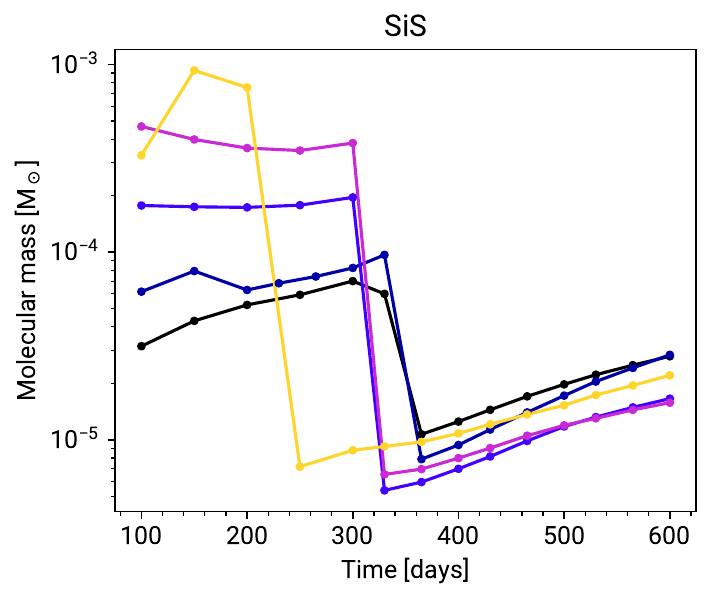}
  \includegraphics[width=0.49\textwidth]{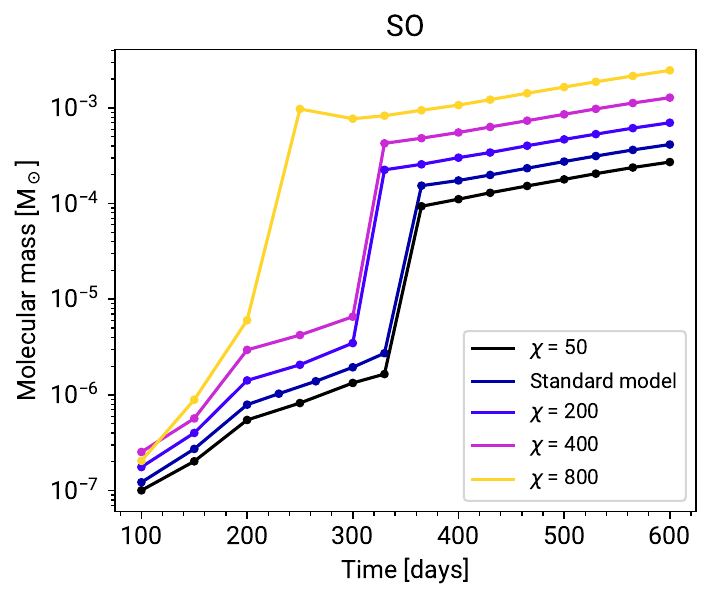}
      \caption{The evolution of the masses of CO, SiO, SiS and SO, for models with different $\chi$ values (higher $\chi$ means higher O-zone density).}
         \label{fig:ff_molmass}
  \end{figure*}

 \begin{figure*}
  \centering
   \includegraphics[width=0.49\textwidth]{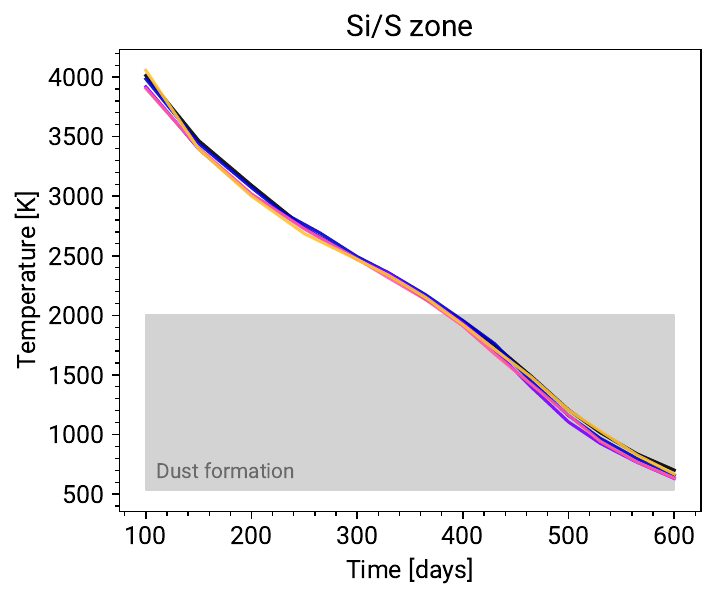}
  \includegraphics[width=0.49\textwidth]{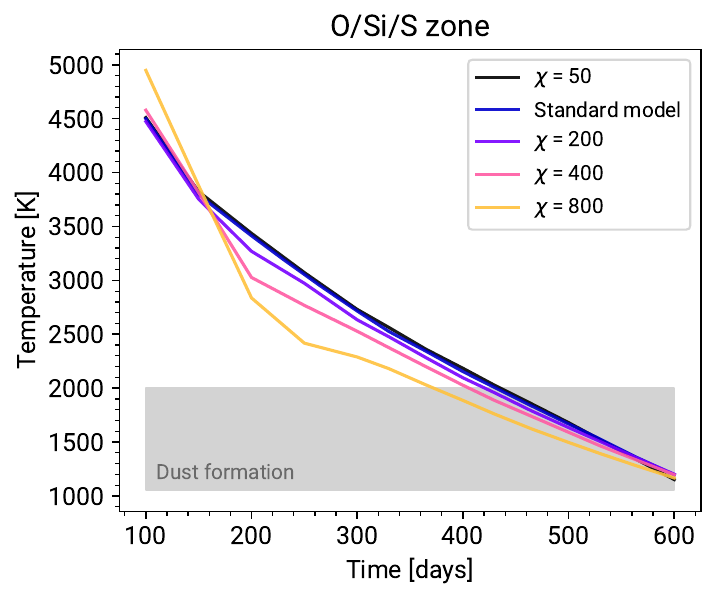}
   \includegraphics[width=0.49\textwidth]{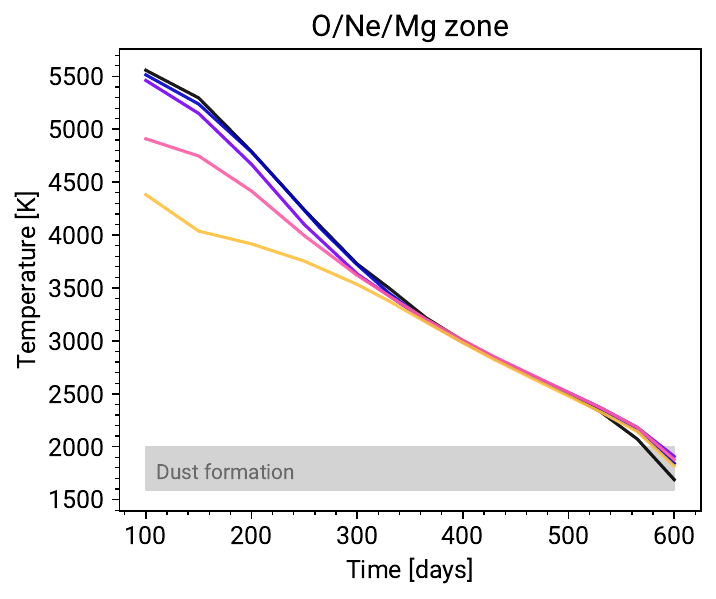}
  \includegraphics[width=0.49\textwidth]{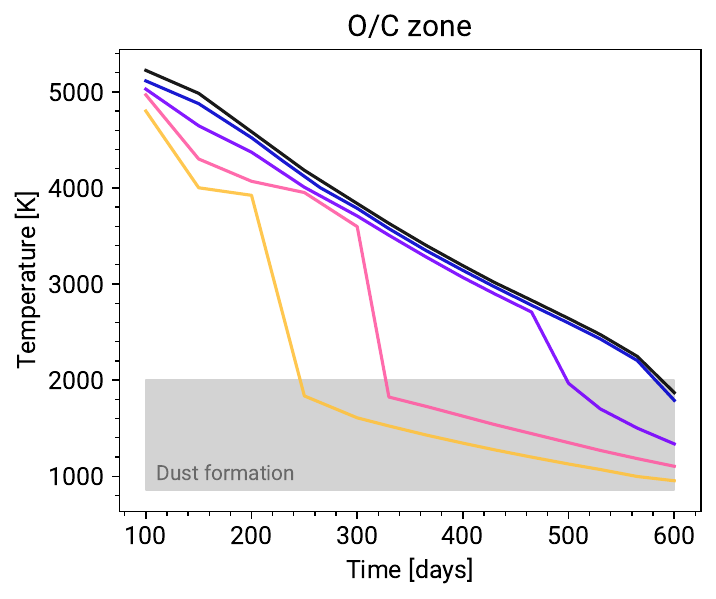}
      \caption{The thermal evolution in the Si/S, O/Si/S, O/Ne/Mg O/C zones, for models with different $\chi$ values. Assuming dust formation to begin at 2000 K, the models predicts an on-set of this sometime in the time interval 250-380 days. At high $\chi$ the first dust formed would be carbonaceous (in the O/C zone), whereas at low $\chi$-values it would be silicate-based (in the Si/S and O/Si/S zones).}
         \label{fig:ff_temp}
  \end{figure*}

From previous studies, it is known that ejecta structure can significantly affect the amount of molecules formed \citep[e.g.,][]{gearhartCarbonMonoxideFormation1999,cherchneffChemistryPopulationIII2009,liljegrenCarbonMonoxideFormation2020}. We perform here a small parameter study aimed to investigate the sensitivity to density for the Type Ibc case. 
This serves as a catch-all for varying either expansion velocity or the clumping of the gas. %here, to investigate how sensitive the models are to different filling factors, which is essentially the same thing as changing either the density in the model or the expansion velocity (larger velocity leads to a lower density, etc). 
We do this by varying $\chi$ in Eq.~\eqref{eq:den_struc}; higher $\chi$ means denser O-zones and less dense Fe/He and Si/S zones. We here test $\chi=[50, 200, 400, 800]$ (the standard model has $\chi=100$.)
These values correspond to O-zone overdensities of 1.8-14.
For each $\chi$ model SUMO is rerun completely, meaning also variations in gamma-deposition per zone, etc.

\subsection{Molecular masses and cooling}
\label{sect:ff_molmass-cool}

Fig.~\ref{fig:ff_molmass} shows how the molecular masses are affected by density variations. %the changes in the molecular masses for the CO, SiO, SiS and SO. 
Overall, the trends and  patterns seen in the standard model (discussed in detail in Sect.~\ref{sect:evol_molmass}) are retained when changing the density. The timings of those trends are, however, shifted due to the change in the physical conditions. %which influence both molecular formation and cooling, and the heating and cooling by the atomic gas. 
With a few exceptions, the molecular abundances increase when the zone densities increase (the Si/S zone density decreases when the O-zone densities increase). 
%The earliest epochs sometimes show a somewhat more complex behaviour, but from $\sim$200 days this trend is uniform and robust XXX Is this really true for SiS  -consider removing the sentence and joining this paragraph with the previous one XXX. 

For CO, %we generally see an increase in the amount of CO formed at almost all time instances, when increasing the density. 
the steep increase of CO mass, which occurs at 600d days in the standard model due to the reduced efficiency of the destruction reaction~\eqref{eq:co+o+} at lower temperatures, occurs at significantly earlier times in higher density models - down to 200 days in the $\chi=800$ case.
The increase in CO abundance leads to a stronger cooling, which lowers the temperature compared to the standard model and shuts off reaction~\eqref{eq:co+o+} earlier. 
This lowering in temperature in the higher density models is shown in the lower right panel of Fig.~\ref{fig:ff_temp}, where the O/C zone temperature evolution for the different models is plotted. 
As seen, the higher density models will have lower temperatures at all stages. Consistently, the lower density model ($\chi=50$) has a lower CO abundance, a later increase in CO mass, and a later decrease in temperature, following the same logic.  At 600 days the CO mass is $\sim 10^{-3}$~\msun~for the lowest density model, and $\sim 5 \times 10^{-2}$~\msun~for the highest density model.

SiO shows a very similar trend for all models at different phases, with an increased SiO molecular mass with increased density. The SiO mass in the highest density $\chi=800$ model is larger at 600 days, and the largest mass of any of the investigated models. 

For SiS and SO, which have strongly connected chemistry (Sect.~\ref{sect:evol_molmass}), the overall trends are also the same as in the standard model. 
% Due to the interaction between molecular formation, cooling and the thermal evolution of models,
The higher density models form more molecules at earlier times, which in turn cools the surroundings more efficiently. 
The behavior of these two species then depends on the temperature-sensitive reaction~\eqref{eq:so2s2}, as previously discussed.
As the temperature is lower in the high density models there is consequently a decline in SiS mass and associated rise in SO mass at earlier phases: 300 days for the $\chi=600$ model, and 200 days for the $\chi=800$ model, compared to the 330 days for the standard model.

The temperature evolution for the O/Si/S zone is shown in the upper right panel of Fig.~\ref{fig:ff_temp}. 
The increased molecular abundance in this zone influences the temperature; a higher density leads more than 1000 K lower temperature at some of the earlier phases, between 200 and 400 days.
This is primarily due to the increase in SiO, which provides $\sim$80\% of the cooling at 200 days for the $\chi=800$ model. 
There are smaller cooling contribution from SiS and SO, as well, for the respective models where they are abundant. 

Similarly, the increase in molecules also affects the temperature in the O/Ne/Mg zone.
% \arjc{Check if molecules actually have any impact, masses are small..} 
As there is very little sulfur in this zone, the most abundantly formed molecules are CO, which is the main molecular coolant, and SiO.
The difference in temperature is up to $\sim1500$ K in the highest-density model, at 100 days.
After 300 days we see little difference in the temperature between the investigated models. 
As seen in Fig.~\ref{fig:ff_temp}, there is no significant cooling either due to more molecules or the increased densities in the Si/S zone, for any of the investigated models.

Dust can possibly start forming once the temperature falls below $\sim$2000K \citep[e.g.,][]{tielens2005,sarangiChemicallyControlledSynthesis2013,sarangiCondensationDustEjecta2015,sluderMolecularNucleationTheory2018}, which is indicated in the figure.
%It should be noted that this is an estimation; the dust formation process is complex and likely out of LTE, and it depends on other environmental factors as well, such as elemental abundances and pressure.
Some commonly found dust species, such as amorphous carbon or corundum dust (\ch{Al2O3} monomers) can condense at higher temperatures than for example fosterite dust (\ch{Mg2SiO4} monomers) or iron dust \citep{gail_2010,lattimer_1978}.
The indicated start of dust formation then represents a lower limit on the time at which dust can \textit{potentially} form, in the different zones. 
Fig.~\ref{fig:ff_temp} shows that, depending on the O-zone densities, such a transition is predicted to begin in the range 250-380 days.

% \fixme{more discussion about dust?}

%  \arjc{We could add a discussion of observational constraints on this. While likely there are no direct MIR observations (however, check 2011dh), changes in optical or NIR photometry, or appearance of optical line distortions can reveal dust formation (as it did for e.g. 87A). Ergon 2015: ``Except for SNe 2011dh and 1993J and the peculiar Type Ibn SN 2008jc (Smith et al. 2008), we have found no reports of dust emission in Type IIb or stripped envelope SNe in the literature.'' Here 1993J was interpreted as CSM dust, an Ibn SNe would involve CSM components as well. I checked Matsuura 2017 (Handbook of SNe): She lists dust detections in four SESNe: 1990I (Ib, Elmhamdi), 2005at (Ic, Kankare 2014),2011dh (Ergon 2014,2015), 2013df (IIb, Szalai 2016). So clearly very little data on detection side.}

% \arj{Assuming that dust forms at $T\lesssim$ 2000 K (REF), Fig. X shows that the models predict a dust-formation onset epoch of 250-380 days. Observationally, indications of dust formation in stripped-envelope SNe have been reported for SN 1990I (ref), SN 2005at (ref), SN 2011dh (ref), and SN 2013df (ref). The data here....}

% \arjc{One should also consider SESNe not showing any signs of dust formation. Here...} \arjc{Discuss some well-observed SNe where there appears to be no dust formation at any epoch} 

% The lower tempeartures means that this reaction becomes inefficient at a earlier time, which in turn leads to the decrease of SiS and increase of SO occurring at earlier times for higher density models. 

\subsection{Changes to the IR spectra}

  \begin{figure*}
  \centering
  \includegraphics[width=\hsize]{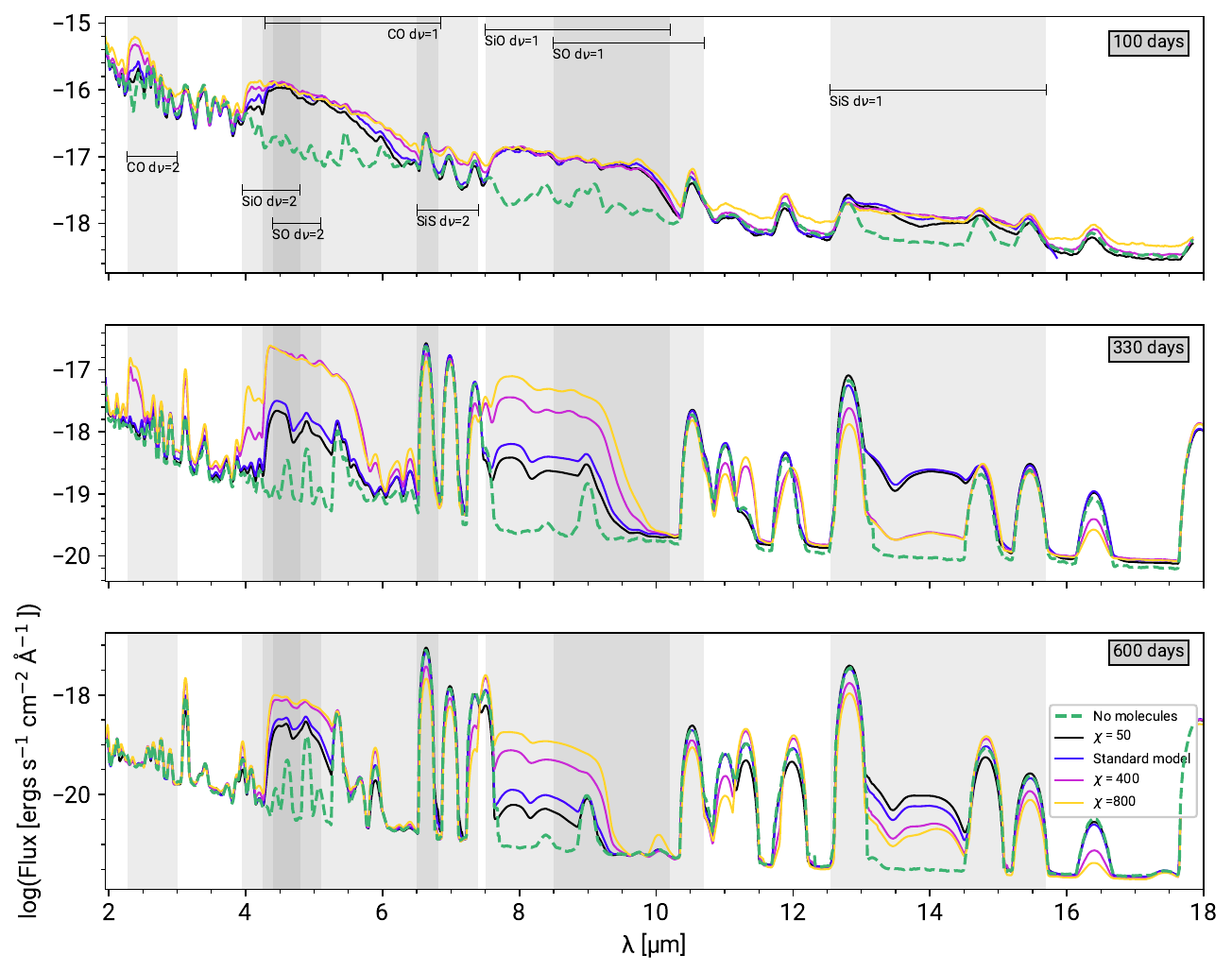}
      \caption{The spectra of models with different densities, at three different times. The fundamental and first overtone emissions of CO, SiO, SiS and SO are marked.}
         \label{fig:ir_spec_ff}
  \end{figure*}

The difference in density, molecular formation, and temperature will also influence the IR spectra, which is shown in 
Fig.~\ref{fig:ir_spec_ff} where the emission is plotted for models with different $\chi$, at three time instances.
It is clear that molecular contribution is important in this spectral domain; for all the models with molecules there is significant molecular emission for the three time instances, for all densities. 

Increasing the density of the models typically leads to a stronger molecular emission, however, the effects are complex. 
At 100 days the emission from the higher-density models are very similar to that of the lower-density models, due to optical depth effects as many lines are optically thick at this epoch.
Consequently, observations of these times could be used to constrain other quantities such as zone volumes.

At 330 and 600 days, the differences are much larger, and the emission of the CO and SiO fundamental bands are more than an order of magnitude stronger at the highest O-zone density ($\chi=800$) compared to the standard model.
Observations at these epochs could therefore be potentially used to deduce the O-zone densities. 

Some of the complex behaviour for the molecular masses also imprints on the spectra; at 330 days the $\chi = 800$ model has very weak SiS fundamental band emission because the SiS mass has decreased significantly in this model at this time (as discussed in Sect.~\ref{sect:ff_molmass-cool}). 
At 500 days the SiS emission is not visible for any of the models. 

For the higher-density models, the CO, SiO and SiS first overtone bands become much stronger, even though at first glance this can seem counter-intuitive as the temperatures of these models are lower. 
However, several different processes influence how strong the molecular emission for a species is, and they can interact in a complex manner, which makes the impact of for example changing the density in the models complex to deduce. 
A higher excited population of molecules will typically lead to a stronger first overtone emission. 
The electron collision excitation rates, the main excitation mechanism for molecules, increase with temperature (which is lower at higher densities) and the electron number density $n_e$ (which is higher at higher density). 
Additionally, the Sobolev self-absorption becomes more prominent at higher densities, which can lead to higher excitation of the molecules. 
The net effect for the mentioned species is a higher excitation, and, thus, more emission in the first overtone band. 
% \arjc{Perhaps this is mostly driven by fundamental band becoming more and more optically thick, so its emission gets weaker..while overtone remains thin?}

For SiS, the band at 7~\micron~is visible at 100 days for both the $\chi=400$ and $\chi=800$ models, which is not the case for the standard model.  
The SiO first overtone band is much stronger at 330 days.
The CO first overtone band is also more prominent; at 100 days this emission dominates the region, with a peak flux similar to the strong CO fundamental band. 

\section{Optical spectra}
\label{sect:optical_spectra}
  \begin{figure*}
  \centering
  \includegraphics[width=\hsize]{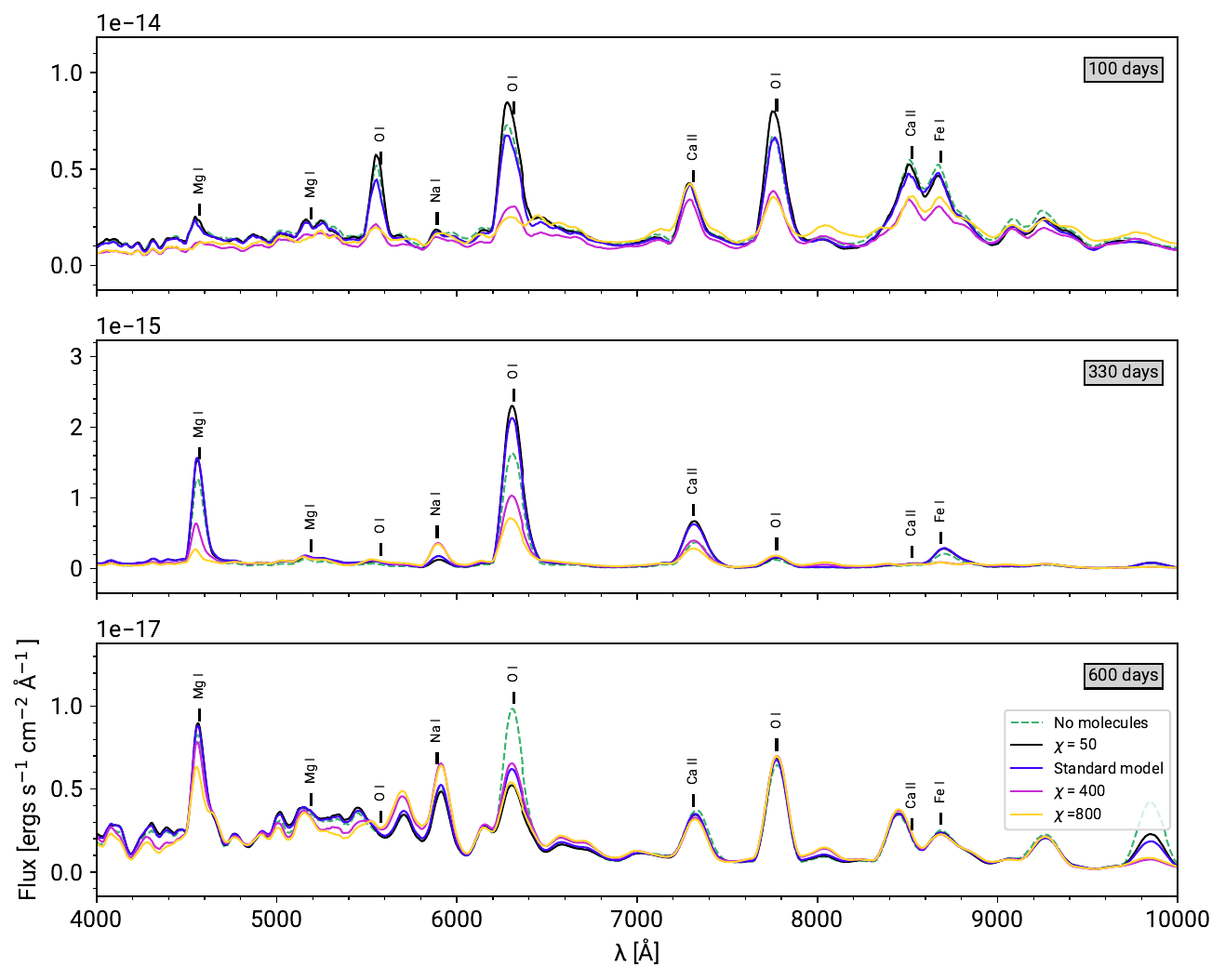}
      \caption{Optical spectra at three different times for different densities as labeled. By comparing to Fig. \ref{fig:optical_nomol}, one can conclude that the [O I] 6364, 6364 reduction for higher $\chi$ is driven by molecular cooling at 330 and 600d, whereas at 100d is it driven by other density-dependent processes.}
         \label{fig:optical_spectra}
  \end{figure*}

While IR observations can enable direct detection of molecular emission, the presence of molecules may also influence the optical spectra.
This can happen by a few different mechanisms. 1) Atoms get locked up into molecules. 2) Cooling by the molecules can decrease the zone temperature. 3) The molecular opacity can change the radiation field. Here we focus on the first two effects. We include rovibrational line opacity in the transfer, however this likely have quite minor effects on the radiation field. 
%We do not include the molecular electronic UV excitations in the radiative transfer. 

% \arjc{Should mention this also in methodology section somewhere} %\arjc{Kör du med simuleringar upp till 2.5 microns i SUMO? Vi borde bena ut lite om transfer genom rovibrational banden har någon signifikans.} %The consequence of either of these effects would be a weaker emission in the optical range for atomic lines. AJ: Det är inte uppenbart att (1) leder till svagare atomic emission, energin måste fortfarande emitteras ut.

In our standard model, the molecules do not form in large enough amounts to act as significant sinks for any atomic species to affect their emission by the first mechanism. % (discussed in Sect.~\ref{sect:evol_molmass}). 
The cooling by the molecules can be significant, however, lowering the temperature by several hundred degrees at specific times. The amount of molecular cooling varies with time and is very zone-dependent, its impact is thus quite complex.  In the O/C zone, there is an effect at early and late times, but not at intermediate times. %are also both initial cooling effects and a large cooling that occurs later in the evolution due to the increased CO mass. 
Fig.~\ref{fig:optical_spectra} shows the optical spectra from 4 000 \AA~to 10 000 \AA~at three times (100, 330 and 600 days) for the standard model (dark blue line), the standard model without any molecules included (green dashed line), and three different density models ($\chi = 50$, black line, $\chi = 400$, pink line, $\chi = 800$, yellow line). 

Comparing the standard model with molecules (blue solid line) and without (green dashed line), there are some differences. At 100 days, some cooling due to molecules occurs in the O/Si/S and C/O zones, which slightly impacts the [O I] 6300, 6364~{\AA} line and the Ca II and [Fe I] blends at 8400-8800{~\AA}.
At 330 days there is no longer significant molecular cooling in any of the zones, and the spectra of the standard model are more or less identical to the spectra of the model with no molecules. 
At 600 days, the O/C zone is cooled by the increased amount of CO. 
This has a large impact on [O I] 6300, 6364~{\AA}~(almost 50\% weaker), which is commonly used as diagnostic for the core mass. 
For the standard model, we can conclude that the optical spectra are only slightly affected by molecule formation and cooling, except for the mentioned [O I] 6300, 6364~{\AA} feature which can become significantly quenched at late times.

An import aspect of how results may vary for other CO core masses is that for this relatively massive one (6.8 \msun), the molecule-forming zones (Si/O and O/C) constitute only 43\% of the total O-zone mass (with 57\% being in the O/Ne/Mg zone). At lower CO core masses, this number is typically higher, for example 68\% in a 4 $M_\odot$ CO core \citep{jerkstrandLatetimeSpectralLine2015}. As such, the spectral impact of molecule formation is almost certainly larger for smaller CO cores. However, in the context of trying to determine whether some Ibc SNe come from relatively massive Wolf-Rayet progenitors, the results here suggest that optical models ignoring molecular chemistry are not likely to severely underestimate the CO core mass - optical emission is not predicted to be strongly quenched by molecular IR cooling - at least not in the early nebular phase.

Fig.~\ref{fig:optical_spectra} also shows the optical spectra of models with different densities; $\chi = 50$ (black line), $\chi =400$ (pink lines) and $\chi = 800$ (yellow line).
The variation in the spectra are due to two effects; the denser models tend to produce more molecules, which in turn leads to a larger cooling, and higher density in itself can change the emission even if the molecular impact stays the same. To disentangle these effects we plot in Fig. ~\ref{fig:optical_nomol} spectra for models without molecules for different $\chi$.
%It should therefore be emphasised that the standard model without molecules will differ somewhat from e.g. the $\chi = 50$ model without molecules (see Fig.~\ref{fig:optical_nomol}, where the models with different densities without molecules are plotted). 

From Fig. \ref{fig:optical_spectra}, the single model with lower O-zone densities ($\chi=50$) has a spectrum is very similar to the standard model at all times. The higher-density models, however, differ significantly. % in partcular the $\chi=400$ and $\chi=800$ ones. 
At 100 days both the $\chi=400$ and the $\chi=800$ models have less distinct O-zone emission lines. Inspection of Fig. \ref{fig:optical_nomol} shows that this is so also with molecules switched off - it is therefore an effect not driven by the chemistry.

At 330d, however, the [O I] 6300, 6364 feature is not sensitive to density in the molecule-free models, but is drastically weakened (factor 2) in the chemical models. Mg I] 4571 is, on the other hand, still density-sensitive also without molecular effects.

At 600d, higher density gives stronger [O I] 6300, 6364 in molecule-free models, but weaker [O I] 6300, 6364 in models with molecule cooling. At any $\chi$, molecular cooling here strongly reduces the strength of the line. Other lines are however little affected.

\section{Comparison with molecular emission observations}
\label{sect:comp_obs}

\begin{table*}
\caption{Overview of the SN observations used in Fig.~\ref{fig:obs_comp}.}             % title of Table
\label{table:co1st_obs}      % is used to refer this table in the text
\centering      % used for centering table
\def\arraystretch{1.2}
\begin{tabular}{c c c c c c}        % centered columns (4 columns)
\hline\hline                 % inserts double horizontal
SN & SN type & Reference & Day after explosion & Distance [Mpc] & \ch{^{56}Ni} mass \\ \hline
SN2000ew & Ic & \citet{gerardyCarbonMonoxideType2002} & 97 &15 & - \\
SN2007gr	&Ic& \citet{hunterExtensiveOpticalNearinfrared2009} &	137&	9.29	&0.076$\pm$0.02 \\
SN2011dh&	IIb & \citet{ergonTypeIIbSN2015} &	206&	7.8	&0.075$\pm$0.025 \\
SN2020oi &Ic	&\citet{rhoNearInfraredOpticalObservations2021}	&63	&114	&0.4
\end{tabular}
\end{table*}

\begin{figure}
\centering
\includegraphics[width=\hsize]{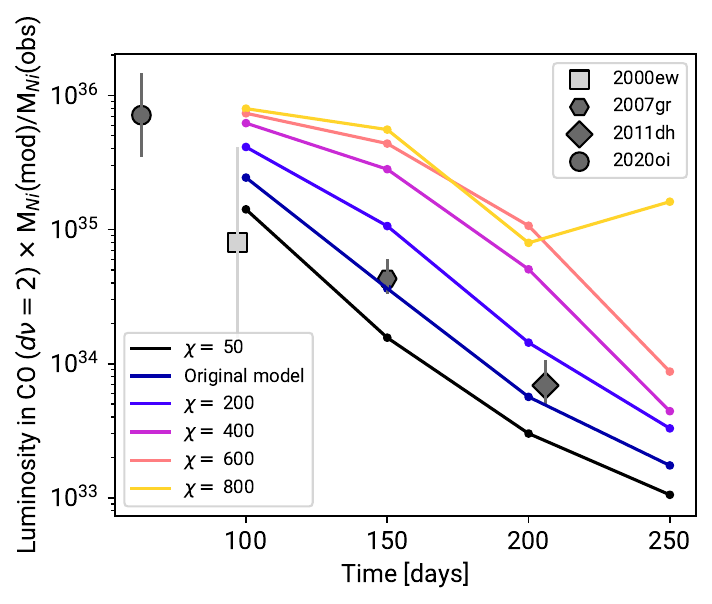}
\caption{Total luminosity in the first overtone of CO, for models with different $\chi$ values (lines), compared to observations of stripped-envelope supernovae (data taken from \citet{gerardyCarbonMonoxideType2002}, \citet{hunterExtensiveOpticalNearinfrared2009}, \citet{ergonTypeIIbSN2015} and \citet{rhoNearInfraredOpticalObservations2021}). The observed values are scaled with the difference in \ch{^{56}Ni} mass between the model and the SN.
For SN 2000ew there is no observational estimate of this so the scaling factor is set to 1, with an error bar of factor 5 attached. The error bars indicate the uncertainties in the \ch{^{56}Ni} mass when such are reported. There is no reported error for SN 2020oi, so it was arbitrarily set to be a factor 2.}
\label{fig:obs_comp}
\end{figure}

%While there are very few observations of molecules in type Ib or Type Ic SNe, and we are not trying the fit any observations with our model in this paper, we here compare the general trends available. 
All spectral observations of molecules in Type Ibc SNe have been of the first CO overtone emission (see Table \ref{table:co1st_obs}), which is observable from the ground. While it should be emphasized that we are not attempting to reproduce any specific SN with our single model here, we nevertheless compare the model luminosity in the CO first overtone to the available observations to gauge if the model is in the right ballpark for this feature. The four observed SNe (three Type Ic and one Type IIb) to which we are comparing the model results are listed in Table~\ref{table:co1st_obs}. 

Fig.~\ref{fig:obs_comp} shows the observations as scatter points and the results for models with different $\chi$ values as the lines.  
The measured luminosities are additionally scaled with the \ch{^{56}Ni} mass of the model divided by the known \ch{^{56}Ni} mass of the SN. This is to take differences in total powering level into account. For SN 2000ew no \ch{^{56}Ni} mass has been estimated, we then put this factor to 1.

As can be seen, the models predict an increase in the degree of O-zone clumping ($\chi$) to give a higher luminosity in the CO first overtone. The growth flattens out as one exceeds $\chi \sim$600. The strong increase in emission seen in the $\chi=800$ model between 200 and 250 days is due to the rapid CO mass growth that occurs in these models at that time.

Comparing the modeling results with the observations we can conclude that the modeling results fall into the same range as the observed CO first overtone luminosities.
At face value, there is a good match between the observations and the standard model, but we do not ascribe any particular significance to this until more detailed modeling where also other properties of each SN are taken into account. Rather, the takeaway point from the figure is that the modeling does not produce results in any glaring conflict with current observations of the CO overtone in stripped-envelope SNe, and this is encouraging for the methodology and continued work.

% the models fall into the observational range of CO first overtone luminosity.

\section{Future possibilities for SNe IR spectra observations}
\label{sect:obs_jwst}
To get more information about different types of molecules observations throughout the infrared region are needed. While these SNe, in their nebular evolution, will be the brightest early (around 100-200 days) observations from two or more epochs would be desirable to investigate the time evolution of the formed molecules. 
A suggestion based on the model results from this work would be to observe at $\sim$100 days and $\sim$350 days. 

At $\sim$100 days, the CO, SiO, and SiS fundamental bands, as well as the CO and SiO first overtones, are predicted to be observable. 
As the molecular fundamental bands are insensitive to densities at these times (discussed in Sect.~\ref{sect:midir_spec}), observations could be used to investigate other properties, such as zone masses or volumes. It is a promising epoch to search for the first SiS detection in an SN.

At $\sim$350 days, SiS becomes weak in the model, but SO has become abundant and its fundamental band contributes significantly to the blend with the SiO fundamental band at 7-10 \micron. 
At these epochs, there are large differences in the molecular emission between models with different densities, which therefore can be constrained. It is a promising epoch for the first SO detection in an SN.
%The combination of the two observations could then be used to probe the densities of the oxygen zones.
%Performing such observations will however require a new generation of telescopes, as the described SN spectrum with interesting molecular emission is either too faint or outside the spectral range of the current generation telescopes.

The IR regions of the spectrum discussed in Sect.~\ref{sect:midir_spec} will fall within the wavelength coverage of JWST. The fundamental and first overtone of CO and the first overtone of SiO are in the range of JWST/NIRSpec %postsubmit \arjc{Specify the NIRSpec range}
, while the bands further to the red will be observable with the Medium Resolution Spectroscopy (MRS) instrument of JWST/MIRI. %postdubmit \arjc{Specify the MRS range}. 

To investigate the feasibility of observing the different molecular lines, at different phases with JWST, the JWST Exposure Time Calculator\footnote{\href{https://jwst.etc.stsci.edu/}{https://jwst.etc.stsci.edu/}}\citep{pontoppidan_etc_2016}
was used, with the model spectra presented in
Fig.~\ref{fig:irspec} as template. 
With the ETC tool, the exposure time needed for the relevant filters to achieve a given S/N for the line peaks can be estimated. We placed the SN at a distance of 10 Mpc for these calculations. %, or if it is possible to achieve some SNR within a reasonable exposure time. 

At 100 days the first overtone of CO and the blend between the fundamental band of CO and the first overtone of SiO are covered by JWST/NIRSpec with the filters G235M and G395M,  respectively, and bright enough that short exposure times ($\sim$1 min per filter) will yield SNR $>$50. 
% \arjc{This seems much too short exposure times?}
The SiO and SiS fundamental bands are covered by JWST/MIRI MRS filters and require exposure times of $\sim$2 min for the SiO line and $\sim$5 min for the SiS line to get an SNR $>$10.
Consequently, at 100 days the discussed molecular lines could with ease be observed with JWST for an SN similar to the standard model in this work. 

At 360 days the SN is significantly dimmer and while longer exposure times are needed observations at these times are still attainable.
The strong CO fundamental band still needs short exposure times only ($\sim$1 min yields an SNR of $\sim20$).
However, the SiO and SO fundamental band blend require a longer exposure time of $\sim$1 h to get an SNR > 10. The SiO emission may also be quenched by silicate dust depletation at these times \citep{sarangiChemicallyControlledSynthesis2013}.
The SiS fundamental band flux predicted by our models is at this time not bright enough to be observable. 

Observations beyond 400 days become challenging; the fundamental band of CO is still observable with JWST/NIRSpec at 600 days at achievable exposure times, however, bands of the other molecules would require a very long exposure. 
% becomes too faint to be observable within reasonable exposure times with the JWST/MIRI. 
Observations at these epochs would realistically require an SN significantly closer than 10 Mpc.  

It is therefore realistically achievable to observe two epochs of an SN at $\sim$10 Mpc at $\sim$100 and $\sim$350d. Such observations would be of great value to extend our understanding of Type Ibc SNe, both in terms of the evolution of molecules and early stages of the dust production, and for pinning down the composition and the ejecta mass of the SN. This would give insight into the possible progenitors of stripped SNe. 
% \arjc{Possibly make a bullet list of some concrete ``firsts'' one could achieve?}

\section{Summary and conclusion}
\label{sec:summary}
We have presented the inclusion of molecular chemistry and cooling into the NLTE spectral synthesis code SUMO for supernova modeling. The treatment involves the calculation of Compton destruction rates for molecules, a large chemical kinetic network, and NLTE ionization, excitation, and cooling by CO, SiO, SiS and SO. We carry out the calculations with fully time-dependent atomic and molecular kinetics, and temperature.

We have applied this new framework to investigate how molecules form and affect physical conditions, and optical and IR emission in a Type Ic supernova (ejecta mass 5 \msun) over the 100-600d time interval.
%In this work, we have modeled the molecular formation and cooling, as well as the resulting spectra, of a Type Ic SN.  The standard model includes 5 zones - Fe/He, Si/S, O/Si/S, O/Ne/Mg, and O/C - and the molecules that form depend on the physical conditions and compositions of the particular zone. 
%Several improvements have been made to the molecular treatment in \textsc{Sumo}; the molecular destruction rate by Compton electron is now directly calculated in the code, the collisional \\(de-)excitation rates for molecules are derived from data instead of estimated, and the chemical kinetic network has been expanded to include sulfur and silicon-bearing molecular species.  
Our main results are as follows:
\begin{itemize}
\item Most molecules have destruction governed by Compton electrons. From our \textsc{Sumo}-calculated destruction rates we provide updated values for the work-per-ion-par quantity $W$ suitable for SESNe: 23 eV for CO, 10 eV for SiO, 9 eV for SiS and 24 eV for SO. 
%postsubmit check SiO and SiS numbers again
%for CO is similar to the parameterized prescription, and the calculated values for $W_i$, described in \citet{liuOxygenTemperatureSN1995}; while in our calculations we get roughly $50\%$ higher value, the estimate captures the behavior quite well. This result holds for the different times and densities tested, which indicates that using the prescription is probably accurate to within a factor 2 for CO. This available prescription constant $W_i$ available was specifically calculated for CO, and when trying to apply it to other molecules the discrepancy is larger. When applied to SiO and SiS there is a difference of up to a factor 4, compared to the values calculated in this work. We provide new optical estimates for $W_i$ values to use.
    \item The five most abundant molecules predicted to form in stripped-envelope SNe are CO, SiO, SO, SiS, and \ch{S2}. 
    The CO, SiO, and SiS abundances increase over time and are most abundant at our last modeled epoch of 600d. SO and \ch{S2}, on the other hand, are most abundant at our first modeled epoch of 100 days. The maximum molecule masses produced are $2\times 10^{-3}$~\msun~of \ch{CO}, $3\times 10^{-3}$~\msun~of SiO, $7\times 10^{-4}$~\msun~of SO, $3\times 10^{-4}$~\msun~of \ch{S2}, and $1\times 10^{-4}$~\msun~of SiS. %Only a small fraction of the atoms are locked into molecules at 600 days; 5\% of C, 1\% of O, 0.4\% of Si, and 0.2\% of S. 
    
    \item CO is mainly formed in the O/C zone, while SiO and SO primarily form in the O/Si/S zone. The sulfur species, SiS and \ch{S2}, initially mostly forms in the O/Si/S zone, however, after 300 days they are most abundant in the Si/S zone. The molecules, therefore, provide tracers of three different nuclear burning layers: helium burning (CO), neon burning (SiO, SO, early SiS and \ch{S2}), and explosive oxygen burning (late SiS and \ch{S2}).
    %\item The final masses at 600 days are $\sim$2$\times 10^{-2}$\msun of \ch{CO}, $\sim$3$\times 10^{-3}$\msun of SiO, $\sim$7$\times 10^{-4}$\msun of SO, and $\sim$3$\times 10^{-5}$\msun of SiS and \ch{S2}. In the standard model only a small fraction of the atoms are locked into molecules at 600 days; 5\% of C, 1\% of O, 0.4\% of Si, and 0.2\% of S. 
     \item Molecular cooling occurs primarily in the O/C zone and in the O/Si/S zone, by CO and SiO, respectively. In our standard model, a modest temperature effect of a few hundred K occurs in these zones.
     \item The impact on optical spectra, such as on the [O I] 6300, 6364 feature commonly used to attempt O-mass determinations, depends on the O-zone density. After some transition time, this line is significantly quenched by molecule cooling. This occurs around 550d in our default density model but already at 200d in the highest density model. 
     \item Temperatures low enough to allow for the onset of dust formation ($\sim$2000 K) are reached at 350-450d for zones where silicate dust can form, and 250-600d for a zone where carbon dust can form. Formation starts earlier in the range for higher density models.
    \item The molecules are predicted to give strong ro-vibrational emission in the infrared, primarily from the fundamental bands of CO at 4.6 \micron, SiO at 7.9 \micron, and SiS at 13.2 \micron. The SO fundamental band, with a peak at 8.7 \micron, is blended with the stronger SiO fundamental band. However, after 350 days, when there is a large increase in SO mass, the models indicate it should be possible to separate the two bands and detect SO. 
    \item The first overtone emissions are predicted to be significantly weaker than the fundamental bands, and clearly distinguishable only for SiO (4.1 $\mu$m) and CO (2.3 $\mu$m). The SO overtone at 4.6 $\mu$m coincides with the much stronger CO fundamental band, and the SiS overtone at 6.8 $\mu$m coincides with much stronger atomic emission by [Ni II] 6.8 $\mu$m and [Ar II] 6.9 $\mu$m - both of these are therefore predicted to be difficult to detect.   We show that the luminosity of the CO first overtone is sensitive to the density of the O/C zone. %We found that increasing the density leads to an increase in the CO luminosity in a predictable manner.
    \item We investigate the observability of the predicted molecular IR emission with JWST. For typical distances (10 Mpc), most molecular emissions can be observed with good S/N with minutes or hours of exposure times. We identify an optimal strategy to observe the SN at 2 or 3 epochs in the 100-400d window. We demonstrate how such observations can give information on molecular masses and zone properties, and give the first detections of SiS and SO in supernovae.
\end{itemize}

The models presented in this study offer valuable insights into the molecular formation process in SESNe, and we predict the observability of different molecules at various stages of a SESN.
By combining these results with future observations obtained from state-of-the-art optical, near-IR, and MIR instruments (e.g., JWST), we have the potential to resolve degeneracies and gain a comprehensive understanding of the evolutionary trajectory of type Ibc SNe. 
Furthermore, these combined efforts may offer clues regarding the origin of SESNe.

\begin{acknowledgements}
SL and AJ acknowledge funding from the Swedish National Space Board  (SNSB/Rymdstyrelsen), grant 2017-R 95/17. AJ acknowledges funding from the European Research Council (ERC) under the European Union’s Horizon 2020 Research and Innovation Program (ERC Starting Grant 803189 – SUPERSPEC).
The computations were enabled by resources provided by the Swedish National Infrastructure for Computing (SNIC) at the PDC Center for High Performance Computing, KTH Royal Institute of Technology, partially funded by the Swedish Research Council through grant agreement no. 2018-05973.
PSB acknowledges support from the Swedish Research Council through an individual project grant with contract No. 2020-03404. 
RB and SNY acknowledge funding by the UK Space Agency and Science and Technology Funding Council (STFC) grant  ST/R000476/1 and ERC Advanced Grant Project No. 883830.
We thank Vincenzo Laporta for clarifications regarding calculations for electron collisions with CO.
\end{acknowledgements}

\bibliographystyle{aa}
\bibliography{ref,ref_manual}

\begin{appendix} %First appendix
\section{Molecular structure}
\label{app:model_mol}

% The only available data for the electron collision transition probabilites are for vibrational transitions (see Fig.\fixme{ref}). 
% However as the molecule bands are collections of rovibrational tranistions, we want to split the available vibrational transition probabilities into rovibrational transitions. 

% Data of cross-sections $\omega$ and, from this, the collisional transition rates $q$ are available for electron collision vibrational transitions $\nu \rightarrow \nu'$. 

% The current experimental and theoretical 

% The current experimental and theoretical works do not provide
% all the rate coeﬃcients needed to model the CO statistical ro-
% vibrational population of the ground X 1 Σ + and excited A 1 Π elec-
% tronic levels. Therefore, we have to resort to extrapolation and
% scaling laws to fill the gaps.

\begin{figure}
\centering
\includegraphics[width=0.9\hsize]{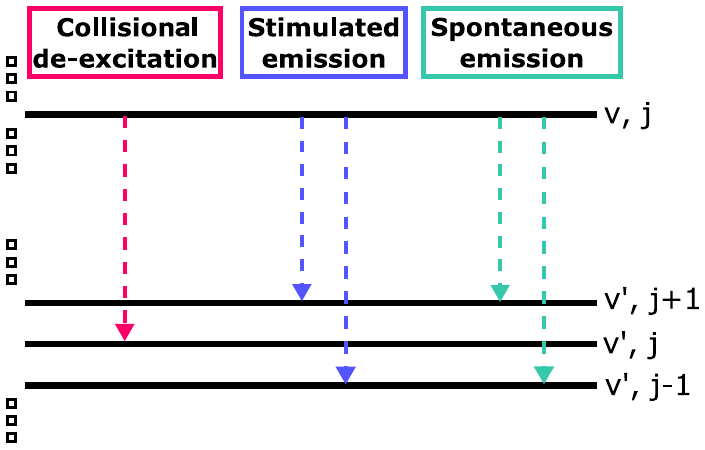}
\caption{The possible de-excitation channels for the molecules. }
%\label{fig:scheme1}
\label{fig:appa1}
\end{figure}

\begin{figure}
\centering
\includegraphics[width=0.9\hsize]{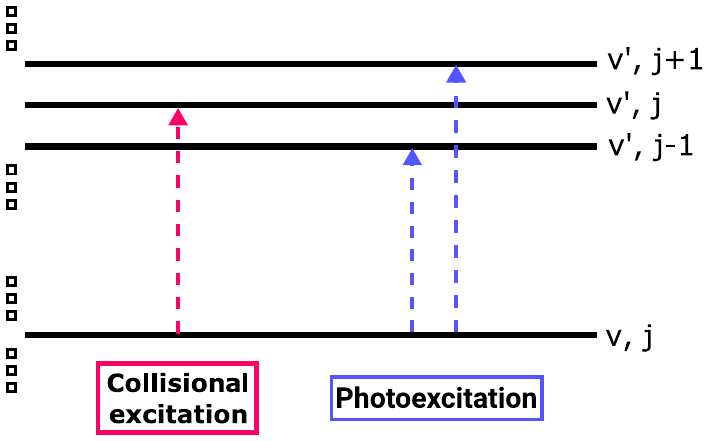}
\caption{The possible excitation channels for the molecules. }
%\label{fig:scheme1}
\label{fig:appa2}
\end{figure}

% \arjc{Start with description of energy levels, A-values, ionization energies? Then move on to collision x-sections of various kinds.}
Currently, neither theoretical nor experimental data is available for all the collision rates needed to model molecule statistical ro-vibrational populations in SNe environments. 
We here detail the assumptions and extrapolations used in this work to fill in the gaps in the data, and describe the setup of the model molecule.  

\subsection{General setup}

For a specific level in a molecule, the possible de-excitation channels are spontaneous emission, stimulated emission, and collisional de-excitation and possible excitation channels are collisional excitation and photoexcitation. 
The molecules included in the radiative transfer are all heteronuclear diatomic molecules, and follow the selection rules for electric dipole allowed rovibrational radiative transitions, which are $\Delta \nu = \pm 1, \pm2, $,..., and, $\Delta j = \pm 1$.
We assume that the dominant channel for $\nu$-changing collisions is rotationally elastic, such that the rotation number $j$ is not changed in such an interaction \citep{laportaElectronimpactResonantVibration2012}.
Finally, we assume that the rotational levels within a vibrational level are in relative thermal equilibrium, so that they are distributed according to a Boltzmann distribution with respect to each other. 
Consequently, if we know the total population of a specific vibrational level and the temperature, we can calculate the population of the rovibrational levels. 
A schematic of the possible excitation and de-excitation channels can be seen in Figs.~\ref{fig:appa1} and \ref{fig:appa2}. 

\subsection{Collisional vibrational transitions}

For collisions with electrons, cross-sections $\sigma(E)$ for vibrational transitions to the ground level, such as $\sigma(\nu=1 \rightarrow \nu =0)$, $\sigma(\nu=2 \rightarrow \nu =0)$ etc, are available for CO \citep{risticRateCoefficientsResonant2007}.
From these cross-sections, the rate coefficients as functions of temperature, $q(\nu \rightarrow \nu =0)(T)$, can then be calculated assuming a Maxwellian distribution for the thermal electrons.  Rate coefficients for excitation can consequently be obtained through detailed balance relations. 

To account for other such transitions in CO where no data are available, scaling laws have to be applied. There are several different implementations of such scaling laws available for collisional transitions, and while physical motivations are typically given for a specific formulation, there is no general consensus for which scheme is most accurate. This issue is discussed in detail in \citet{thiRadiationThermochemicalModels2013}, where they compared three different scaling laws: from \citet{scoville_collisional_1980},  from \citet{elitzur_vibrational_1983}, and from \citet{chandraCollisionalRatesVibrotational2001}. 

% Scoville et al. (1980) argued that the probabilities of collision-
% induced vibrational transitions are proportional to the corre-
% sponding radiative transition matrix elements (Born-Coulomb
% long range interaction). Therefore, the vibrational de-excitation
% rate coeﬃcients from v can be derived from the v = 1 0
% rate coeﬃcients:

\citet{scoville_collisional_1980}, basing their reasoning upon the Bethe approximation, suggests that the collisional transition probabilities are proportional to the spontaneous radiative transition probabilities $A$, such that $q(\nu \rightarrow \nu') / q(1 \rightarrow 0) \propto A(\nu \rightarrow \nu') / A(1 \rightarrow 0)$. 
Therefore, 
\begin{equation}
    q(\nu \rightarrow \nu') = \frac{A(\nu \rightarrow \nu')}{A(1 \rightarrow 0)} q(1 \rightarrow 0).
\end{equation}

\citet{elitzur_vibrational_1983} base their expression on the initial work by \citet{procaccia_vibrational_1975}, in which vibrational collisions were investigated and analyzed. 
They suggest that:
\begin{equation}
\begin{aligned}
    q(\nu \rightarrow \nu') = &(\nu - \nu') \times q(1 \rightarrow 0) \\
    &\times \exp \left(  - (\nu - \nu' -1) \frac{1.5 E_\nu / kT}{1 + 1.5 E_\nu / kT} \right),
\end{aligned}
\end{equation}
where $E_\nu$ is the energy level of $\nu$ with respect to the ground energy level, $T$ is the temperature, and $k$ is the Boltzmann constant. 

Based on the Landau-Teller relationship for transitions between adjacent states, \citet{chandraCollisionalRatesVibrotational2001} proposed the following relationship:
\begin{equation}
    q(\nu \rightarrow \nu') = \frac{\nu\left(2\nu' + 1\right)}{2\nu-1} \times q(1 \rightarrow 0).
\end{equation}

Another scaling law of interest is from \citet{liuOxygenTemperatureSN1995}, where they, similar to this work, investigate molecules in SNe environments.
Their physical motivation is in part similar to that of \citet{scoville_collisional_1980}, however, instead of scaling with the $A$-value, they use the absorption oscillator strength $f_{\nu \nu'}$ for a radiative transition from level $\nu$ to $\nu'$, namely
\begin{equation}
    q(\nu \rightarrow \nu') = \frac{f(\nu \rightarrow \nu')}{f(1 \rightarrow 0)} q(1 \rightarrow 0).
\end{equation}      

While the individual estimates for a specific collision rate can vary widely depending on which rule is used, we found through experimentation that the final level populations and, in particular, the final spectra are not very sensitive to this choice. This indicates that the excitation channels described with the scaling laws are secondary to those where data are available. These tests were done for both high-density and low-density models.

Throughout this work, we then chose to use the \citet{chandraCollisionalRatesVibrotational2001} scaling law as this is the most modern such scheme, and it is the one deemed most appropriate and used in \citet{thiRadiationThermochemicalModels2013}. 

\subsection{Collisional rovibrational transitions}

\begin{figure}
\centering
\label{fig:vibex}
\includegraphics[width=0.9\hsize]{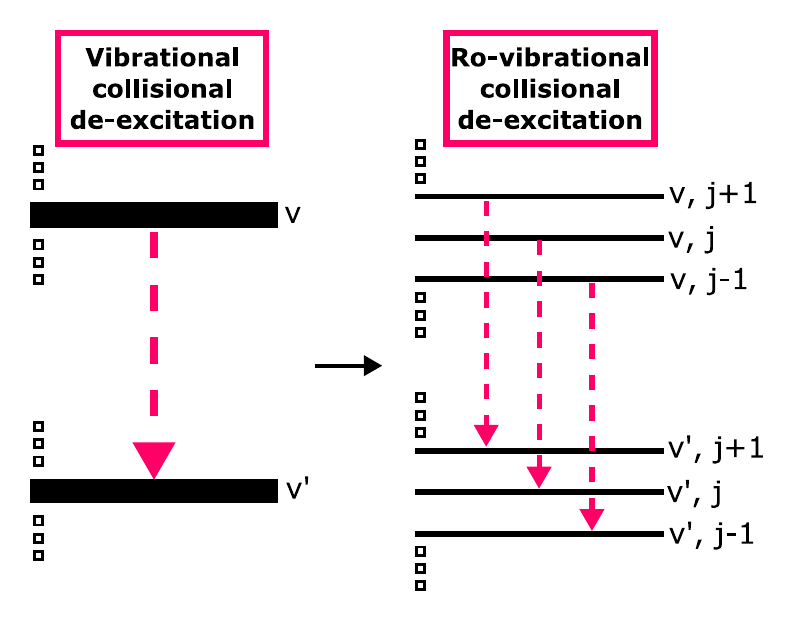}
\caption{Schematic of the vibrational collisional de-excitation through collisions with electrons. The left part shows the collisional transition rate coefficients available from experiments and theoretical work. The right part shows how we divide these into rovibrational rates.}
\label{fig:scheme1}
\end{figure}

From the data and scaling laws we obtain the collisional transition rate coefficient $q(\nu \rightarrow \nu')$, for vibrational transitions $\nu \rightarrow \nu'$. 
For the purpose of this work, however, we are interested in the rovibrational transitions, and, therefore, need to divide the available vibrational transition rate coefficients $q(\nu \rightarrow \nu')$ further into rovibrational transition rate coefficients $q(\nu, j \rightarrow \nu', j')$, as illustrated in Fig.~\ref{fig:vibex}. 
% To do this we want to know how $q$ changes with excitation level, i.e. the internal population distribution in the $j$ quantum number of a vibrational line.  
$q$ is likely independent, or only very weakly dependent on the $j$-number in the case of CO \citep{laportaElectronimpactResonantVibration2012}.
% \fixme{ref} showed that the $\sigma(\nu = 1, j = 0 \rightarrow \nu' = 0, j = 0)$ cross-section is essentially identical to the $\sigma(\nu = 1, j = 100 \rightarrow \nu' = 0, j = 100)$ cross-section, for CO-electron collisions . 

We therefore assume that the collisional rate coefficient $q$ is independent of the $j$-number, such that $q(\nu, j=0 \rightarrow \nu', j=0)=q(\nu, j \rightarrow \nu', j)=q(\nu \rightarrow \nu)$, for any $j$.
%Here, we assume that the internal levels in a vibrational band are in thermal equilibrium and that the dominant channel of collisional transitions is between rovibrational levels with the same rotational number $j$, i.e. $\nu, j \rightarrow \nu' j$.
This further ensures that the collisional transition rates between vibrational levels are conserved with respect to the vibrational transition rate coefficient from previous work, as the sum of all individual transition rates then is equal to the total given rate, such that 
\begin{equation}
\begin{aligned}  
   n_\nu \times q(\nu \rightarrow \nu') = &\sum_j n_{\nu, j} \times q(\nu, j \rightarrow \nu', j)  \\
   =& \sum_j n_{\nu, j} \times q(\nu \rightarrow \nu' ),
\end{aligned}
\end{equation}
for a vibrational state $\nu$ with rotational levels $j$. 

Data for molecule-electron collisional excitation rates, are only available for CO, to the authors' knowledge. 
For lack of any better alternative, the collisional transitions rate coefficients derived from both data and scaling laws for CO, were also used for SiO, SiS and SO.

\subsection{Rotational levels}

The typical energies of pure rotational transitions are at least an order of magnitude smaller than transitions between rovibrational levels, and the rotational levels within a vibration level will be in relative LTE in the investigated environment, as mentioned. 
This means that the internal distribution of rotational levels, within a vibration level, can be described by the Boltzmann distribution.
In practice, the relative LTE is enforced by assuming that 1) all rotational levels within a vibrational level are connected through collisional rate coefficients $C_{\nu, j \rightarrow \nu, j'}$, and 2) these collisional rates are large enough to ensure relative LTE. 
We arbitrarily picked $C_{\nu, j \rightarrow \nu, j'}$=1000 s$^{-1}$, which is about 1000 times larger than the largest collision-induced transitions between vibrational levels.
Through testing, we have deduced that this value always enforced relative LTE in the rotational levels within a vibrational level, for the parameter space used in this work. 

% Due to the smaller transition energies between pure rotation levels, should be large enough to always  

% a large collisional rate for pure rotation collision transitions $C_{\nu, j \rightarrow \nu, j'}$, within a vibrational band. 
% We here use a value of 1000 times larger than the largest collision-induced transitions between vibrational levels.  

\section{Compton electron collision cross-sections}
\label{app:el_col}

\begin{table}
\caption{Constants of fits to electron collision cross-sections, with references to the original data. The units of both $\sigma_{Compton}$ and a, b, c, d are cm$^{2}$.} title of Table
\label{table:cs_compton}      % is used to refer this table in the text
\centering                          % used for centering table
\begin{tabular}{cccccc}
\hline
Species & a          & b          & c         & d & Ref           \\ \hline
CO      & -5.92e-13 & -9.29e-17 & 3.64e-13 & 2.40e-13 & (1) \\
SiO     & -9.39e-13 & 2.43e-16  & 7.45e-13 & 2.52e-13 & (2)   \\
SiS     & -1.00e-12 & 1.06e-14  & 8.27e-13 & 4.22e-13   & (2) \\
SO      & -1.44e-12 & 4.77e-15  & 8.98e-13 & 6.30e-13   & (3)
\end{tabular}
\tablebib{(1)\citet{itikawa_cross_2015} (2) \citet{joshipuraElectronImpactIonization2007} (3) \citet{joshipura_electron_2008-1}
}
\end{table}

\begin{figure}
\centering
\includegraphics[width=\hsize]{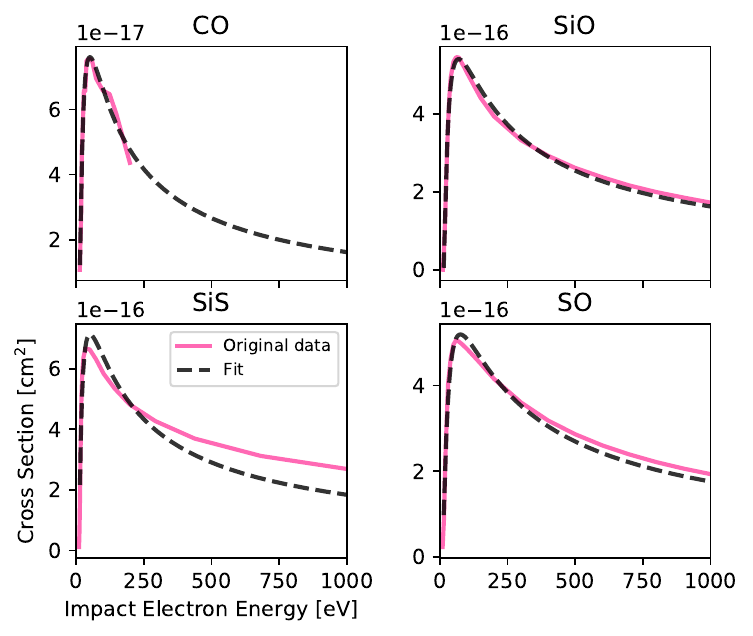}
\caption{Overview of the cross-sections of included molecular species, with the original data and the fits, as described in the text. }
\label{fig:cs_compton}
\end{figure}

We use the form defined in \citet{Arnaud1985} for the Compton electron collision cross-sections $\sigma_{Compton}$[cm$^2$] in \textsc{Sumo} as
\begin{equation}
    \sigma_{Compton} = \frac{1}{u \times i^2}  \left (a \left(1 - \frac{1}{u}\right) + b \left(1 + \frac{1}{u}\right)^2 + c \log (u) + \frac{d \log{u}}{u}\right),
\end{equation}
where $u$ is the impact electron energy $e$ normalized by the ionization energy $i$ such that $u = e/i$, and $a$, $b$, $c$, and $d$ are constants specific for the cross-section for each species. 
We fit this function to the cross-sections given in \citet{itikawa_cross_2015} for CO, \citet{joshipuraElectronImpactIonization2007} for SiO and SiS, and \citet{joshipura_electron_2008-1} for SO. 
The values for $a$, $b$, $c$, and $d$ for the molecular species included in the radiative transfer calculations are given in Table \ref{table:cs_compton}
These fits and the original data are plotted in Fig. \ref{fig:cs_compton}.

\section{Ionization energies}
\label{app:ion_en}

\begin{table}
\caption{Ionization energies for molecules. }             % title of Table
\label{tab:ionization_energies}      % is used to refer this table in the text
\centering      % used for centering table
\def\arraystretch{1.2}
\begin{tabular}{l l c}        % centered columns (4 columns)
\hline\hline                 % inserts double horizontal
Species & I.E. [eV] & Ref\\ \hline
\ch{CO} & 14.0&(1)\\
\ch{CO+} & 41.8&(2)\\
\ch{SiO} & 11.6&(1)\\
\ch{SiO+} & 20.6$^\star$&(3)\\
\ch{SiS} & 10.5&(1)\\
\ch{SiS+} & 17.4$^\star$&(3)\\
\ch{SO} & 10.3&(1)\\
\ch{SO+} & 19.4$^\star$&(3)\\
\end{tabular}
\tablefoot{$^\star$ Ionization energies calculated in this work.}
\tablebib{(1)~NIST Chemistry WebBook (\href{https://webbook.nist.gov/}{https://webbook.nist.gov/}) (2)~\citet{hille_cross_1978} (3)~See text
}
\end{table}

The ionization energies of CO, SiO, SiS, SO, \ch{CO+}, \ch{SiO+}, \ch{SiS+} and \ch{SO+} are needed for the calculation of ionization rates.  Literature values have been used for CO, \ch{CO+}, SiO, SiS, and SO. For \ch{SiO+}, \ch{SiS+} and \ch{SO+}
the values have been calculated here using the MOLPRO package \cite{MOLPRO2012, MOLPRO2020}. Vertical ionization energies were obtained by performing full valence MRCI calculations using the aug-cc-pV5Z basis sets. Cluster corrected, Davidson relaxed, energies are reported.  

We first calculated the equilibrium bond lengths for the \ch{SiO+}, \ch{SiS+} and \ch{SO+} doublet spin states. These bond lengths were then used to calculate the energies for the corresponding doubly ionized species. The lower energies, at the given bond lengths, of the singlet and triplet states for SiO$^{++}$, SiS$^{++}$ and SO$^{++}$ were used to obtain the ionization energies, i.e. the singlet state for SO$^{++}$ and the triplet states for SiO$^{++}$ and SiS$^{++}$. In this way the vertical ionization energies were found to be 20.6, 17.4 and 19.4 eV for \ch{SiO+}, \ch{SiS+} and \ch{SO+} respectively. The ionization energies used in \textsc{Sumo} are listed in Table~\ref{tab:ionization_energies}.

%\fixme{Methods}

\section{Reactions}
\label{app:reactions}
The following tables contain the parameters used to calculate the reaction rates used in this work, divided into thermal collision reactions in Tables \ref{table:rc1}, \ref{table:rc2} and \ref{table:rc3}, and recombination reactions in Table \ref{table:recomb1}. When possible we refer to the original work, in other cases the database from which the value is from is indicated.

Using the Eq.~\eqref{eq:arr}, the reaction rate $k(T)$ for a specific reaction can be retained. 
The units of both $\alpha$ and $k(T)$ depend on the number of reactants: s$^{-1}$ for unimolecular reactions and cm$^{3}$s$^{-1}$ for bimolecular reactions.
The units of $\gamma$ is K, and $\beta$ is a dimensionless number.

%XXX I am not able to figure out why you have gamma zero for RC92-RC94 or how I understand the three temperature ranges where the parameters apply - but maybe that will become clear when text is added in this section XXX

\begin{table*}
\caption{Reaction rate coefficients for thermal collisions. }
\label{table:rc1}      
\centering       
\def\arraystretch{1.2}
\begin{tabular}{llrclrrrl}
\hline \hline
 ID   & Reaction type         &            &     Reaction      &              &   $\alpha$ &   $\beta$ &   $\gamma$ & Ref.    \\
\hline
RC1&Charge Exchange&\ch{C2 + O2+ }&\ch{->}&\ch{C2+  + O2}&4.1E-10&0&0&(1) \\
RC2&Charge Exchange&\ch{C2 + O+ }&\ch{->}&\ch{C2+  + O}&4.8E-10&0&0&(1) \\
RC3&Charge Exchange&\ch{C2+  + S}&\ch{->}&\ch{C2 + S+ }&5.8E-10&0&0&(1) \\
RC4&Charge Exchange&\ch{C + C2+ }&\ch{->}&\ch{C+  + C2}&1.1E-10&0&0&(1) \\
RC5&Charge Exchange&\ch{C + CO+ }&\ch{->}&\ch{C+  + CO}&1.1E-10&0&0&(1) \\
RC6&Charge Exchange&\ch{C + He+ }&\ch{->}&\ch{C+  + He}&6.3E-15&0.75&0&(2) \\
RC7&Charge Exchange&\ch{C + O2+ }&\ch{->}&\ch{C+  + O2}&5.2E-11&0&0&(1) \\
RC8&Charge Exchange&\ch{C+  + C2O}&\ch{->}&\ch{C + C2O+ }&1E-09&-0.5&0&(3) \\
RC9&Charge Exchange&\ch{C+  + S}&\ch{->}&\ch{C + S+ }&5E-11&0&0&(4) \\
RC10&Charge Exchange&\ch{C+  + SO}&\ch{->}&\ch{C + SO+ }&2.6E-10&-0.5&0&(1) \\
RC11&Charge Exchange&\ch{C+  + Si}&\ch{->}&\ch{C + Si+ }&2.1E-09&0&0&(1) \\
RC12&Charge Exchange&\ch{C+  + SiS}&\ch{->}&\ch{C + SiS+ }&2.3E-09&-0.5&0&(1) \\
RC13&Charge Exchange&\ch{CO2 + He+ }&\ch{->}&\ch{CO2+  + He}&1.21E-10&0&0&(5) \\
RC14&Charge Exchange&\ch{CO2+  + O2}&\ch{->}&\ch{CO2 + O2+ }&5.3E-11&0&0&(6) \\
RC15&Charge Exchange&\ch{CO2+  + O}&\ch{->}&\ch{CO2 + O+ }&9.62E-11&0&0&(7) \\
RC16&Charge Exchange&\ch{CO + O+ }&\ch{->}&\ch{CO+  + O}&4.9E-12&0.5&4580&(8) \\
RC17&Charge Exchange&\ch{CO+  + C2}&\ch{->}&\ch{CO + C2+ }&8.4E-10&0&0&(1) \\
RC18&Charge Exchange&\ch{CO+  + CO2}&\ch{->}&\ch{CO + CO2+ }&1E-09&0&0&(9) \\
RC19&Charge Exchange&\ch{CO+  + O2}&\ch{->}&\ch{CO + O2+ }&1.2E-10&0&0&(9) \\
RC20&Charge Exchange&\ch{CO+  + O}&\ch{->}&\ch{CO + O+ }&1.4E-10&0&0&(7) \\
RC21&Charge Exchange&\ch{CO+  + S}&\ch{->}&\ch{CO + S+ }&1.1E-09&0&0&(1) \\
RC22&Charge Exchange&\ch{He+  + O2}&\ch{->}&\ch{He + O2+ }&3.3E-11&0&0&(5) \\
RC23&Charge Exchange&\ch{He+  + Si}&\ch{->}&\ch{He + Si+ }&3.3E-09&0&0&(1) \\
RC24&Charge Exchange&\ch{O2+  + S}&\ch{->}&\ch{O2 + S+ }&5.4E-10&0&0&(1) \\
RC25&Charge Exchange&\ch{O2+  + Si}&\ch{->}&\ch{O2 + Si+ }&1.6E-09&0&0&(1) \\
RC26&Charge Exchange&\ch{O+  + O2}&\ch{->}&\ch{O + O2+ }&1.9E-11&0&0&(10) \\
RC27&Charge Exchange&\ch{S+  + Si}&\ch{->}&\ch{S + Si+ }&1.6E-09&0&0&(1) \\
RC28&Charge Exchange&\ch{S+  + SiS}&\ch{->}&\ch{S + SiS+ }&3.2E-09&-0.5&0&(1) \\
RC29&Ion-Neutral&\ch{C2 + He+ }&\ch{->}&\ch{C + C+  + He}&1.6E-09&0&0&(1) \\
RC30&Ion-Neutral&\ch{C2 + O2+ }&\ch{->}&\ch{CO + CO+ }&4.1E-10&0&0&(1) \\
RC31&Ion-Neutral&\ch{C2 + O+ }&\ch{->}&\ch{C + CO+ }&4.8E-10&0&0&(1) \\
RC32&Ion-Neutral&\ch{C2+  + O2}&\ch{->}&\ch{CO + CO+ }&8E-10&0&0&(1) \\
RC33&Ion-Neutral&\ch{C2+  + O}&\ch{->}&\ch{C + CO+ }&3.1E-10&0&0&(11) \\
RC34&Ion-Neutral&\ch{C2O + He+ }&\ch{->}&\ch{C+  + CO}&1E-09&-0.5&0&(3) \\
RC35&Ion-Neutral&\ch{C + O2+ }&\ch{->}&\ch{CO+  + O}&5.2E-11&0&0&(1) \\
RC36&Ion-Neutral&\ch{C + SiO+ }&\ch{->}&\ch{CO + Si+ }&1E-09&0&0&(12) \\
RC37&Ion-Neutral&\ch{C+  + CO2}&\ch{->}&\ch{CO + CO+ }&1.1E-09&0&0&(13) \\
RC38&Ion-Neutral&\ch{C+  + O2}&\ch{->}&\ch{CO + O+ }&4.54E-10&0&0&(14) \\
RC39&Ion-Neutral&\ch{C+  + O2}&\ch{->}&\ch{CO+  + O}&3.42E-10&0&0&(14) \\
RC40&Ion-Neutral&\ch{C+  + SO}&\ch{->}&\ch{CO + S+ }&2.6E-10&-0.5&0&(1) \\
RC41&Ion-Neutral&\ch{C+  + SO}&\ch{->}&\ch{CO+  + S}&2.6E-10&-0.5&0&(1) \\
RC42&Ion-Neutral&\ch{C+  + SiO}&\ch{->}&\ch{CO + Si+ }&5.4E-10&-0.5&0&(12) \\
RC43&Ion-Neutral&\ch{CO2 + He+ }&\ch{->}&\ch{C + O2+ }&1.1E-11&0&0&(5) \\
RC44&Ion-Neutral&\ch{CO2 + He+ }&\ch{->}&\ch{C+  + O2}&4E-11&0&0&(15) \\
RC45&Ion-Neutral&\ch{CO2 + He+ }&\ch{->}&\ch{CO + O+ }&1E-10&0&0&(5) \\
RC46&Ion-Neutral&\ch{CO2 + He+ }&\ch{->}&\ch{CO+  + He + O}&8.7E-10&0&0&(5) \\
RC47&Ion-Neutral&\ch{CO2 + O+ }&\ch{->}&\ch{CO + O2+ }&9.4E-10&0&0&(10) \\
RC48&Ion-Neutral&\ch{CO2+  + O}&\ch{->}&\ch{CO + O2+ }&1.64E-10&0&0&(7) \\
\hline
\end{tabular}
\end{table*}

\begin{table*}
\caption{Reaction rate coefficients for thermal collisions, continued. }
\label{table:rc2}      
\centering       
\def\arraystretch{1.2}
\begin{tabular}{llrclrrrl}
\hline \hline
 ID   & Reaction type         &            &     Reaction      &              &   $\alpha$ &   $\beta$ &   $\gamma$ & Ref.    \\
\hline
RC49&Ion-Neutral&\ch{CO + He+ }&\ch{->}&\ch{C+  + He + O}&1.6E-09&0&0&(15) \\
RC50&Ion-Neutral&\ch{CO + SiO+ }&\ch{->}&\ch{CO2 + Si+ }&7.9E-10&0&0&(12) \\
RC51&Ion-Neutral&\ch{He+  + O}&\ch{->}&\ch{He + O+ }&7.6-15&-0.05&-4.3&(16) \\
RC52&Ion-Neutral&\ch{He+  + O2}&\ch{->}&\ch{He + O + O+ }&1.1E-09&0&0&(5) \\
RC53&Ion-Neutral&\ch{He+  + S2}&\ch{->}&\ch{He + S + S+ }&2E-09&0&0&(3) \\
RC54&Ion-Neutral&\ch{He+  + SO}&\ch{->}&\ch{He + O + S+ }&8.3E-10&-0.5&0&(1) \\
RC55&Ion-Neutral&\ch{He+  + SO}&\ch{->}&\ch{He + O+  + S}&8.3E-10&-0.5&0&(1) \\
RC56&Ion-Neutral&\ch{He+  + SiO}&\ch{->}&\ch{He + O + Si+ }&8.6E-10&-0.5&0&(1) \\
RC57&Ion-Neutral&\ch{He+  + SiO}&\ch{->}&\ch{He + O+  + Si}&8.6E-10&-0.5&0&(1) \\
RC58&Ion-Neutral&\ch{He+  + SiS}&\ch{->}&\ch{He + S + Si+ }&3.8E-09&-0.5&0&(1) \\
RC59&Ion-Neutral&\ch{He+  + SiS}&\ch{->}&\ch{He + S+  + Si}&3.8E-09&-0.5&0&(1) \\
RC60&Ion-Neutral&\ch{O2 + S+ }&\ch{->}&\ch{O + SO+ }&1.5E-11&0&0&(17) \\
RC61&Ion-Neutral&\ch{O2 + SiS+ }&\ch{->}&\ch{SO + SiO+ }&2.67E-11&0&0&(18) \\
RC62&Ion-Neutral&\ch{O2 + SiS+ }&\ch{->}&\ch{SO+  + SiO}&6.23E-11&0&0&(18) \\
RC63&Ion-Neutral&\ch{O2+  + S}&\ch{->}&\ch{O + SO+ }&5.4E-10&0&0&(1) \\
RC64&Ion-Neutral&\ch{O + SiO+ }&\ch{->}&\ch{O2 + Si+ }&2E-10&0&0&(19) \\
RC65&Ion-Neutral&\ch{S + SiO+ }&\ch{->}&\ch{SO + Si+ }&1E-09&0&0&(1) \\
RC66&Neutral-Neutral&\ch{C2 + O2}&\ch{->}&\ch{CO + CO}&1.5E-11&0&4300&(20) \\
RC67&Neutral-Neutral&\ch{C2 + O}&\ch{->}&\ch{C + CO}&2E-10&-0.12&0&(4) \\
RC68&Neutral-Neutral&\ch{C2O + O}&\ch{->}&\ch{CO + CO}&8.59E-11&0&0&(21) \\
RC69&Neutral-Neutral&\ch{C + C2O}&\ch{->}&\ch{CO + C2}&2E-10&0&0&(22) \\
RC70&Neutral-Neutral&\ch{C + CO2}&\ch{->}&\ch{CO + CO}&1E-15&0&0&(23) \\
RC71&Neutral-Neutral&\ch{C + CO}&\ch{->}&\ch{C2 + O}&2.94E-11&0.5&58025&(24) \\
RC72&Neutral-Neutral&\ch{C + O2}&\ch{->}&\ch{CO + O}&5.56E-11&0.41&-26.9&(22) \\
RC73&Neutral-Neutral&\ch{C + SiO}&\ch{->}&\ch{CO + Si}&1E-16&0&0&(25) \\
RC74&Neutral-Neutral&\ch{C + SO}&\ch{->}&\ch{CO + S}&3.5E-11&0&0&(22) \\
RC75&Neutral-Neutral&\ch{CO2 + O}&\ch{->}&\ch{CO + O2}&2.46E-11&0&26567&(26) \\
RC76&Neutral-Neutral&\ch{CO2 + Si}&\ch{->}&\ch{CO + SiO}&2.72E-11&0&282&(27) \\
RC77&Neutral-Neutral&\ch{CO + O}&\ch{->}&\ch{C + O2}&1E-16&0&0&(25) \\
RC78&Neutral-Neutral&\ch{CO + O2}&\ch{->}&\ch{CO2 + O}&5.99E-12&0&24075&(28) \\
RC79&Neutral-Neutral&\ch{CO + S}&\ch{->}&\ch{C + SO}&1E-16&0&0&(25) \\
RC80&Neutral-Neutral&\ch{CO + Si}&\ch{->}&\ch{C + SiO}&1.3E-09&0&34513&(27) \\
RC81&Neutral-Neutral&\ch{CO + SiO}&\ch{->}&\ch{CO2 + Si}&1E-16&0&0&(25) \\
RC82&Neutral-Neutral&\ch{O2 + S}&\ch{->}&\ch{O + SO}&1.76E-12&0.81&-30.8&(29) \\
RC83&Neutral-Neutral&\ch{O2 + Si}&\ch{->}&\ch{O + SiO}&1.72E-10&-0.53&17&(30) \\
RC84&Neutral-Neutral&\ch{O + S2}&\ch{->}&\ch{S + SO}&1.7E-11&0&0&(31) \\
RC85&Neutral-Neutral&\ch{O + SO}&\ch{->}&\ch{O2 + S}&6.6E-13&0&2760&(32) \\
RC86&Neutral-Neutral&\ch{O + SiO}&\ch{->}&\ch{O2 + Si}&1E-16&0&0&(25) \\
RC87&Neutral-Neutral&\ch{S + SiS}&\ch{->}&\ch{S2 + Si}&5.75E-11&0.1&200&(33) \\
RC88&Neutral-Neutral&\ch{S + SO}&\ch{->}&\ch{O + S2}&1.73E-11&0.5&11500&(24) \\
RC89&Neutral-Neutral&\ch{S2 + Si}&\ch{->}&\ch{S + SiS}&7E-11&0&0&(22) \\
RC90&Radiative Association&\ch{C + C}&\ch{->}&\ch{C2}&4.36E-18&0.35&161.3&(34) \\
RC91&Radiative Association&\ch{C + C+ }&\ch{->}&\ch{C2+ }&4.01E-18&0.17&101.5&(34) \\
RC92&Radiative Association&\ch{C + O}&\ch{->}&\ch{CO}&1.48E-17&0.5&0&(35) \\
RC93&Radiative Association&\ch{C+  + O}&\ch{->}&\ch{CO+ }&1.1E-17&0.11&-121.5&(36) \\
RC94&Radiative Association&\ch{O + O}&\ch{->}&\ch{O2}&4.9E-20&1.6&0&(1) \\
RC95&Radiative Association&\ch{O + S}&\ch{->}&\ch{SO}&1.1E-19&0.28&1298&(37) \\
RC96&Radiative Association&\ch{O + Si}&\ch{->}&\ch{SiO}&3.24E-17&0.31&-21.2&(38) \\
\hline
\end{tabular}
\end{table*}

\begin{table*}
\caption{Reaction rate coefficients for thermal collisions, continued. }
\label{table:rc3}      
\centering       
\def\arraystretch{1.2}
\begin{tabular}{llrclrrrl}
\hline \hline
 ID   & Reaction type         &            &     Reaction      &              &   $\alpha$ &   $\beta$ &   $\gamma$ & Ref.    \\
\hline
RC97&Radiative Association&\ch{O + Si+ }&\ch{->}&\ch{SiO+ }&9.22E-19&-0.08&-21.2&(38) \\
RC98&Radiative Association&\ch{S + S}&\ch{->}&\ch{S2}&1.37E-19&0.3&-79&(37) \\
RC99&Radiative Association&\ch{S + Si}&\ch{->}&\ch{SiS}&1.047E-16&0.3&66&(37) \\
\hline
\end{tabular}
\tablefoot{The original work is referenced when possible. The main sources are the rate coefficient compilation databases UMIST Database for Astrochemistry (\citealt{mcelroyUMISTDatabaseAstrochemistry2013}, \href{http://udfa.ajmarkwick.net/}{www.astrochemistry.net}), Kinetic Database for Astrochemistry KIDA (\citealt{wakelamKINETICDATABASEASTROCHEMISTRY2012}, \href{http://kida.obs.u-bordeaux1.fr}{http://kida.obs.u-bordeaux1.fr}) and the chemical kinetic database of NIST (\citealt{NIST}, \href{https://kinetics.nist.gov/}{https://kinetics.nist.gov/}). 
The reaction network was additionally supplemented by reaction rates from \citet{sluderMolecularNucleationTheory2018} and \citet{cherchneffChemistryPopulationIII2009}, and by updated coefficient rates found by literature search. The chemical network used is also available in a .csv format on \href{https://github.com/sliljegren/chemical_network}{https://github.com/sliljegren/chemical\_network}.}

\tablebib{(1)~\citet{Prasad_1980};
(2)~\citet{Smith_1993};
(3) UMIST ~\citet{mcelroyUMISTDatabaseAstrochemistry2013};
(4) KIDA ~\citet{wakelamKINETICDATABASEASTROCHEMISTRY2012};
(5)~\citet{Adams_1976};
(6)~\citet{Copp_1982};
(7)~\citet{Fehsenfeld_1972};
(8)~\citet{petuchowskiCOFormationMetalrich1989};
(9)~\citet{Adams_1978};
(10)~\citet{Adams_1980};
(11)~\citet{Viggiano_1980};
(12)~\citet{herbst_chemistry_1989};
(13)~\citet{Fahey_1981};
(14)~\citet{Martinez_Jr__2008};
(15)~\citet{Anicich_1977};
(16)~\citet{zhao_radiative_2004};
(17)~\citet{smith_1981};
(18)~\citet{wlodek_gas-phase_1989};
(19)~\citet{Fehsenfeld_1969};
(20)~\citet{fontijn_2001};
(21)~\citet{bauer_laser_1985};
(22)~\citet{Smith_2004};
(23)~\citet{husain_kinetic_1975};
(24)~\citet{Mitchell_1984};
(25)~\citet{cherchneffChemistryPopulationIII2009};
(26)~\citet{tsang_1986};
(27)~\citet{mick_shock_1994};
(28)~\citet{koike_shock_1991};
(29)~\citet{lu_experimental_2004};
(30)~\citet{Le_Picard_2002};
(31)~\citet{Singleton_1988};
(32)~\citet{Leen_1988};
(33)~\citet{mick_high-temperature_1994};
(34)~\citet{andreazzaFormationSi2C21997};
(35)~\citet{gustafssonRadiativeAssociationRate2015};
(36)~\citet{zamecnikovaFormationCORadiative2019};
(37)~\citet{andreazzaFormationS2Radiative2005};
(38)~\citet{andreazzaRadiativeAssociationSi1995}}
\end{table*}

\begin{table*}
\caption{Coefficients for the molecular recombination reactions.}             
\label{table:recomb1}      
\centering       
  
  \begin{tabular}{llrclrrrl}
\hline
 ID   & Reaction type         & &Reaction&                                    &   $\alpha$ &   $\beta$ &   E$_a$ & Ref.    \\
\hline
 RR1  & Dissociate recombination       & \ch{CO+ + e- } & \ch{->} & \ch{C + O + h$\nu$}   &   2.36e-12  &      -0.29   &     -17.6   & (1) \\
 RR2  & Dissociate recombination       & \ch{CO2+ + e- }& \ch{ ->} & \ch{CO + O + h$\nu$}   &   3.8e-07  &      -0.5    &     0   & (1) \\
 RR3  & Dissociate recombination       & \ch{C2+ + e-  }& \ch{->} & \ch{C + C + h$\nu$}   &   3e-07  &      -0.5    &     0   & (1) \\
 RR4  & Dissociate recombination       & \ch{C2O+ + e- }& \ch{ ->} & \ch{CO + O + h$\nu$}   &   3e-07  &      -0.5    &     0   & (2) \\
 RR5  & Dissociate recombination       & \ch{O2+ + e- } & \ch{->} & \ch{O + O + h$\nu$}   &   1.95e-07 & -0.7    &     0   & (3) \\
RR6  & Dissociate recombination       & \ch{S2+ + e- } & \ch{->} & \ch{S + S + h$\nu$} & 2e-07 & -0.5 & 0& (4)\\
RR7 & Dissociate recombination       & \ch{SiO+ + e- } & \ch{->} & \ch{Si + O + h$\nu$} & 2e-07 & -0.5 & 0& (4) \\
RR8  & Dissociate recombination       & \ch{SiS+ + e- } & \ch{->} & \ch{Si + S + h$\nu$} & 2e-07 & -0.5 & 0& (4) \\
RR9  & Dissociate recombination       & \ch{SO+ + e- } & \ch{->} & \ch{S + O + h$\nu$} & 2e-07 & -0.5 & 0& (4)\\
\hline
\end{tabular}
\tablefoot{When possible the original works are referenced.
The coefficients were obtained from the UMIST database (\citealt{mcelroyUMISTDatabaseAstrochemistry2013}, \href{http://udfa.ajmarkwick.net/}{www.astrochemistry.net}).}
\tablebib{(1)~\citet{brianDissociativeRecombinationMolecular1990};
(2) \citet{algeMeasurementsDissociativeRecombination1983}; \citet{mcelroyUMISTDatabaseAstrochemistry2013}; (4) \citet{Prasad_1980}\
}

\end{table*}

\clearpage

\section{Reaction rate plots}
\label{app:reaction_rate_plots}
 \begin{figure*}
  \centering
   \includegraphics[width=0.49\textwidth]{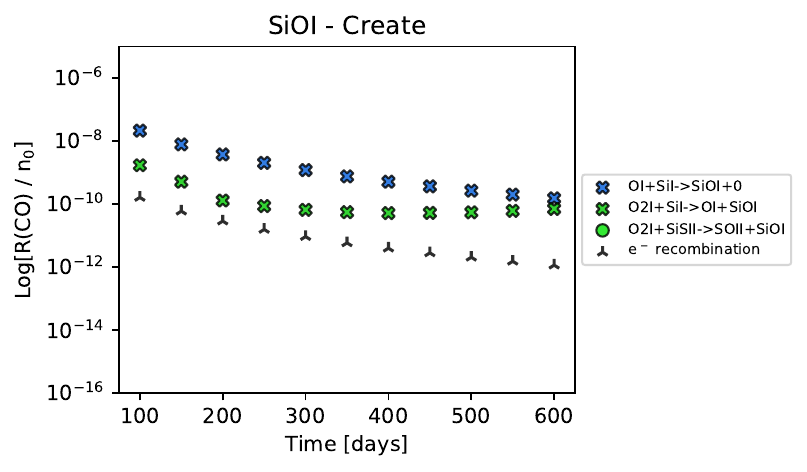}
   \includegraphics[width=0.49\textwidth]{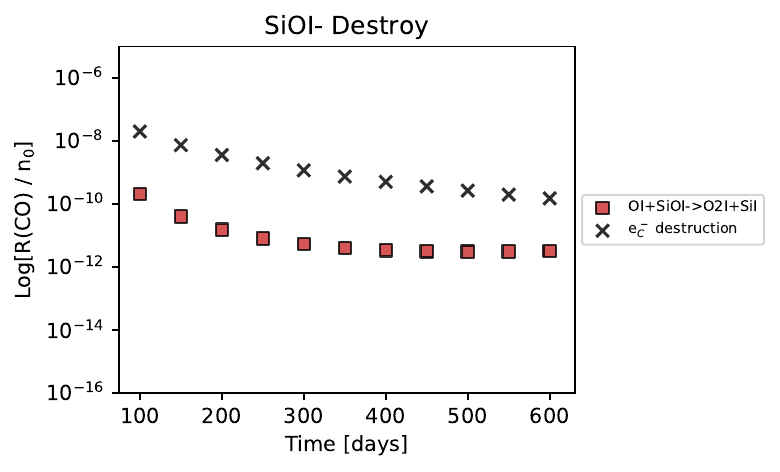}
   \includegraphics[width=0.49\textwidth]{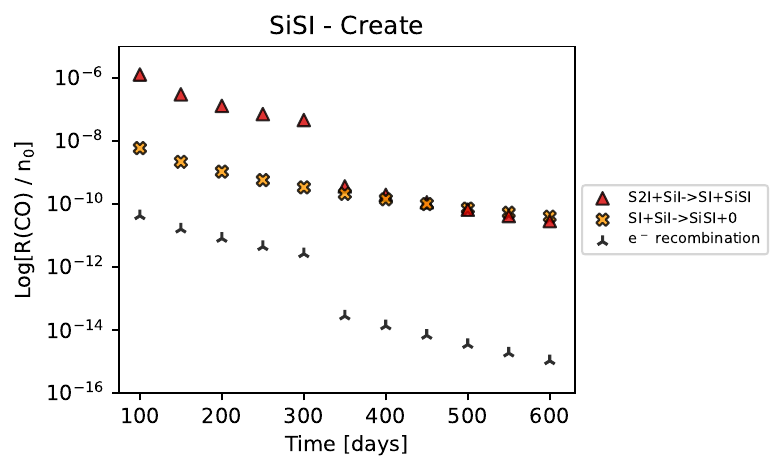}
   \includegraphics[width=0.49\textwidth]{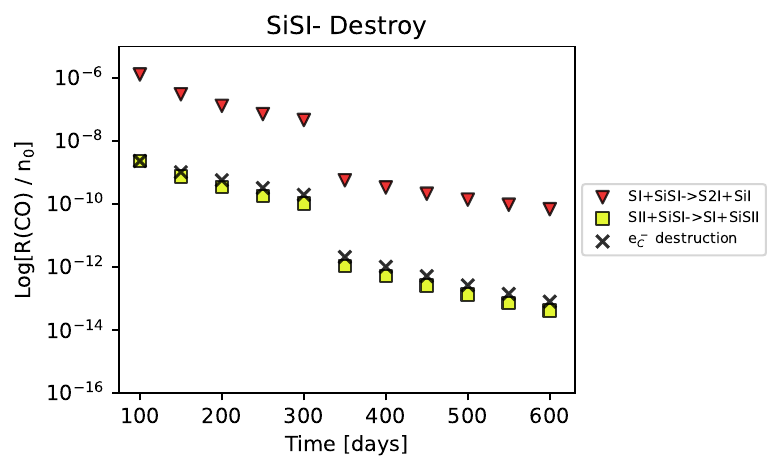}
   \includegraphics[width=0.49\textwidth]{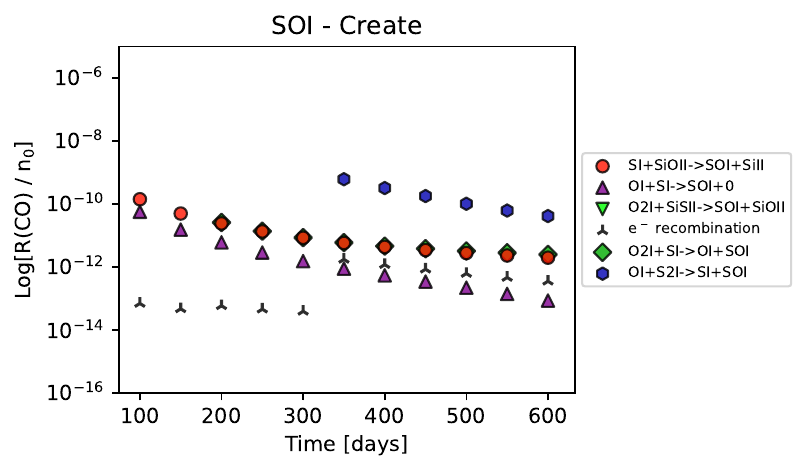}
\includegraphics[width=0.49\textwidth]{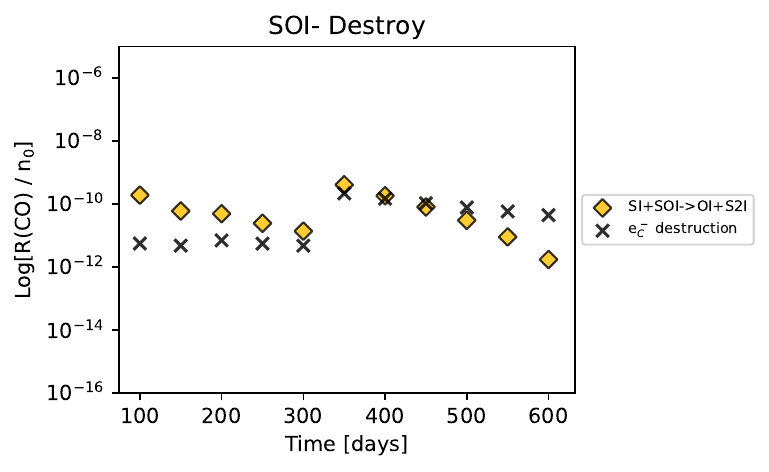}
\includegraphics[width=0.49\textwidth]{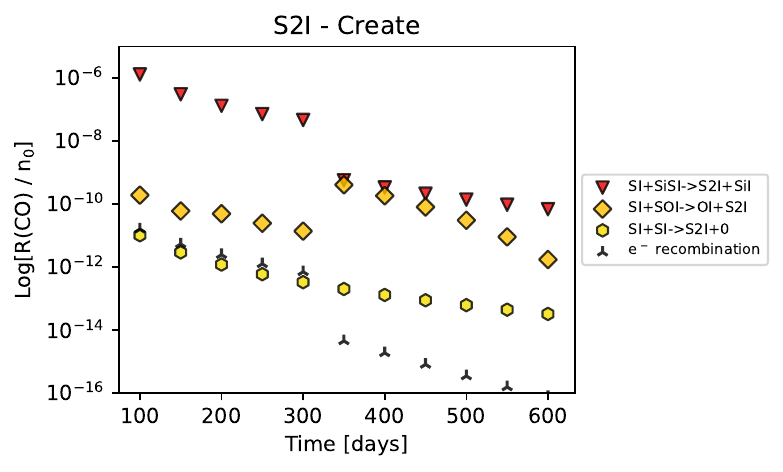}
\includegraphics[width=0.49\textwidth]{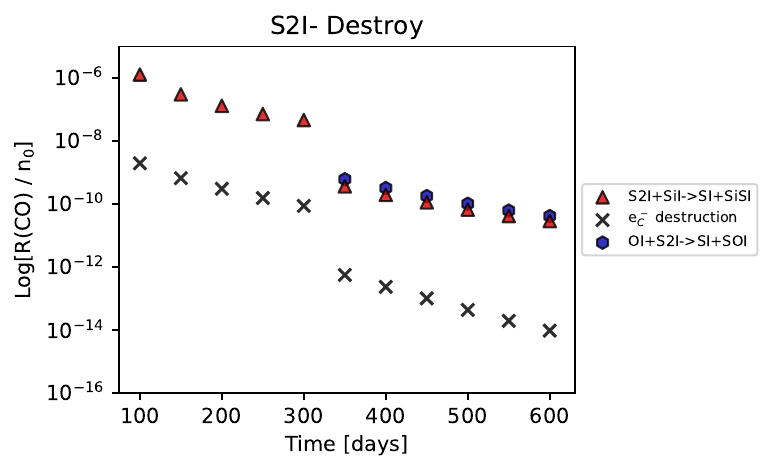}
\caption{Log of the reaction rates, divided by the density in the O/Si/S zone, with time, for the most abundant molecules in this zone. This indicates which rates are important for the formation and destruction of different molecules, with time. In these plots XI means the neutral atom or molecule, XII represent the once-ionized ion of an atom or molecule. }
         \label{fig:s_reaction_rates}
  \end{figure*}

 \begin{figure*}
  \centering
   \includegraphics[width=0.49\textwidth]{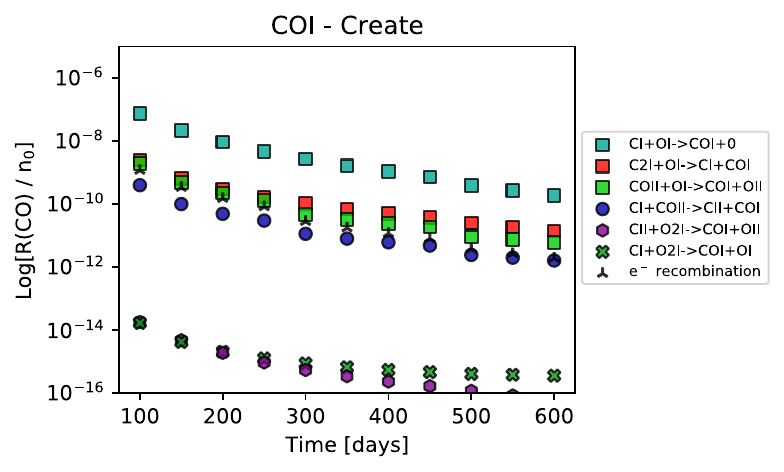}
   \includegraphics[width=0.49\textwidth]{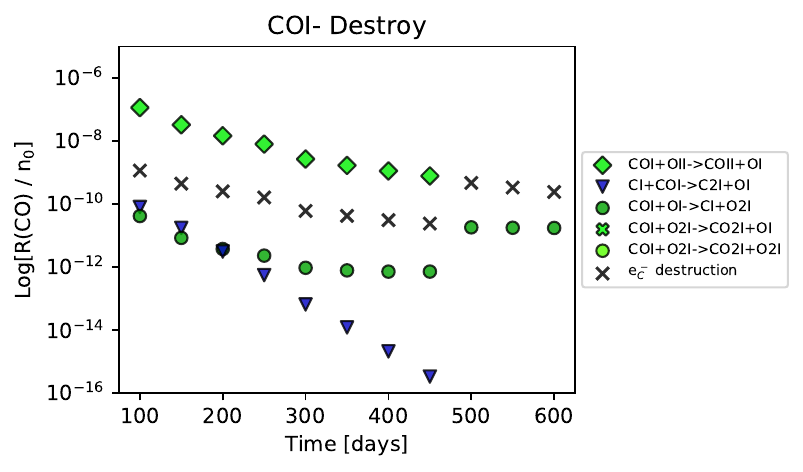}
\caption{The reaction rates divided with the density in the O/C zone, for formation and destruction processes of CO. In these plots XI means the neutral atom or molecule, XII represent the once-ionized ion of an atom or molecule.  }
         \label{fig:co_reaction_rates}
  \end{figure*}

The reaction rates the most abundant molecules in the standard model shown in Figs.~\ref{fig:s_reaction_rates} and \ref{fig:co_reaction_rates}. 
The rates of the sulfur and silicon species are from the O/Si/S zone, and the carbonaceous species are from the O/C zone, which are the zones where they are predominantly formed. 
It should be noted that these rates depend on several different properties; the rate coefficients, which can be temperature dependent, and the amount of available reacting species, which can change with time. 
These plots consequently do not necessarily reflect the changes of the physical conditions, but indicate which reactions are important for the formation and destruction of a specific species. 
For a discussion about the interpretation of these plots, see \citet{liljegrenCarbonMonoxideFormation2020}. 

\clearpage

\section{Spectra with different $\chi$, without molecules}

\begin{figure*}
% \label{fig:nomol_opt}
  \centering
   \includegraphics[width=0.99\textwidth]{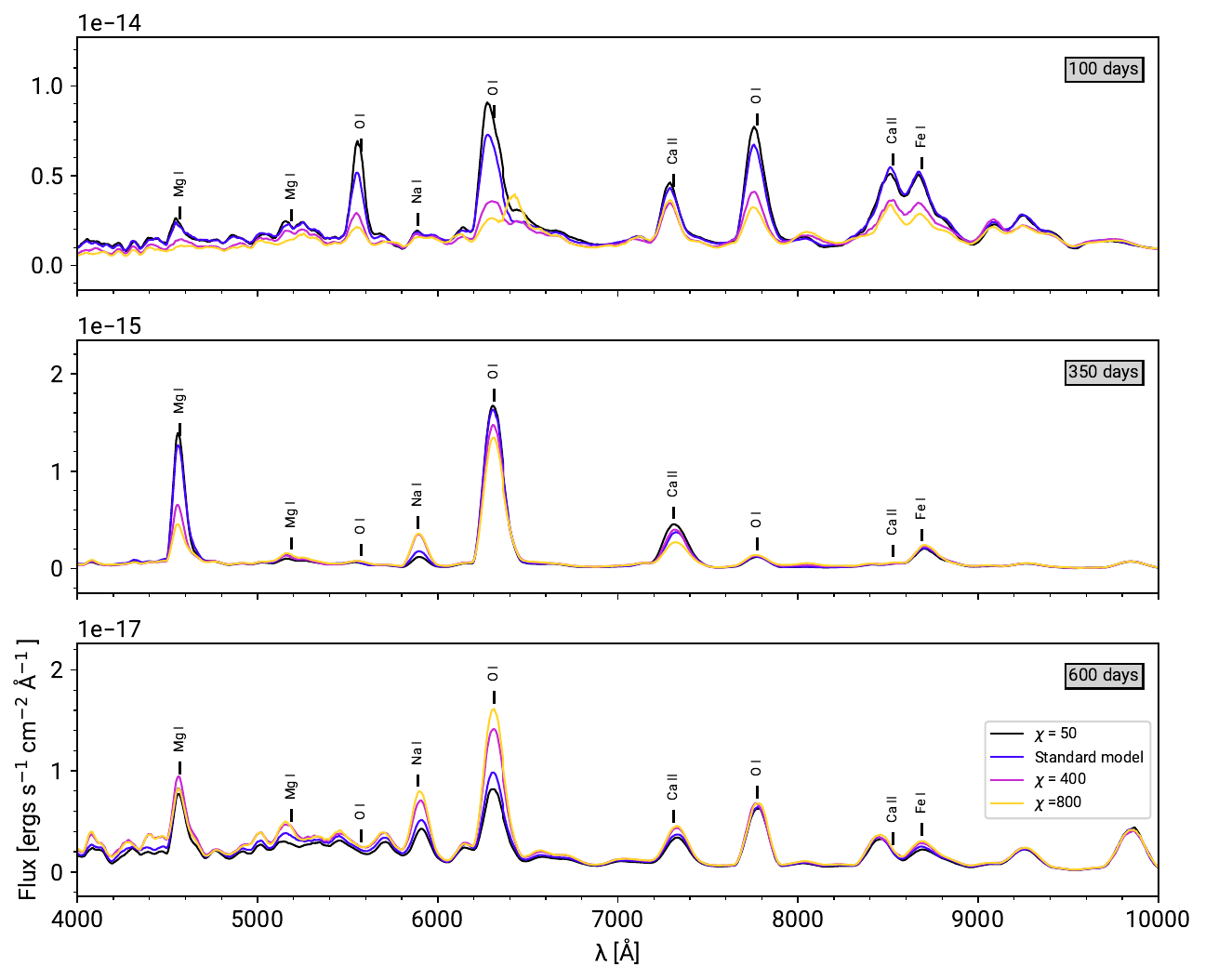}
\caption{The optical spectra for models with different densities and with no molecules included. }
         \label{fig:optical_nomol}
  \end{figure*}

\begin{figure*}
% \label{fig:nomol_ir}
  \centering
   \includegraphics[width=0.99\textwidth]{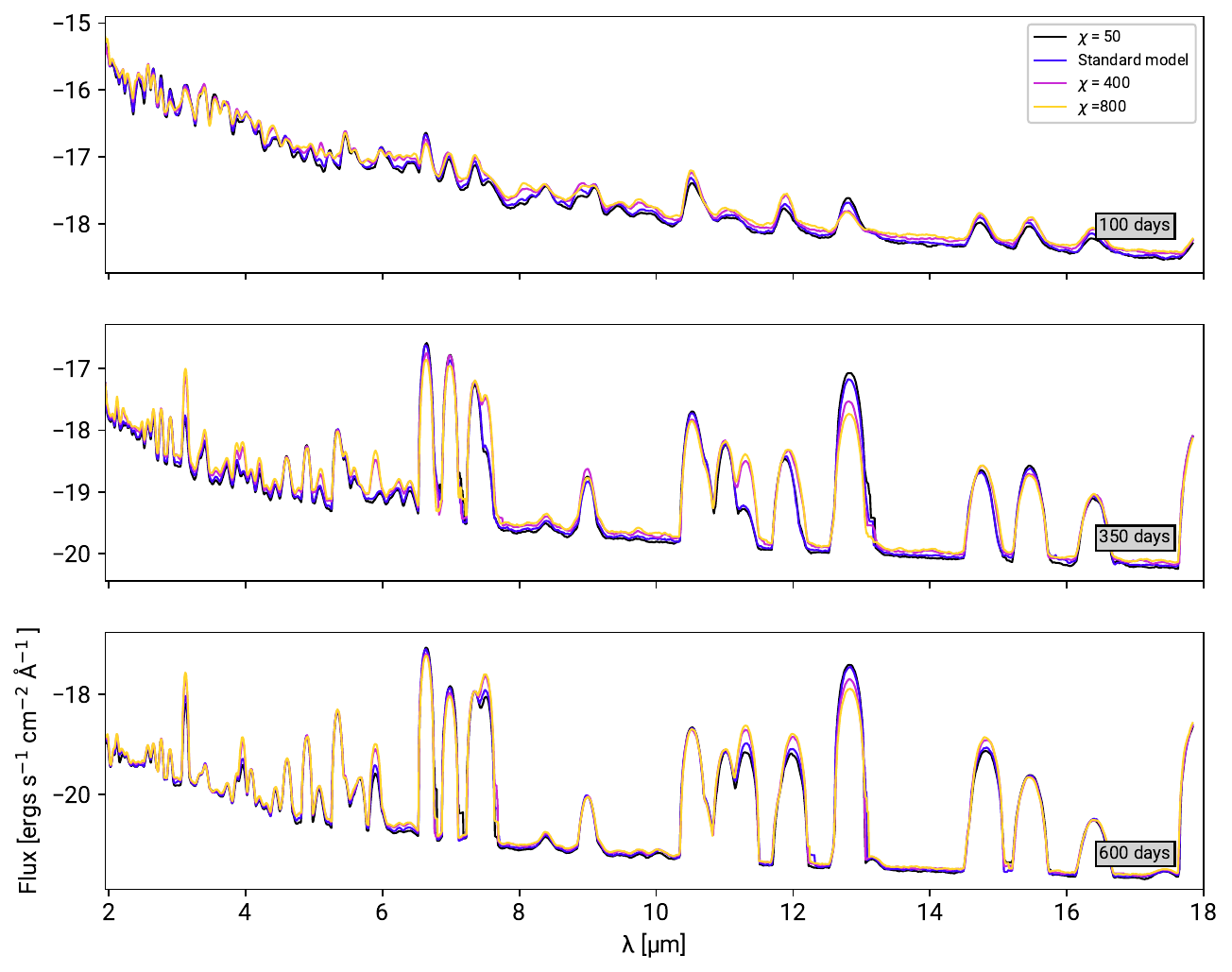}
\caption{The IR spectra for models with different densities when no molecules are included.}
%postsubmit \arjc{Maybe we can skip this?}
         \label{fig:ir_nomol}
  \end{figure*}

The differences in the spectra between models with varying densities can be either due to more or less molecules forming, which influences the cooling or due to higher ddensities themselves, which influences the emission. This is shown in Figs.~\ref{fig:optical_nomol} and \ref{fig:ir_nomol}, where models with different densities, but with no molecules included, are plotted.

\end{appendix}

\end{document}